\documentclass[a4paper]{article}
\pdfoutput=1
\usepackage{jheppub}
\usepackage[utf8]{inputenc}
\usepackage[T1]{fontenc}
\newcommand{\la}{\lambda_1}
\newcommand{\lb}{\lambda_2}
\newcommand{\lc}{\lambda_3}
\newcommand{\g}{\,\mathrm{GeV}}
\newcommand{\be}{\begin{equation}}
\newcommand{\ee}{\end{equation}}

\newcommand{\Vtree}{V^{(0)}}
\newcommand{\Vone}{V^{(1)}}
\newcommand{\f}{\varphi}
\newcommand{\ms}{\overline{\textrm{MS}}}
\newcommand{\og}{\mathcal{O}(g^2)}
\newcommand{\ogg}{\mathcal{O}(g^4)}
\newcommand{\mgw}{\mu_{\textrm{GW}}}
\newcommand{\gcw}{g_{X}}
\newcommand{\order}{\mathcal{O}}
\usepackage{amsmath}
\usepackage{amssymb}
\usepackage{amsthm}
\usepackage{amsfonts}
\usepackage[mathscr]{eucal}
\allowdisplaybreaks[3]
\usepackage{xcolor}
\usepackage{slashed}
\usepackage{siunitx} 

\usepackage{hyperref}			

\title{Systematic analysis of radiative symmetry breaking\\ in models with extended scalar sector}
\author[a,c]{Leonardo Chataignier,}
\author[a]{Tomislav Prokopec,}
\author[b]{Michael G. Schmidt}
\author[a,d]{and Bogumi\l a~\'Swie\.zewska}

\emailAdd{lcmr@thp.uni-koeln.de}
\emailAdd{t.prokopec@uu.nl}
\emailAdd{M.G.Schmidt@thphys.uni-heidelberg.de}
\emailAdd{b.swiezewska@uu.nl}

\affiliation[a]{Institute for Theoretical Physics, Spinoza Institute \& EMME$\Phi$, Utrecht University, \\Princetonplein 5, 3584 CC Utrecht, The Netherlands}
\affiliation[b]{Institute for Theoretical Physics, Universit\"{a}t Heidelberg, \\Philosophenweg 16, D-69120 Heidelberg, Germany}
\affiliation[c]{Institute for Theoretical Physics, University of Cologne, \\Z\"{u}lpicher Stra\ss e 77, 50937 K\"{o}ln, Germany}
\affiliation[d]{Faculty of Physics, University of Warsaw, \\Pasteura 5, 02-093 Warsaw, Poland}

\abstract{
Radiative symmetry breaking (RSB) is a theoretically appealing framework for the generation of mass scales through quantum effects. It can be successfully implemented in models with extended scalar and gauge sectors. We provide a systematic analysis of RSB in such models: we review the common approximative methods of studying RSB, emphasising their limits of applicability and discuss the relevance of the relative magnitudes of tree-level and loop contributions as well as the dependence of the results on the renormalisation scale.  The general considerations are exemplified within the context of the conformal Standard Model extended with a scalar doublet of a new SU(2)$_X$ gauge group, the so-called SU(2)cSM. We show that various perturbative methods of studying RSB may yield significantly different results due to renormalisation-scale dependence. Implementing the renormalisation-group (RG) improvement method recently developed in ref.~\cite{Chataignier:2018}, which is well-suited for multi-scale models, we argue that the use of the RG improved effective potential can alleviate this scale dependence providing more reliable results.}
\arxivnumber{1805.09292}

\begin{document}
\maketitle


\section{Introduction}

Dynamical generation of mass scales in models with classical scale symmetry via the mechanism of dimensional transmutation has been an important complement to the Brout--Englert--Higgs mechanism~\cite{Englert:1964, Higgs:1964, Higgs:1964-2, Higgs:1966} since the seminal papers of S.~Coleman and E.~Weinberg~\cite{Coleman:1973} and W.~Bardeen~\cite{Bardeen:1995}. If an interplay between the tree-level and loop effects appears  the gauge symmetry can be radiatively broken by a vacuum expectation value (VEV) of a scalar field, without the need of introducing any mass terms to the theory.

This, however,  requires introducing classical scale invariance to the model, the status of which has been treated in different ways in the literature. In ref.~\cite{Meissner:2006} the scale symmetry is realised by forbidding tree-level mass terms and using dimensional regularisation to avoid quadratic divergencies. In a cut-off regularisation scheme with a~regulator $\Lambda$ this requires counter terms exactly cancelling $\Lambda^2$ divergencies~\cite{Meissner:2007}. The conformal symmetry is compared to supersymmetry and follows from some yet unknown theory of gravity. Alternatively, in refs.~\cite{Chankowski:2014, Lewandowski:2017}, it was proposed to have an effective theory around the Planck scale with a fixed cut-off $\Lambda\sim M_{\textrm{Pl}}$ arising from some unknown fundamental theory with small (or zero) mass terms which do not receive further radiative $\Lambda^2$ contributions in perturbation theory by some conspiracy of the introduced couplings, as in the old proposal by Veltman~\cite{Veltman:1980} (see also~\cite{Antipin:2013}). Looking for a concept of conformal invariance not present in a fundamental theory, the existence of a UV fixed point near $M_{\textrm{Pl}}$ based on inclusion of gravity was discussed in ref.~\cite{Shaposhnikov:2009} and related to the concept of ``asymptotic safety'', see e.g.\ refs.~\cite{Litim:2014, Pelaggi:2017, Litim:2015, Gorsky:2014}.

The attractive idea of generating all  mass scales via dimensional transmutation is not realised in nature in its minimalistic version which would be the standard model (SM) with classical conformal symmetry (cSM)~\cite{Coleman:1973}.\footnote{In ref.~\cite{Coleman:1973} too small mass of the Higgs boson was given as the reason why cSM is not realised in nature. However, in this reference the contribution from the top quark was not included since it had not yet been discovered at that time. With the inclusion of the top quark it appears impossible to obtain a stable radiative minimum with perturbative values of the Higgs self-coupling. Nonetheless, the authors of refs.~\cite{Steele:2012, Elias:2003} argue that the conformal SM with the correct value of the Higgs boson mass is still possible in a scenario where the Higgs self-coupling is rather large and resummation of higher order contributions to the effective potential is necessary. However, in these scenarios one may expect that the running Higgs self-coupling develops a Landau pole at rather low energy scales, which is considered problematic.} However, the SM needs to be extended in order to address issues like dark matter, neutrino masses, inflation and electroweak baryogenesis, not to mention gravity. Therefore radiative effects may still be an important component of models beyond the SM. Substantial amount of  work has been done in this direction, in the literature models featuring RSB with an extra scalar singlet (real or complex)~\cite{Meissner:2006, Meissner:2008, Foot:2007s, AlexanderNunneley:2010, Farzinnia:2013, Gabrielli:2013, Allison:2014, Sannino:2015, Ghorbani:2017}, more complicated scalar sectors~\cite{Foot:2007-3,Espinosa:2007, Espinosa:2008, Lee:2012, Helmboldt:2016, Fink:2018, Davoudiasl:2014} and new gauge groups~\cite{Hempfling:1996,Foot:2007,Chang:2007, Hambye:2008, Iso:2009, Dermisek:2013, Hambye:2013, Carone:2013, Khoze:2013-1, Khoze:2013-2,Khoze:2013-3,Heikinheimo:2013,Khoze:2014, Altmannshofer:2014, Benic:2014,  Karam:2015, Karam:2016, Plascencia:2015, DiChiara:2015, Plascencia:2016, Oda:2017, Guo:2015} (often also including extra fermionic degrees of freedom) were considered.\footnote{Of course also in models with mass terms present already at tree level, featuring fields beyond the SM (see e.g. refs.~\cite{Barger:2008,Gonderinger:2012, Profumo:2014}, among many others), radiative effects, RG running and improvement of the effective potential can be essential and the discussion of the present paper can be relevant.}

The issue of radiative symmetry breaking (RSB) is a subtle one since it relays on a balance between the tree-level and loop contributions and thus special care has to be taken of the validity of the perturbative treatment, especially in the presence of multiple scales related to more than one scalar fields or couplings. In the present article we discuss various methods to handle RSB in models with additional scalars. We review and assess the existing approximative methods as well as implement a recently developed method of RG improvement of the effective potential~\cite{Chataignier:2018} which allows to improve the accuracy of perturbation theory.

A common method of dealing with multi-scalar effective potentials is the one devised by E.~Gildener and S.~Weinberg (GW)~\cite{Gildener:1976}. It is very convenient since it allows to reduce the problem of minimising a multivariable function to a study of a function of just one variable. However, it relies on the assumption that along a generic direction in the field space the tree-level potential dominates over the one-loop correction which is expressed in terms of an assumption for the scalar couplings being of the order of typical gauge couplings squared. Then the loop corrections are only relevant along a direction where the tree-level potential vanishes.  Another scenario analysed in the literature is that of the coupling linking the SM to the new sector (the so-called portal coupling) being very small. In such a case RSB can occur in the new sector independently of the SM, inducing an effective mass term for the SM Higgs field. We refer to this scenario as sequential symmetry breaking. These two, however, do not exhaust all the possible configurations in the parameter space. Therefore, studies beyond the approximate schemes mentioned above are desirable.

An important issue, when comparing different methods of studying RSB, is the scale dependence of the method. In the archetypical one-field case of scalar electrodynamics (QED)~\cite{Coleman:1973}, the computations are carried out at the scale of the vacuum expectation value (VEV) of the single scalar field. When more scalar fields acquire VEVs, the renormalisation scale can be fixed to be equal to any of the VEVs, not necessarily the SM one. On the contrary, within the GW method computations are performed at a scale at which the tree-level potential develops a flat direction, which is different from the scales of the VEVs of the scalar fields.  While physical quantities do not depend on the renormalisation scale, some frequently used approximations, like e.g.\ running masses, do. It is thus relevant to understand whether the differences in the value of the reference scale between the common approaches to RSB can cause discrepancies between the obtained results.

Perturbativity of the loop expansion is another issue relevant to RSB studies. In the case of one-field models, if correct hierarchy of couplings is present one can choose the renormalisation scale such that there is no threat from the side of large logarithmic corrections. However, if more scalar fields are present this is more complicated.  If a minimum appears for vastly differing values of the fields it might happen that the loop corrections are significant even if a suitable hierarchy of couplings is present. This points towards the issue of improving the effective potential using the renormalisation group (RG) equation. Again, with more scalar fields the issue of RG improvement is not straightforward. There have been some proposals on this issue in the past~\cite{Einhorn:1983, Ford:1994,  Ford:1996, Ford:1996-2,Casas:1998} and recently we developed a new approach~\cite{Chataignier:2018}. In this paper we apply this method and compare the obtained results to the unimproved cases.

In the present paper we first discuss various issues related to RSB in general models with extended scalar sectors and then exemplify the general considerations by studying a concrete model in detail. The model studied in this paper is a conformal SM  extended with an extra scalar field which is a singlet under the SM gauge group while being a doublet of an extra SU(2)$_X$ group, which acts only in the non-SM sector. The model has been studied before in refs.~\cite{Hambye:2013, Carone:2013, Khoze:2014},  and in an extended version in refs.~\cite{Karam:2015, Plascencia:2016}. Various phenomenological implications, such as LHC phenomenology, dark matter (DM) relic abundance from the extra gauge bosons as well as possibilities for baryogenesis have been worked out. In refs.~\cite{Karam:2015, Plascencia:2016} the computations were performed using the GW method, in ref.~\cite{Hambye:2013, Khoze:2014} the sequential approach was used, whereas in ref.~\cite{Carone:2013} the one-loop effective potential was studied directly using numerical methods (with the contributions from the scalar fields to the effective potential neglected).  In the present paper we aim at comparing different approaches to studying RSB, focusing on the hierarchy of different contributions and the scale dependence. Moreover, we study the RG-improved effective potential of ref.~\cite{Chataignier:2018} and compare the results with the ones obtained from the simple one-loop approach. 

The paper is organised as follows. We start from explaining in section~\ref{sec:hierarchy-of-RSB} how to systematically look for radiatively generated minima. We go from the simple one-field examples of $\f^4$-theory and scalar QED to models with extended scalar sector. We give a general scheme for minimising the potential and discuss the range of applicability of the frequently employed methods. Moreover, we explain the issue of scale dependence and introduce the method of RG improvement of the effective potential. Next, in section~\ref{sec:SU2} we apply those considerations to the SU(2)cSM. We provide a systematic study of RSB in this model for different ranges of parameters and we compare different methods, focusing on the scale dependence and the usefulness of RG improvement. We present our conclusions and outlook in section~\ref{sec:summary}.

\section{Radiative symmetry breaking in multifield models\label{sec:hierarchy-of-RSB}}

In this section we discuss possible patterns of symmetry breaking in classically conformal models and explain how to systematically look for a global minimum of the effective potential.

\subsection{Single-field case\label{sec:simple-examples}}

To prepare the ground for the more general discussion let us start from briefly reviewing the archetypical examples of RSB presented in the seminal paper of S.~Coleman and E.~Weinberg (CW)~\cite{Coleman:1973}. The simplest classically conformal model is the massless $\f^4$ theory with the following  one-loop effective potential
\be
V(\f)=\frac{1}{4}\lambda \f^4+\frac{9\lambda^2\f^4}{64\pi^2}\left[\log\frac{3\lambda\f^2}{\mu^2}-\frac{3}{2}\right], \label{eq:V-phi4}
\ee
where $\mu$ is the renormalisation scale. Looking at eq.~\eqref{eq:V-phi4} we see that in order for the potential to have a nontrivial minimum the one-loop term, which is of the order $\lambda^2$, should be of comparable magnitude as the tree-level term, which scales as $\lambda$. For perturbative values of $\lambda$, $\lambda^2\ll \lambda$ so this can only happen for large values of the logarithm, which invalidates the loop expansion. The use of RG-improved potential, which is valid also for large values of the logarithm, shows that RSB does not occur in this model.

Another example, which in turn provides a viable case of RSB, is massless scalar quantum electrodynamics (QED) with the following lagrangian being a function of a complex scalar field $\phi$ and the vector field $A_{\mu}$
\be
\mathcal{L}=V_{\textrm{tree}}(\phi)+\mathcal{L}_{\textrm{int}}\supset\frac{1}{4}\lambda|\phi|^4+\frac{1}{4} e^2 A_{\mu} A^{\mu} |\phi|^2,
\ee
where we displayed only the relevant terms. 
Due to the symmetry of the model we can write the effective potential in terms of the real component of the complex scalar field $\phi$, denoted by $\f$.  Then the tree-level potential acquires the same form as in eq.~\eqref{eq:V-phi4}
and the one-loop correction to the effective potential is given by
\be
\Vone(\f)=\frac{\f^4}{64\pi^2}\left[9\lambda^2\left(\log\frac{3\lambda\f^2}{\mu^2}-\frac{3}{2}\right)+\lambda^2\left(\log\frac{\lambda\f^2}{\mu^2}-\frac{3}{2}\right)+\frac{3e^{4}}{16}\left(\log\frac{e^{2}\f^2}{4\mu^2}-\frac{5}{6}\right)\right] .
\ee
Again, in order to obtain RSB the one-loop term has to be comparable to the tree-level potential. We already know that in the perturbative regime the term proportional to $\lambda^2$ can be neglected, and we see that $\Vtree$ and $\Vone$ can be comparable within the range of applicability of perturbative expansion as long as $\lambda=\mathcal{O}(e^4)$ (this relation is explicit at the scale $\mu=\frac{e}{2}\langle\f\rangle$).\footnote{In the original ref.~\cite{Coleman:1973} due to a different choice of the renormalisation scheme this scale was set to $\mu=\langle\f\rangle$. For this reason and for the sake of simplicity, in what follows we also set the scale at which the CW minimum is computed to the VEV of the scalar field. Moreover, when more particles with different masses are present in a given model, it is not possible to set the scale to the value of field-dependent mass of all the particles.} In this case the fact that the one-loop contribution is of the same size as the tree-level one does not infringe perturbation theory, but rather it signals that the loop expansion is not the correct scheme for perturbation theory. The expansion should be reorganised such that the sum $\Vtree + \Vone$ is treated as the leading term, while subleading terms bring higher powers of $e^4$ (with the $\lambda=\mathcal{O}(e^4)$ scaling). Thus, in scalar QED RSB is a viable mechanism and it requires a special relation between the couplings.\footnote{In ref.~\cite{Coleman:1973} it was shown that this  relation is rather generic in scalar QED.}

Therefore we may conclude that for perturbative RSB to occur it is crucial that there exists a hierarchy of couplings which allows to equate the tree-level and one-loop terms without introducing large logarithmic terms and entering non-perturbative regime.

It is worth noting that the hierarchy of couplings and the resulting reorganisation of perturbation theory is of profound importance for the study of gauge dependence of the minima of the effective potential~\cite{Nielsen:1975, Nielsen:1987, Metaxas:1995, Andreassen:2014, Andreassen:2014prl, Espinosa:2016}. It is known that the effective potential is a gauge dependent quantity~\cite{Jackiw:1974}, however it was shown~\cite{Nielsen:1975, Fukuda:1975} that the value of the potential at an extremal point does not depend on the choice of gauge. This is an important issue since the value of the effective potential at a minimum determines whether it is a global or only a local minimum, and a stable theory should be built  around a global minimum. However, the aforementioned results are non-perturbative, i.e. they make a statement about the full effective potential which is usually not known. It appears that, if one performs perturbation theory naively, gauge independence of the value of the effective potential at the minimum is not maintained at a fixed order in loop expansion. On the contrary, if the hierarchy of couplings is taken into account properly, and perturbation theory is performed as an expansion in powers of the relevant coupling, then gauge independence is maintained (up to the relevant order in the couplings)~\cite{Andreassen:2014, Andreassen:2014prl, Espinosa:2016}.

\subsection{Multifield case\label{sec:multifield-intro}}
The situation becomes more involved when there are more scalar fields present in the theory, and the~effective potential becomes a function of more variables, with more couplings. In general, the one-loop effective potential in a multifield case reads (using the $\ms$ renormalisation scheme and the Landau gauge)
\be
\Vone(\f_i)=\frac{1}{64 \pi^2}\sum_{a}n_a M_a^4(\f_i)\left(\log\frac{M_a^2(\f_i)}{\mu^2}-C_a\right), \label{eq:one-loop}
\ee
where the sum runs over all particle species.  $M_a(\f_i)$ denotes field-dependent mass of a particle (for a scalar particle it is an eigenvalue of the second derivative of the tree-level potential), which corresponds to (a tree-level approximation to) the physical mass only at the minimum of the tree-level potential. $n_a$ counts the number of degrees of freedom associated with each species and $C_a=\frac{5}{6}$ for vector bosons and $C_a=\frac{3}{2}$ for other particles.\footnote{These constants depend on the regularisation and renormalisation schemes chosen. In the present work we use dimensional regularisation and $\ms$. If dimensional reduction was used, the $C_a$ would be equal $\frac{3}{2}$ for all particle species.} For a particle of spin $s_a$ the factor $n_a$ is given by
\be
n_a=(-1)^{2s_a} Q_a N_a (2s_a+1),\notag
\ee
where $Q_a=1$ for uncharged particles, and $Q_a=2$ for charged particles, $N_a=1,\;3$ for uncoloured and coloured particles, respectively. We also define a related quantity, which will be useful for the following discussion,
\be
\mathbb{B}(\mu, \lambda_j,\f_i)=\frac{1}{64 \pi^2}\sum_{a}n_a M_a^4(\f_i),\label{eq:B}
\ee
where $a$ again runs over all particle species.

There are several difficulties associated with studying minima of multifield effective potentials. Firstly, the effective potential is now a function of many arguments, and can develop minima along different directions. Those minima can coexist and it is important to identify the global minimum of the potential in order for the theory to be stable and predictive.\footnote{A local minimum can constitute a ground state for a model, if it is sufficiently stable (metastable). Moreover, local minima are relevant for the thermal history of the Universe, since tunnelling between local minima is possible. In this work, however, we focus on identifying the global minimum of the effective potential.} Secondly, the scalar field-dependent masses are eigenvalues of the Hessian of the tree-level potential, whose dimension is equal to the number of scalar fields. Thus, in general, analytical expressions for the field-dependent masses become complicated. Moreover, when RSB is concerned, with more scalar couplings the simple arguments derived from the relation between the scalar coupling and the gauge coupling can no longer be applied in a straightforward manner. Furthermore, there no longer exists a single natural scale, being the VEV of the single scalar field --- the VEVs of different scalar fields can vary vastly leading to possible large logarithms, making the question of which scale should be chosen as the reference scale for the computations more relevant. While the first two problems are rather technical complications, the latter two require developing some understanding. We address these issues in the following sections.

\subsection{Hierarchy of different contributions \label{sec:hierarchy}}
 The simple examples of the previous section show that for perturbative RSB to occur it is crucial that the one-loop correction to the effective potential can compete with the tree-level term. This was expressed in terms of scaling of the scalar coupling as $\lambda\sim\mathcal{O}(e^4)$.  In the multifield case the relative magnitudes of couplings are still important,\footnote{With more particle species $\mathcal{O}(e^4)$ is no longer unambiguous. Here we take as the relevant quantity the fourth power of the maximal gauge or Yukawa coupling divided by the loop factor of $\frac{1}{64\pi^2}$ and multiplied by the number of degrees of freedom $n_a$ and denote it as $\mathcal{O}(g^4)$. Therefore, $\mathcal{O}(g^4)=\mathcal{O}(\frac{n_a}{64\pi^2}\max{g_i^4})$, where the maximum is taken over gauge and Yukawa couplings. For the SM at $\mu=246\g$ this means $\mathcal{O}(g^4)=\order(0.01)$.} however there are other factors that influence the interplay between the tree-level and the one-loop terms. It is not possible to factor out all the scalar fields from the effective potential, therefore also ratios of the scalar fields matter for the order of different contributions. Furthermore, there is no single natural candidate for the renormalisation scale, thus its choice can change the relative magnitude of the two contributions to the effective potential.

To understand the interplay between the tree-level and one-loop terms let us analyse the stationary-point equations for the one-loop effective potential,
\be
\frac{\partial V}{\partial \f_i}=\frac{\partial \Vtree}{\partial \f_i} +\frac{\partial \Vone}{\partial \f_i}=0.\label{eq:stat}
\ee
There are three possible scenarios of the interplay between the $\Vtree$ and $\Vone$ terms
\begin{enumerate}
\item $\mathbf{\Vtree\sim \Vone}.$ If generically the tree-level contribution is of the order of the one-loop term, then the sum of the two constitutes the leading term in perturbation theory and it is mandatory to study the full one-loop effective potential (this is the analog of the $\lambda\sim\ogg$ one-field case). One may expect that the scalar contributions to $\Vone$ are suppressed since they contain $\lambda_i^2$ and therefore they can be neglected.\label{point:g4}

\item $\mathbf{\Vtree\gg\Vone}.$ If generically the tree-level potential dominates over the one-loop correction,\footnote{Of course if one of the scalar fields is large enough the one-loop potential becomes dominant because of the large logarithms, regardless of the magnitudes of the couplings. Here we refer to the region in the field space where the logarithms remain small.} the tree-level terms constitute the leading order expression and since $\Vtree$ is conformally symmetric there is no symmetry breaking (this is the analog of the $\lambda\sim\og$ scaling in the one-field case). Then RSB can only be realised if the tree-level potential vanishes along a flat direction, which is lifted by quantum corrections. This is the basic assumption of the E.~Gildener and S.~Weinberg (GW) method~\cite{Gildener:1976}, which is discussed in section~\ref{sec:GW} below.\label{point:g2}

\item It can also happen that along different directions in the field space (for different $i$ in eq.~\eqref{eq:stat}), different, i.e.\ tree-level or one-loop terms dominate. Then, along some directions there might be no symmetry breaking at leading order, and then the VEVs generated by RSB along other direction may induce traditional spontaneous symmetry breaking in the remaining directions. This scenario will be referred to as sequential symmetry breaking.\label{point:mixed}
\end{enumerate}

These scenarios are studied in detail in section~\ref{sec:SU2} for the SU(2)cSM model. Below we discuss the applicability of the well-known GW method and comment on the sequential approach to symmetry breaking.

\subsubsection{Gildener--Weinberg method \label{sec:GW}}

E.~Gildener and S.~Weinberg developed a method of studying RSB in models with more scalar fields~\cite{Gildener:1976}, which can be briefly summarised as follows. Assuming all the scalar couplings are of the order of $g^2$ the tree-level effective potential is of the order $g^2$, whereas the one-loop correction is of the order $g^4$.\footnote{The dependence of the relative magnitude of $\Vtree$ and $\Vone$ on the ratios of the fields is not discussed in ref.~\cite{Gildener:1976}.} Thus, the one-loop term is too small to change the behaviour of the tree-level term, unless the latter vanishes. Since the field space is multidimensional, the tree-level potential does not need to vanish everywhere, it is enough that it develops a flat direction at a certain renormalisation scale, i.e. it should form a valley with a minimal value (equal zero) along a ray in the field space. Then, along this ray quantum corrections can play a role, and induce a local minimum, whereas in the other directions the tree-level term dominates and thus there are no non-trivial minima. This allows to reduce the multidimensional problem to studying the potential along one direction in the field space and thus is very convenient and widely used in the literature, see e.g.\ \cite{Foot:2007, AlexanderNunneley:2010, Farzinnia:2013, Sannino:2015, Foot:2007s, Foot:2007-3,Helmboldt:2016, Benic:2014, Guo:2015, Karam:2015, Karam:2016,Plascencia:2016,Fink:2018}. 

It is important to underline, however, that this method is guaranteed to find a global minimum of the potential under the assumption that the scalar couplings are of the order of $g^2$, or in other words, that the tree-level potential dominates over the one-loop contribution generically in the field space. Assume that the one-loop potential can compete with the tree-level one everywhere in the parameter space, see case~\ref{point:g4} above. Then, even if the tree-level potential has a flat direction at a certain scale, as in the GW case, the local minimum that is found along this direction need not be the global one (the one that is sought) because a minimum can form along any direction in the field space, its depth depending on the values of the parameters.  Thus, the GW method can also be used in the case in which $\Vtree\sim\Vone$, however then one should check whether the minimum that was found is indeed the global minimum of the potential. This issue is not commonly acknowledged in the literature.

\subsubsection{Sequential symmetry breaking\label{sec:sequential}}
An important class of classically conformal models are the so-called Higgs portal models, where the additional scalar fields are coupled to the SM field content only through an interaction with the Higgs doublet. In such models another approximate scheme for the study of symmetry breaking exists, the aforementioned sequential symmetry breaking. It is realised when the terms representing the coupling between the Higgs doublet and the extra scalar multiplet in eq.~\eqref{eq:stat} are very small (compared to the contributions of the new scalar field) and only enter at subleading level. In this case the problem can be viewed as follows --- first we consider the ``hidden'' sector only. If RSB appears in the hidden sector (e.g.\ due to the presence of a hidden gauge group), then the VEV of the scalar produces an effective mass term for the Higgs boson, and the symmetry breaking in the SM sector proceeds in the usual way (see\ e.g. refs.~\cite{Gabrielli:2013, Iso:2009, Hambye:2013, Khoze:2013-1,Khoze:2013-2, Khoze:2013-3, Khoze:2014, Altmannshofer:2014, Plascencia:2015, Guo:2015}). This method simplifies the analysis of RSB considerably, however one needs to assure the smallness of the mixing terms. As we will show on the example of SU(2)cSM, this depends not only on the magnitude of the portal coupling, but also on the ratio of the fields.

\subsection{Scale dependence\label{sec:scale-dep}}
In the previous section we emphasised the importance of the interplay between $\Vtree$ and $\Vone$ for RSB. Closely related to the latter is the issue of the renormalisation-scale dependence of the results. Since the coupling constants run with the scale $\mu$ the hierarchy between them depends on the scale at which they are evaluated. Moreover, the logarithmic terms in $\Vone$ can change substantially with the RG scale. In light of that, it is important to emphasise that when different methods for studying RSB are used, different values of the scale $\mu$ are most suitable. A~natural scale for the computations is the EW scale, in this paper taken to be $\mu=246\g$, corresponding to the VEV of the SM scalar field. The masses of the SM particles are not too far from this scale so the logarithms related to these masses should not be too large at $\mu=246\g$. Similarly natural is a scale $\mu$ equal to any other VEV present in the model. For example, if the sequential approach to symmetry breaking is used, the natural scale is the scale of the VEV of the new scalar field since then RSB occurs in the hidden sector, and the symmetry breaking in the SM sector proceeds in the usual SM way, with the mass term for the Higgs field generated by the VEV of the new scalar field. However, when the GW method is implemented, the computations have to be performed at the $\mgw$ scale at which the tree-level potential develops a flat direction.  The scales listed above can be quite different. This alone can lead to differences in the results obtained when various methods are used --- it is known that the VEVs of scalar fields obtained from the one-loop effective potential run with the renormalisation scale. Therefore, the location of the minimum of the one-loop effective potential depends on the renormalisation scale at which it is computed. Moreover, it is common to use the running masses (for the scalar particles they correspond to the eigenvalues of the matrix of second derivatives of the effective potential evaluated at the minimum) as approximations for the physical (pole) masses of particles. This approximation is expected to give best results when the running masses are evaluated at energy scales close to the physical (pole) mass. This is because to obtain the physical masses one has to add self-energy corrections to the running masses. The self-energy corrections depend on the logarithms of ratios of some functions of the masses and the renormalisation scale. Therefore, if the scale is far away from the scale of the masses, the running-mass approximation might not be accurate.

The arguments presented above show that the choice of the renormalisation scale is not without an effect on the results of RSB analyses, in particular on the location of the minimum as well as the predictions for the masses. Since different methods of studying RSB used in the literature entail different choices for the renormalisation scale, one should ask how this influences the obtained results.

\subsection{RG improvement of the effective potential\label{sec:RG-improvement}}
Another issue that is relevant for RSB analyses is the applicability of the loop expansion, which is valid as long as the logarithmic terms containing ratios of the field-dependent masses and the renormalisation scale $\mu$ are not too large. In case of a one-field problem this can be easily achieved by fixing $\mu$ to a value close to the energy scale we are interested in (in the case of RSB that would be the scale of the VEV). However, with more scalar fields this is not, in principle, possible since the fields can have vastly varying values and we have only one renormalisation scale on our disposal. This could in some cases infringe validity of loop expansion. Whether this happens or not depends on the values of the fields for which a minimum occurs, and on the values of the couplings. The larger the splitting between different field-dependent masses evaluated at the minimum, the less accurate  the result given by a one-loop approximation is.

The tool that is used to save perturbation theory from the threat of large logarithms is the RG improvement of the effective potential. It exploits the invariance of the effective potential under the change of the renormalisation scale.  The issue of the improvement is simple only when it comes to a  theory with one scalar field and one field-dependent mass. With more scalar fields present in the model it becomes more complicated, since resumming different logarithmic terms with just one scale $\mu$ is not a straightforward task.

The methods introduced to cope with such issues are the multiscale techniques~\cite{Einhorn:1983, Ford:1994,  Ford:1996, Ford:1996-2}, a use of the decoupling theorem~\cite{Casas:1998, Iso:2018} and a method recently developed by us in ref.~\cite{Chataignier:2018}. Below we briefly introduce that method of RG improvement while all the details and examples can be found in ref.~\cite{Chataignier:2018}. In section~\ref{sec:SU2} we employ this formalism to study RSB in SU(2)cSM and compare the obtained results with the one-loop approximations.

The RG (Callan--Symanzik) equation for an effective potential being a function of real scalar fields $\f_i$ and parametrised by the couplings $\lambda_j$ reads\footnote{In principle vacuum energy should also contribute to this equation, however since we are interested in models which are conformally invariant at tree level we can disregard this contribution since it is equal to zero.}
\be
\mu \frac{d}{d\mu} V = \left(\mu \partial_{\mu} + \sum_j \beta_j \partial_{\lambda_j}-\frac{1}{2}\sum_i\gamma_i \f_i \partial_{\f_i}\right)V=0,\label{eq:RGE-V}
\ee
where $\beta_j$ and $\gamma_i$ are the standard $\beta$ functions and anomalous dimensions which account for the running of the couplings and field normalisation $Z_i$ with the renormalisation scale $\mu$. They are defined as follows
\begin{align}
\beta_j&=\mu\frac{\partial\lambda_j}{\partial\mu}\label{eq:beta},\\
\gamma_i&=\mu\frac{\partial\log Z_i}{\partial\mu}\label{eq:gamma}.
\end{align}
Equation~\eqref{eq:RGE-V} implies that that the value of the effective potential does not change even  if it is evaluated at a different renormalisation scale $\mu(t)$ as long as we change the values of the couplings and fields' normalisation according to their RGE flow. Namely,
\be
V(\mu, \lambda_j,\f_i)=V(\bar{\mu}(t),\bar{\lambda}_j(t),\bar{\f}_i(t)),\notag
\ee
where $\bar{\mu}(t)=\mu e^t$ and $\bar{\lambda}_j(t)$ is a solution of eq.~\eqref{eq:beta} with the boundary condition $\bar{\lambda}_j(0)=\lambda_j$, and $\bar{\f}_i^2(t)=Z_i(t)\f_i^2$, with $Z_i(t)$ being the solution of eq.~\eqref{eq:gamma} with $Z_i(0)=1$. 

If we work to one-loop order then our potential is the sum of the tree-level and one-loop terms. It~might happen that in some region of the parameter space the loop expansion is not well-behaved and the one-loop potential is untrustworthy. We can then exploit the scale invariance and evaluate $V$ at some other scale, where the perturbative behaviour is more reliable, 
\begin{align}
V(\mu, \lambda_j,\f_i)&=\Vtree(\lambda_j,\f_i)+\Vone(\mu, \lambda_j,\f_i)+ \mathcal{O}(\hbar^2)\nonumber\\
&=\Vtree(\bar{\lambda}_j(t),\bar{\f}_i(t))+\Vone(\bar{\mu}(t),\bar{\lambda}_j(t),\bar{\f}_i(t))+ \mathcal{O}(\hbar^2).\notag
\end{align}
An important observation is that we can now choose $t=t_*$ such that\footnote{In fact, it is not always possible to exactly cancel the one-loop contribution, for details see ref.~\cite{Chataignier:2018}.}
\be
\Vone(\bar{\mu}(t_*),\bar{\lambda}_j(t_*),\bar{\f}_i(t_*))=0.\label{eq:t*-def}
\ee
 Then we obtain
\be
V(\mu, \lambda_j,\f_i)=\Vtree(\bar{\lambda}_j(t_*),\bar{\f}_i(t_*)),\notag
\ee
which is free from large logarithmic terms and valid as long as the running couplings stay perturbative. The condition for $t_*$, eq.~\eqref{eq:t*-def} is an implicit one, and thus it is not possible to give an analytic expression for $t_*$. Nonetheless, it can be solved numerically or an analytic approximation can be used. The first approximation to $t_*$  reads
\be
t_*^{(0)}=\frac{\Vone(\mu, \lambda_j,\f_i)}{2\mathbb{B}(\mu, \lambda_j,\f_i)}\label{eq:t0}
\ee
and is usually a fairly accurate estimate of $t_*$. $\mathbb{B}(\mu, \lambda_j,\f_i)$ is given by eq.~\eqref{eq:B}.\footnote{For this method to work, i.e.\ for a finite $t_*$ to exist, the surface where the one-loop correction vanishes has to be a non-characteristic surface of the RG equation for $V$, eq.~\eqref{eq:RGE-V}, when viewed in the space spanned by $(\mu, \lambda_j,\f_i)$. This is not the case if $\mathbb{B}(\mu, \lambda_j,\f_i)=0$ (at least to first approximation), which is also suggested by eq.~\eqref{eq:t0}. Further discussion of this point can be found in ref.~\cite{Chataignier:2018}.} We refer the interested reader to ref.~\cite{Chataignier:2018} for more details.

\section{Application: conformal SM with an extra scalar and SU(2)\label{sec:SU2}}

In the following section we study RSB in a concrete model, applying the considerations of the previous sections in practice. In what follows we compare different approximative schemes discussing scale dependence of the results and the RG improvement. The aim of the present section is to exemplify our previous general discussion. Nonetheless we believe that the general conclusions are valid also for other models.

It is well known that the conformal SM (i.e.\ the SM without the tree-level mass term for the Higgs boson) cannot account for RSB given the measured values of masses of the top quark and the Higgs boson. In its simplest extension with one extra scalar singlet in principle the correct masses of physical particles can be generated via dimensional transmutation, however this requires a very large coupling between the Higgs doublet and the singlet. This destabilises the RG running of the couplings and generates a Landau pole close to the electroweak scale. Thus, implementation of the RSB mechanism requires more sophisticated models, allowing extensions in the scalar sector or in the gauge sector. 

In this work we study the conformal SM supplemented with an additional SU(2)$_X$ symmetry and a scalar field which transforms as a singlet under the SM gauge group but forms a doublet under the extra SU(2)$_X$. The SM fields are neutral under the SU(2)$_X$ symmetry group. We refer to the model as SU(2)cSM for short. As mentioned in the introduction, this model has been studied previously in refs.~\cite{Hambye:2013, Carone:2013, Khoze:2014} and in extended version also in refs.~\cite{Karam:2015, Plascencia:2016}. It has been studied using the GW method~\cite{Karam:2015, Plascencia:2016}, the sequential approach~\cite{Hambye:2013, Khoze:2014} and numerically, using the full one-loop effective potential (with scalar contributions neglected)~\cite{Carone:2013}.

In this section we study the pattern of symmetry breaking in the SU(2)cSM. First we discuss various approximative schemes to study RSB in this model, following the discussion of section~\ref{sec:hierarchy}. Next we use the full one-loop effective potential (with scalar contributions neglected) to study the parameter space of SU(2)cSM, find radiatively generated minima and masses of particles. Moreover, we find a region where the model is valid up to the Planck scale, i.e.\ a region where the scalar couplings have no Landau poles up to the Planck scale (appearance of a Landau pole would suggest an existence of some fields not described by SU(2)cSM) and the potential is bounded from below (this is necessary for a stable vacuum state to exist).  We also discuss the hierarchy between different contributions. Later we compare the results obtained from the full one-loop potential with the results of analysis with the use of GW method and discuss scale dependence of the results. Our aim is to evaluate validity and complementarity of different methods. Most importantly, we confront the results obtained using perturbative methods with the results derived with the use of the RG improved potential~\cite{Chataignier:2018}. 

\subsection{Introducing SU(2)cSM}

The SU(2)cSM model, consists of the conformal SM extended with an extra SU(2)$_X$ symmetry and a scalar field which is a singlet under the SM gauge group and a doublet under the extra SU(2)$_X$.  The SM field content transforms trivially under SU(2)$_X$. The sector composed of the SU(2)$_X$ doublet and gauge fields is sometimes referred to as the dark or hidden sector. The scalar tree-level potential reads
\be
\Vtree(\Phi, \Psi)=\la \left(\Phi^{\dagger}\Phi\right)^2 + \lb \left(\Phi^{\dagger}\Phi\right) \left(\Psi^{\dagger}\Psi \right)+ \lc \left(\Psi^{\dagger}\Psi\right)^2,\notag
\ee
where $\Phi$ is the SM-gauge-group doublet, and $\Psi$ is a doublet under the hidden SU(2)$_X$ gauge group.
For the potential to be bounded from below it is necessary that (see e.g. ref.~\cite{Kannike:2012} for a derivation)
\be
\la \geqslant 0 \quad\mathrm{and}\quad \lc\geqslant 0 \quad\mathrm{and}\quad \lb\geqslant -2\sqrt{\la\lc}.\label{eq:positivity}
\ee
We can  define the radial fields as follows
\be
h^2=\sum_{i=1}^4 h_i^2=2\Phi^{\dagger}\Phi, \quad \f^2=\sum_{i=1}^4 \f_i^2=2 \Psi^{\dagger}\Psi.\notag
\ee
The effective potential inherits its symmetry properties from the tree-level one~\cite{Peskin} hence we can write it in terms of the so-called classical (or average) fields $\overline{h}$ and $\overline{\f}$ corresponding to vacuum expectation values of $h$ and $\f$ in the presence of an external source $J$,\footnote{Here we do not present a detailed derivation of the effective potential which would justify the appearance of the source field $J$. For a more pedagogical derivation see e.g. refs.~\cite{Coleman:1973, Cheng}.}
\be
\overline{h}=\left[\frac{\langle 0|h|0\rangle}{\langle 0|0\rangle}\right]_J,\quad \overline{\f}=\left[\frac{\langle 0|\f|0\rangle}{\langle 0|0\rangle}\right]_J.\notag
\ee
The real vacuum expectation values (VEVs) of the quantum fields (with $J=0$) correspond to stationary points of the effective potential. In what follows we drop the bars for brevity and use $\f$ and $h$ as the names of the variables appearing in the effective potential. 

The zeroth order effective potential written in terms of $h$ and $\f$ thus reads
\be
\Vtree(h,\f)=\frac{1}{4}\left(\la h^4 + \lb h^2 \f^2 + \lc \f^4\right)\label{eq:Vtree}
\ee
and the one-loop contribution is given by the standard formula of eq.~\eqref{eq:one-loop}, with $\f_i=h,\;\f$. The tree-level field-dependent masses of the scalars are given by
\begin{align}
m^2_{\pm}& = \frac{1}{2}\left(3 \lambda_1 + \frac{\lambda_2}{2}\right)h^2 + \frac{1}{2}\left(\frac{\lambda_2}{2} + 3\lambda_3\right)\varphi^2 \notag\\[2pt]
& \ \ \ \pm\frac{1}{2}\sqrt{\left[\left(3\lambda_1-\frac{\lambda_2}{2}\right)h^2-\left(3\lambda_3-\frac{\lambda_2}{2}\right)\varphi^2\right]^2+4\lambda_2^2h^2\varphi^2}, \label{eq:mass-scalar1}\\
m^2_{G,G^{\pm}}&= \lambda_1 h^2 + \frac{\lambda_2}{2}\varphi^2, \\
m^2_{\tilde{G},\tilde{G}^{\pm}} &= \frac{\lambda_2}{2}h^2+\lambda_3 \varphi^2,
\end{align}
of the gauge bosons by
\be
M_W(h)=\frac{gh}{2},\quad M_Z(h)=\frac{\sqrt{g^2+g'^2}h}{2},\quad M_X(\f)=\frac{g_{X}\f}{2}\notag
\ee
and of the top quark by
\be
M_t(h)=\frac{y_t h}{\sqrt{2}},\label{eq:mass-top}
\ee
where $g$, $g'$ and $g_X$ are the SM SU(2), U(1) and the new SU(2)$_X$ gauge couplings, and $y_t$ corresponds to the top Yukawa coupling. We do not include the other quark flavours in the analysis since due to their small masses their contribution is negligible.

The model possesses a classical conformal symmetry which implies that at tree level all the particles are massless. The inclusion of the one-loop correction to the scalar potential may induce VEVs for the scalar fields and generate masses by dimensional transmutation. This possibility is studied in detail in the following subsections.

\subsection{Review of possible approximations\label{sec:systematic-SU2}}
In this section we discuss how to search for minima of the SU(2)cSM model exploiting the possible hierarchies between various contributions and so we apply the general considerations of section~\ref{sec:hierarchy} to the particular case of the SU(2)cSM. 

The stationary-point equations read
\begin{align}
\left.\frac{\partial V}{\partial h}\right|_{h=v, \f=w}&=\la v^3 +\frac{1}{2} \lb v w^2 +\left.\frac{\partial \Vone}{\partial h}\right|_{h=v, \f=w}=0,\label{eq:stat1}\\
\left.\frac{\partial V}{\partial \f}\right|_{h=v, \f=w}&=\lc w^3 +\frac{1}{2} \lb v^2 w +\left.\frac{\partial \Vone}{\partial \f}\right|_{h=v, \f=w}=0,\label{eq:stat2}
\end{align}
where $\Vone$ is given by the general formula of eq.~\eqref{eq:one-loop}. We should analyse these equations in different regimes separately (see points~\ref{point:g4}--\ref{point:mixed} on page~\pageref{point:g4}).

\subsubsection{$\Vtree\sim\Vone$}

In this case the tree-level potential is comparable to the one-loop contribution.\footnote{In fact, the reasoning presented in this subsection is valid as long as the scalar contributions to $\Vone$ can be neglected. Since we take into account both $\Vtree$ and $\Vone$, the formulas derived are applicable also in the case in which the tree-level contribution is dominant and the one-loop term is subleading.} The scalar contributions to $\Vone$ should be dropped, but apart from that no further simplifications occur.\footnote{It is interesting to note, as was pointed out within the context of a similar model in ref.~\cite{Loebbert:2018} which appeared one day after the first version of this article, that if one performs the calculation of the one-loop effective potential in an arbitrary Fermi gauge all the gauge dependence is contained in the scalar contributions to the one-loop effective potential. Thus dropping the scalar terms, we obtain an effective potential that is gauge independent (to leading order).}  Nonetheless, with this approximation the SM and SU(2)$_X$ one-loop contributions to the effective potential decouple (the former depends only on $h$ and the latter only on $\f$) and the stationary point equations, eqs.~\eqref{eq:stat1}--\eqref{eq:stat2} can be written as (we follow here ref.~\cite{Carone:2013}),
\begin{align}
\la&=-\frac{1}{2} \lb \left(\frac{w}{v}\right)^2 - \frac{1}{16\pi^2}\sum_{a}n_a \alpha_a^4\left[2\log \alpha_a - C_a +\frac{1}{2}\right],\label{eq:stat1-g4}\\
\lc&=-\frac{1}{2} \lb \left(\frac{v}{w}\right)^2 - \frac{9}{256\pi^2}g_X^4\left[2\log\left(\frac{g_X}{2}\frac{w}{v}\right) -\frac{1}{3}\right],\label{eq:stat2-g4}
\end{align}
where in the first equation the sum runs over the top quark and the $W$ and $Z$ gauge bosons and for simplicity of notation we have written the masses of these particles as $\alpha_a v$. We assume that all the couplings are evaluated at the scale corresponding to the minimum, i.e.\ $\mu=\langle h\rangle=v$. These relations resemble the scalar QED condition for RSB, $\lambda\sim \ogg$. However, here there are the additional terms proportional to $\lb$ which couple the SM and the SU(2)$_X$ sectors and introduce the dependence on the ratio of the VEVs. It is clear that in this case not only the hierarchy of couplings is important for RSB, but also the hierarchy of the VEVs. Later, when we present the results of the numerical analysis, it will become clear that $\frac{w}{v}\sim\mathcal{O}(10)$. Therefore, with a fixed value of $\lb$, the term proportional to this coupling in eq.~\eqref{eq:stat1-g4} is $\mathcal{O}(10^4)$ greater than the corresponding one in eq.~\eqref{eq:stat2-g4}. This can cause the tree-level contributions to be dominant in eq.~\eqref{eq:stat1-g4} and subleading in eq.~\eqref{eq:stat2-g4}.

With these relations between the couplings we can proceed to compute the mass matrix for the scalar sector, which corresponds to the second derivative of the effective potential. Using the relations~\eqref{eq:stat1-g4}--\eqref{eq:stat2-g4} we can simplify it to the following form
 \be
 M^2=\left(\begin{array}{cc}
-\lb w^2 + 8 \mathbb{B}^{\textrm{SM}}/v^2 & \lb v w\\
\lb v w & -\lb v^2 + 8\mathbb{B}^{X}/w^2\\
 \end{array}
 \right),\label{eq:mass-matrix-g4}
 \ee
 where $\mathbb{B}^{\textrm{SM}}$ and $\mathbb{B}^{X}$ are defined as in eq.~\eqref{eq:B}, however the sums run only over the SM gauge and Yukawa or dark sector, respectively, and the fields are set to their VEVs. Diagonalising this matrix one obtains the masses of two scalar mass-eigenstates 
 \begin{align}
M_{\pm}^2=\;&\frac{1}{2}\bigg[8\left(b^{\textrm{SM}} +b^{X}\right)-\lb\left(v^2+w^2\right)\notag\\
&\left.\pm \sqrt{64\left(b^{\textrm{SM}} -b^{X}\right)^2+\lb^2\left(v^2+w^2\right)^2+16\lb\left(b^{\textrm{SM}} -b^{X}\right)\left(v^2-w^2\right)}\right],\label{eq:masses-g4}
 \end{align}
 where $M_{+}>M_{-}$. Above, for the sake of brevity, we have used the following definitions $b^{\textrm{SM}}=\mathbb{B}^{\textrm{SM}}/v^2$, $b^{X}=\mathbb{B}^{X}/w^2$. The mass eigenstates are obtained from gauge eigenstates by a rotation matrix as follows\footnote{The fields $h$ and $\f$ in eq.~\eqref{eq:mixing} should be understood as translated by the respective VEVs, $h\to h-v$, $\f\to\f-w$ so that the physical fields (mass eigenstates) have zero VEVs.}
 \be
\left(\begin{array}{c}\phi_{-}\\ \phi_{+}\end{array}\right)=
\left(\begin{array}{rl}
\cos\theta & \sin\theta\\
-\sin\theta & \cos\theta
\end{array}\right)\left(\begin{array}{c}h \\ \f\end{array}\right),\label{eq:mixing}
\ee
where $\phi_{-}$ corresponds to the lower mass $M_{-}$ and $\phi_{+}$ to $M_{+}$. The mixing angle $\theta$ is in the range between $-\frac{\pi}{2}$ and $\frac{\pi}{2}$ and is given by the following formula
$$
\tan 2 \theta = \frac{2\lb v w}{\lb(v^2-w^2)+8(b^{\textrm{SM}}-b^{X})}.
$$

In what follows the SM-like state with the mass of $125\g$ is denoted by $H$, while the new scalar particle by $S$. Here we do not specify which of these fields is heavier (which corresponds to $\phi_{+}$ and which to $\phi_{-}$). In section~\ref{sec:g4} we analyse both cases,  $M_H<M_S$ and $M_H>M_S$. In the first case, the couplings of the Higgs boson to the SM particles are rescaled by $\cos\theta$, while in the latter by $-\sin\theta$.

\subsubsection{$\Vtree\gg\Vone$ \label{sec:GW-SU2}}
In this case, at the leading order we can drop the one-loop contributions, since the tree-level part dominates. Solving eqs.~\eqref{eq:stat1} and~\eqref{eq:stat2} we obtain the following conditions which should be understood as being realised at some scale $\mgw$
\begin{align}
&4 \la\lc - \lb^2 = 0,\label{eq:GW-det}\\
 &w^2=-\frac{\lb}{2\lc}v^2=-\frac{2\la}{\lb}v^2,\label{eq:GW-min}
\end{align}
from which it is clear that $\lb<0$.
We can complete the squares in the tree-level potential to obtain the following form
\be
\Vtree(h,\f)=\frac{1}{4}\left[\la\left(h^2+\frac{\lb}{2\la}\f^2\right)^2+\left(\lc-\frac{\lb^2}{4\la}\right)\f^4\right].\label{eq:complete-squares}
\ee
From eq.~\eqref{eq:complete-squares} it is clear that the tree-level potential is a function of just one variable $h^2+\frac{\lb}{2\la}\f^2$ as long as eq.~\eqref{eq:GW-det} is fulfilled, and thus is flat in the perpendicular direction. Moreover, the VEVs lie exactly along the flat direction and $\Vtree(v,w)=0$, irrespectively of the values of $v$ and $w$.  This shows that this is the GW case.   

The potential along the direction perpendicular to that given in eq.~\eqref{eq:GW-min} is not flat, and thus the field corresponding to this direction in the field space acquires mass at tree level. This mass is given by the nonvanishing eigenvalue of the Hessian of the tree-level potential evaluated at the minimum
\be
M_1^2=(2\la-\lb)v^2=-\lb\rho^2, \label{eq:GW-tree-mass}
\ee
where $\rho^2=v^2+w^2$.

The one-loop effective potential can be considered solely along the flat direction, since along the perpendicular direction the tree-level potential dominates. The one-loop terms lift the flat direction, and also change this direction a little, however this effect is of subleading order. The field dependent Goldstone masses vanish along the flat direction, since it corresponds to the minimum of the tree-level potential, therefore they do not contribute to the one-loop potential.\footnote{This leads, as in the previous case, to an effective potential that is gauge independent to leading order~\cite{Loebbert:2018}.} The potential along the flat direction can be written in terms of the radial coordinate measured along this direction $\phi^2=\f^2+h^2=(1-\frac{\lb}{2\lc})h^2$
\be
\Vone(\phi)=\mathbb{A} + \mathbb{B} \log\frac{\phi^2}{\mu^2_{\textrm{GW}}},\notag
\ee
where $\mathbb{B}$ was defined in eq.~\eqref{eq:B} and $\mathbb{A}$ is given by\footnote{Note that it is common in the literature to factor $\phi^4$ out of $\mathbb{A}$ and $\mathbb{B}$ which we do not do in the present work.} (the notation is the same as in eq.~\eqref{eq:one-loop}) 
\be
\mathbb{A}=\frac{1}{64 \pi^2}\sum_{a}n_a M_a^4(\phi)\left(\log\frac{M_a^2(\phi)}{\phi^2}-C_a\right).\notag
\ee
The scale $\mgw$ is the scale at which eq.~\eqref{eq:GW-det} is satisfied.
 The extremum condition for $V$ gives the following condition for $\rho$
 \be
 \log\frac{ \rho^2}{\mgw^2}=-\frac{1}{2}-\frac{\mathbb{A}}{\mathbb{B}}.\notag
 \ee
 Computing the second derivative of the effective potential one finds that the loop-generated mass of  the scalar particle that was massless at tree level reads
 \be
 M_2^2=
 \frac{8\mathbb{B}}{\rho^2}
 \label{eq:GW-loop-mass}
 \ee
 and thus the radiatively generated extremum is a minimum if 
 \be
 \mathbb{B}>0,\notag
 \ee
 which is the well-known GW condition.  Note that above all the masses and couplings are computed at the scale $\mgw\neq v$. The mass eigenstates are thus given by
 \be
\left(\begin{array}{c}\phi_{2}\\ \phi_{1}\end{array}\right)=
\left(\begin{array}{rl}
\cos\alpha & \sin\alpha\\
-\sin\alpha & \cos\alpha
\end{array}\right)\left(\begin{array}{c}h \\ \f\end{array}\right),\label{eq:mixing-GW}
\ee
where the eigenstate $\phi_2$ corresponds to the loop-generated mass $M_2$, while $\phi_1$ to the tree-level mass $M_1$. According to the GW method~\cite{Gildener:1976, Loebbert:2018} the mixing angle $\alpha$ should be accurately approximated by the tree-level prediction (see eqs.~\eqref{eq:GW-det}--\eqref{eq:GW-min})
\be
\tan\alpha=\frac{w}{v}=\sqrt[4]{\frac{\la}{\lc}}\label{eq:tan-alpha}.
\ee
 
It is interesting to see that the above results for masses can be rederived from eq.~\eqref{eq:masses-g4}  if one assumes that the loop terms proportional to $b^{\textrm{SM}}$ and $b^{X}$ are subleading with respect to the terms proportional to $\lambda_2$ in that equation. If we neglect the first term under the square root, which is quadratic in the loop terms, and expand the remaining square root assuming that 
 $$
 \frac{16\left(b^{\textrm{SM}} -b^{X}\right)\left(v^2-w^2\right)}{\lb\left(v^2+w^2\right)^2}
 $$
 is small we obtain the two following mass eigenvalues,
 \begin{align}
 M^2_{-}&=8\frac{(b^{\textrm{SM}}v+b^{X}w)}{v^2+w^2}=8\frac{\mathbb{B}}{v^2+w^2}=8\frac{\mathbb{B}}{\rho^2}\notag\\
 M^2_+&=-\lb(v^2+w^2)=-\lb\rho^2,\notag
 \end{align}
which are the same as eqs.~\eqref{eq:GW-tree-mass} and~\eqref{eq:GW-loop-mass}, with the difference that within GW method the minimum is assumed to lie along the tree-level flat direction (unless the loop-induced shift is taken into account). Moreover, in the GW approach one can incorporate the scalar contributions to $\mathbb{B}$ without complications. 

We note, as was already mentioned before, that the GW method can be used also when the tree-level potential is of the order of the one-loop corrections. In such cases it gives a local minimum, however one must check that no other deeper minima appear due to the interplay between $\Vtree$ and $\Vone$.
\subsubsection{Contributions of different orders\label{sec:seq+}}
 There is also a possibility that the scalar couplings are of different orders.  
 \begin{enumerate}
 \item Assume that to leading order we can neglect $\la$  and $\Vone$ contributions in the stationary-point equations~\eqref{eq:stat1}, \eqref{eq:stat2} (the terms proportional to $\lambda_2$ and $\lc$ dominate). Then, the dominant part of the tree-level potential (which consists only of the terms proportional to $\lb$ and $\lc$ if  we can neglect the $\la$ term in the potential) has a minimum and is equal zero along the $\f=0$ direction, and $v$ remains undetermined. This means that  the leading-order potential develops a  flat direction along the $h$ field so this is in fact a GW case and CW mechanism along this direction can be studied. We can study the subdominant part of the potential solely along the flat direction, which gives $V(h)=\frac{1}{4}\la h^4+\Vone(h,\f=0)$. This is the SM effective potential and it is known that this potential does not have a stable radiatively generated minimum (it is not possible to obtain positive second derivative of the effective potential). Therefore, this scenario is not realistic.
 
 \item Assume that the term proportional to $\lc$ is of the order of the one-loop contribution and negligible in comparison with the term proportional to $\lb$ and $\la$. This case is symmetric to the previous one, now the GW flat direction is along $\f$. Therefore, the effective potential can be studied in the following form $V(\f)=\frac{1}{4}\lc\f^4+\Vone(h=0,\f)$, which is the pure hidden sector effective potential. If now $\lc\sim\mathcal{O}(g_X^4)$, RSB can occur and the $\f$ field may acquire a nonzero VEV. However, in such a case the VEV of $h$ is zero at leading order and the negligible loop terms can at most induce a small shift of this VEV, as in the GW case. Therefore, it is not possible to reproduce the observed masses of gauge bosons and fermions this way.
 
 \item Let us now focus on the contributions proportional to $\lambda_2$, which couples the two scalar sectors of SU(2)cSM. If they are negligible with respect to both the $\la,\lc$ and one-loop contributions, the two sectors are decoupled and there is no viable RSB scenario. However, since $\lb$ is multiplied by different powers of the VEVs in both equations, and moreover $\la$ and $\lc$ can be of different orders, one of the $\lb$ terms can belong to leading while the other to subleading order. The only viable option is such that the $\lb$ term in eq.~\eqref{eq:stat2} is negligible, compared to the $\lc$ and one-loop terms, whereas it belongs to leading order in eq.~\eqref{eq:stat1}, together with the $\la$ term. This can happen when $\lb$ is small and $w/v>1$.\footnote{Note that this can only be checked after one computes the VEVs. The importance of the hierarchy of the VEVs was also pointed out in ref.~\cite{Guo:2015}.} Then, if $\lc\sim\mathcal{O}(g_X^4)$, the $\f$ field can acquire a nonzero VEV, which contributes to generation of an effective mass term for the $h$ field in eq.~\eqref{eq:stat1}. Then, symmetry breaking in the SM sector proceeds in the usual way, and the loop terms only contribute as subleading corrections to the mass. This is the aforementioned sequential approach.

 Within this scenario, assuming that the SM contributions to the one-loop potential do not shift the location of the minimum, i.e.\ exploiting the fact that the second (one-loop) term on the right-hand side of eq.~\eqref{eq:stat1-g4} is subleading with respect to the first one, we obtain the tree-level relation between the VEVs 
 $$
 \frac{v^2}{w^2}=-\frac{\lb}{2\la},
 $$
 as in the GW case. Then  the mass matrix in eq.~\eqref{eq:mass-matrix-g4} acquires the following form (see e.g.\ ref.~\cite{Khoze:2014})
 \be
 M^2=\left(\begin{array}{cc}
2 \la v^2 + 8 \mathbb{B}^{\textrm{SM}}/v^2  & \sqrt{-2\la\lb}v^2\\
\sqrt{-2\la\lb}v^2 & -\lb v^2 + 8\mathbb{B}^{X}/w^2\\
 \end{array}
 \right),\notag
 \ee
where the mixing terms and $8 \mathbb{B}^{\textrm{SM}}/v^2$ represent subleading corrections. The difference with respect to the general approach lies in the assumption the tree-level relation between the two VEVs holds.
  \end{enumerate}

\subsection{Analysis of the one-loop effective potential\label{sec:CW-analysis}}
\subsubsection{Method of numerical analysis\label{sec:method-numerical}}
In the following we perform a numerical study of RSB in the SU(2)cSM, using the one-loop effective potential. Our aim is to study the possibility of RSB in this model in the available parameter space. Moreover, we determine the region where no Landau poles of the running couplings appear below the Planck scale  and the potential is bounded from below. In such a region SU(2)cSM can be considered a viable and self-consistent model. We refer to this region as the stable region.

In the analysis we use the one-loop effective potential 
\be
V(h,\f)=\Vtree(h,\f) + \Vone(h,\f),\notag
\ee
where $\Vtree$ is given by eq.~\eqref{eq:Vtree}, and $\Vone$ by eqs.~\eqref{eq:one-loop} and~\eqref{eq:mass-scalar1}--\eqref{eq:mass-top}. For computational simplicity we neglect the scalar contributions to the one-loop correction, later checking the applicability of such approximation. 

The computational procedure is as follows. The initial set of free parameters consists of $\la$, $\lb$, $\lc$ and $\gcw$.\footnote{For the SM parameters we use the following values: $g=0.653$, $g'=0.358$ at $\mu=M_Z=91.1876\g$. To evolve the couplings to other energy scales we use one-loop SM $\beta$ function and anomalous dimensions.} We start from solving the stationary-point equations and expressing $\la$ and $\lc$ in terms of $\lb$, $\gcw$, $w$ and $v$. Then, since we want to obtain masses of the gauge bosons that are in agreement with the experimentally measured values, we set $v=246\g$. We also fix $\mu=246\g$, therefore all the couplings are evaluated at $\mu=v$. In the next step of the procedure we fix the values of $\lb$ and $\gcw$ to randomly chosen values. For $\gcw$ we consider the interval
$$
\gcw \in [0.1,\;1.1]
$$
and for $\lb$ three different regimes
$$
\lb \in [-0.01,\;-0.001], \quad \lb \in [-0.1,\;-0.01],\quad \lb \in [-1,\;-0.1].
$$
Then we numerically look for $w$ such that we reproduce the correct Higgs mass, $M_H=125\g$, assuming that the running mass (which corresponds to an eigenvalue of the Hessian of the effective potential, see eq.~\eqref{eq:masses-g4}) can be used as a reliable approximation of the physical mass.  A better accuracy would be obtained if self-energy corrections to the running mass were taken into account, however this is beyond the scope of the present work.  The running-mass approximation is also most frequently used in the literature and since we want to compare different commonly employed methods for studying RSB, we should follow the conventions. 

Using the values of the parameters found with this procedure we compute the mass of the other scalar, $M_S$ and of the new gauge boson $M_X$. There are two possible scenarios --- either the Higgs boson or the other scalar is lighter. We consider both of these scenarios. We also check whether the minimum found is the global minimum of the potential. In this way we obtain the region of the parameter space where viable RSB mechanism can be realised.

Further on we use the values of the scalar couplings and $\gcw$ as boundary conditions for the running of the couplings and solve their RGE equations. We check at which scales  Landau poles arise in the scalar couplings and treat the lowest of such scales as a limit of applicability of a model with such couplings.\footnote{We run our computations with \textit{Mathematica}~\cite{Mathematica} and in evaluating the location of the Landau pole we rely on its ability to detect the region where a function is too steep to be evaluated in the process of numerically solving the differential equation that defines it.} Finally, if there are no poles up to the Planck scale, we check the stability conditions, eq.~\eqref{eq:positivity} evaluated at the Planck scale to verify that the potential is stable~\cite{Sher:1989, Chataignier:2018}. 

We also check what the mixing between $h$ and $\f$ in the mass-eigenstate with $M_H=125\g$ is, since it determines how the couplings of the Higgs boson to the gauge bosons and fermions are rescaled with respect to the SM. This rescaling is given by $\cos \theta$ in the case with $M_H<M_S$ and by $-\sin\theta$ when the ordering is opposite (see eq.~\eqref{eq:mixing}). This mixing is constrained by the LHC and LEP, through the measurements of the $W$ boson mass, Higgs signal rates and LHC SM Higgs searches. Since only $h$ couples to the gauge bosons and their couplings are SM-like, a rule of thumb is that  $\cos\theta$ should be close to one when $M_H<M_S$, and $-\sin\theta$ should be close to one in the opposite case. We can obtain more concrete constraints by reinterpreting the results obtained for a singlet extension of the SM~\cite{Robens:2016, Robens:2015, Ilnicka:2018}, since the scalar sectors of the two models are very similar (in the absence of $H\to XX$, which, as we will see later, are kinematically forbidden, the decay channels of the Higgs boson as well as its couplings are the same in both models). In this analysis we use the bounds from Higgs signal strengths as an approximate indicator of the allowed magnitude of the mixing (see refs.~\cite{Robens:2016, Robens:2015, Ilnicka:2018} for a more detailed discussion). This leads to the following bounds: $|\cos\theta|\geqslant 0.93$ for $M_H<M_S$ and $|\sin\theta|\geqslant 0.87$ for $M_H>M_S$.\footnote{These bounds do not apply when the two masses do not differ much, however discussion of this case is beyond the scope of the present paper, for more details see refs.~\cite{Robens:2016, Robens:2015, Ilnicka:2018}.}

\subsubsection{Small portal coupling\label{sec:g4}}

Using the method described in the previous section with $\lb \in [-0.01,\;-0.001],$ we generated $2\cdot 10^5$ points in the parameter space for each of the two possible mass orderings --- the Higgs boson $H$ can be either lighter or heavier than the additional scalar $S$. 

\begin{figure}[h!]
\center
\includegraphics[height=.35\textwidth]{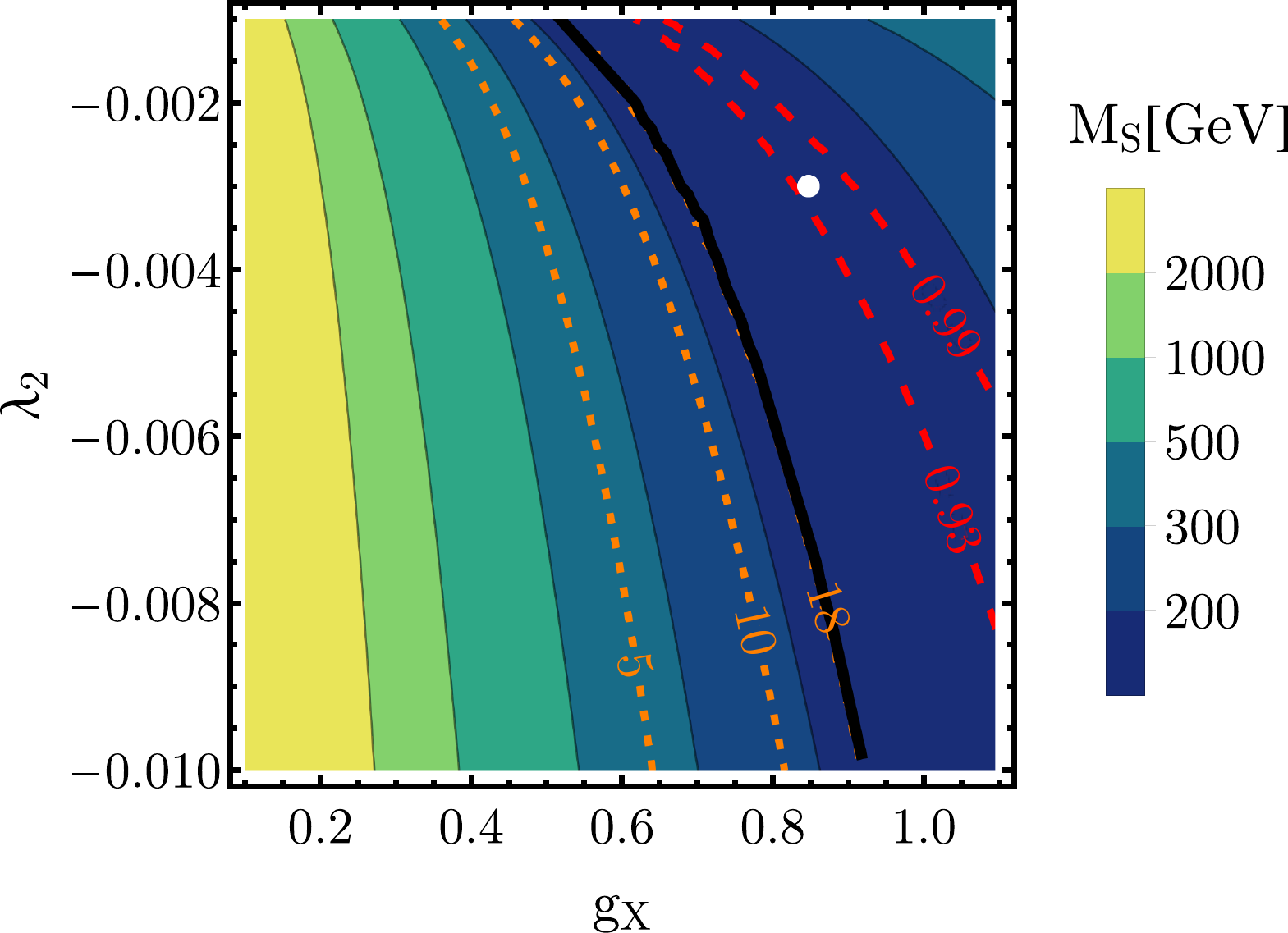}\hspace{.3cm}
\includegraphics[height=.35\textwidth]{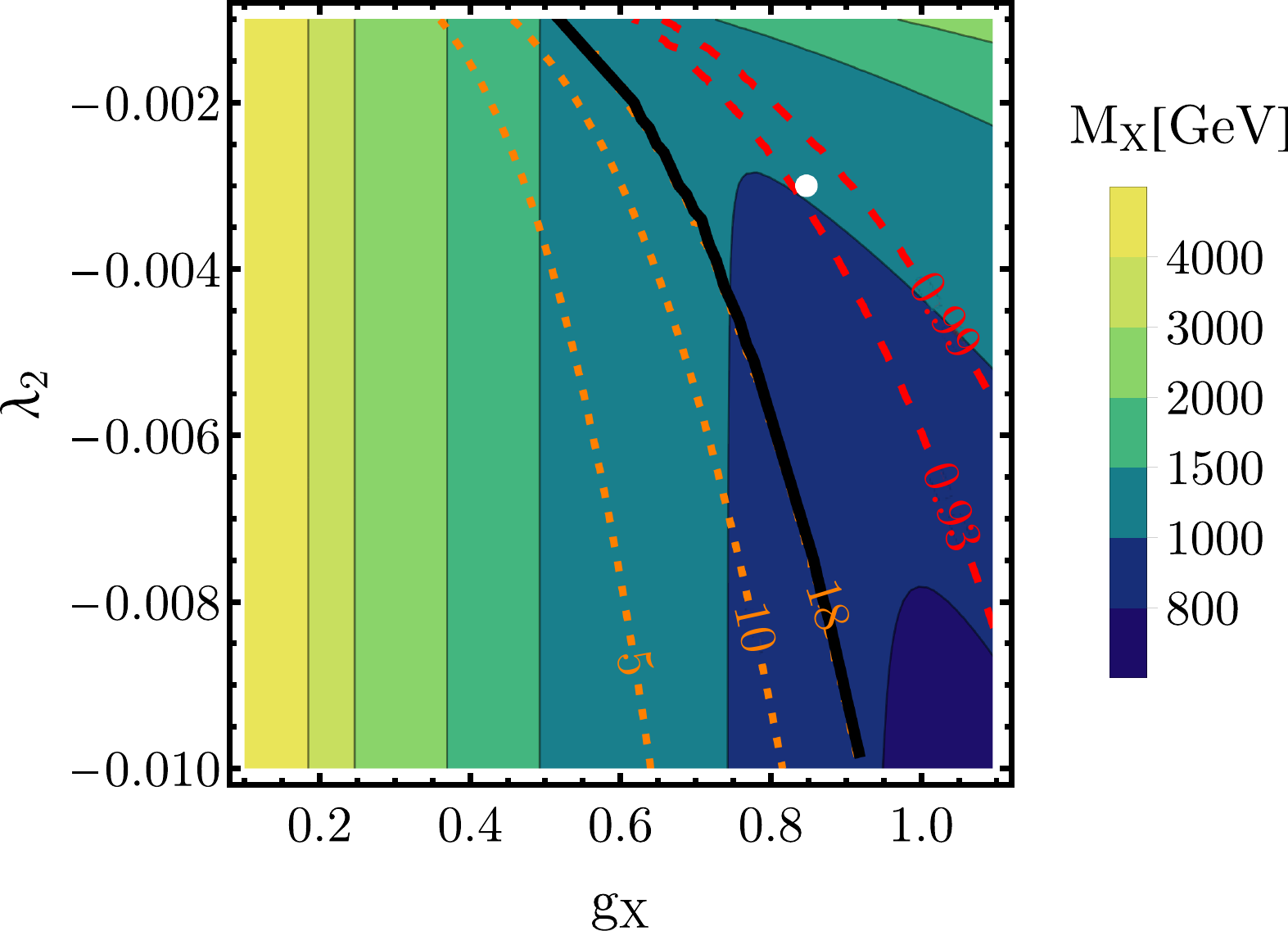}
\\[5pt]
\includegraphics[height=.35\textwidth]{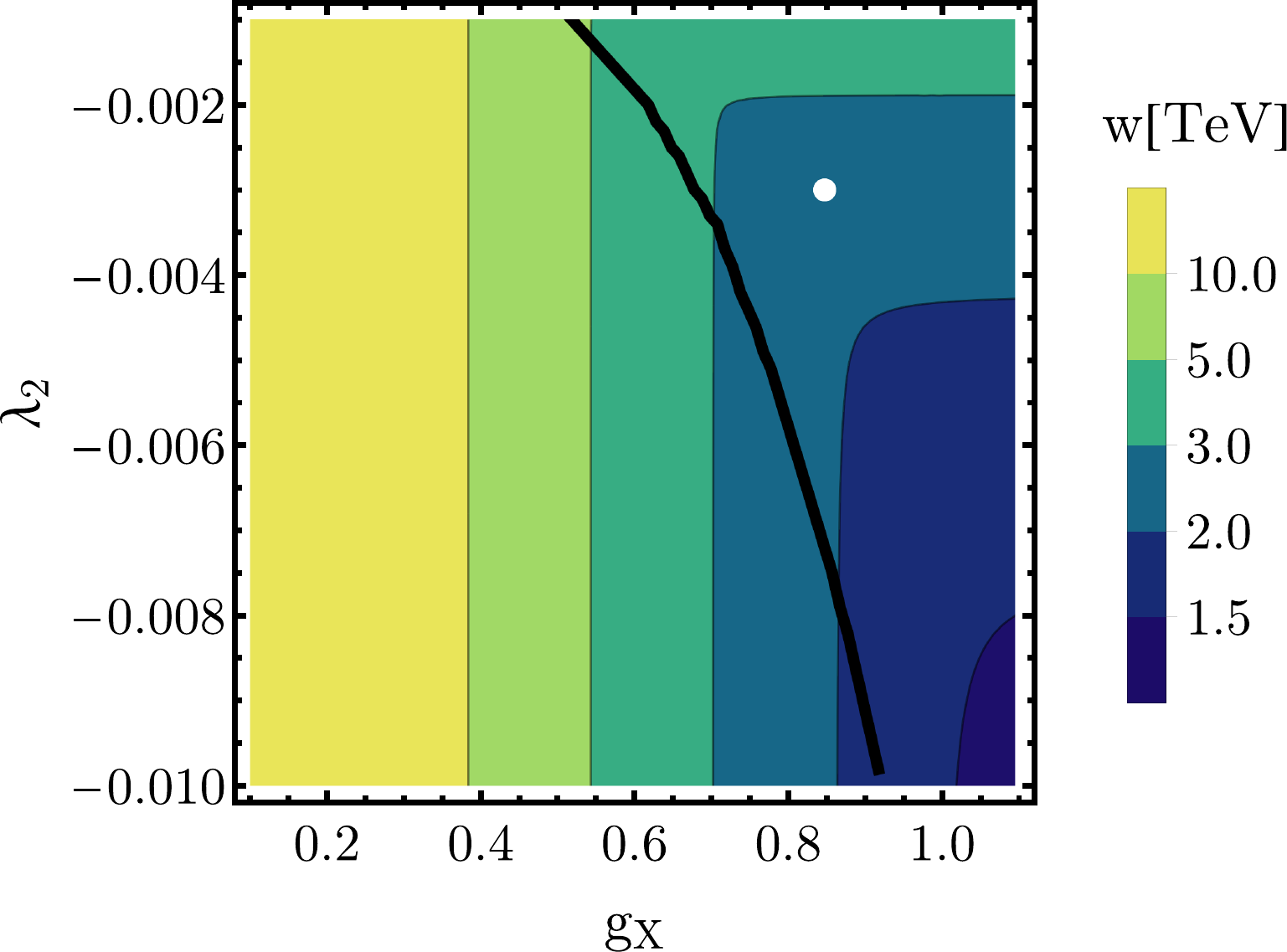}\\[5pt]
\includegraphics[height=.35\textwidth]{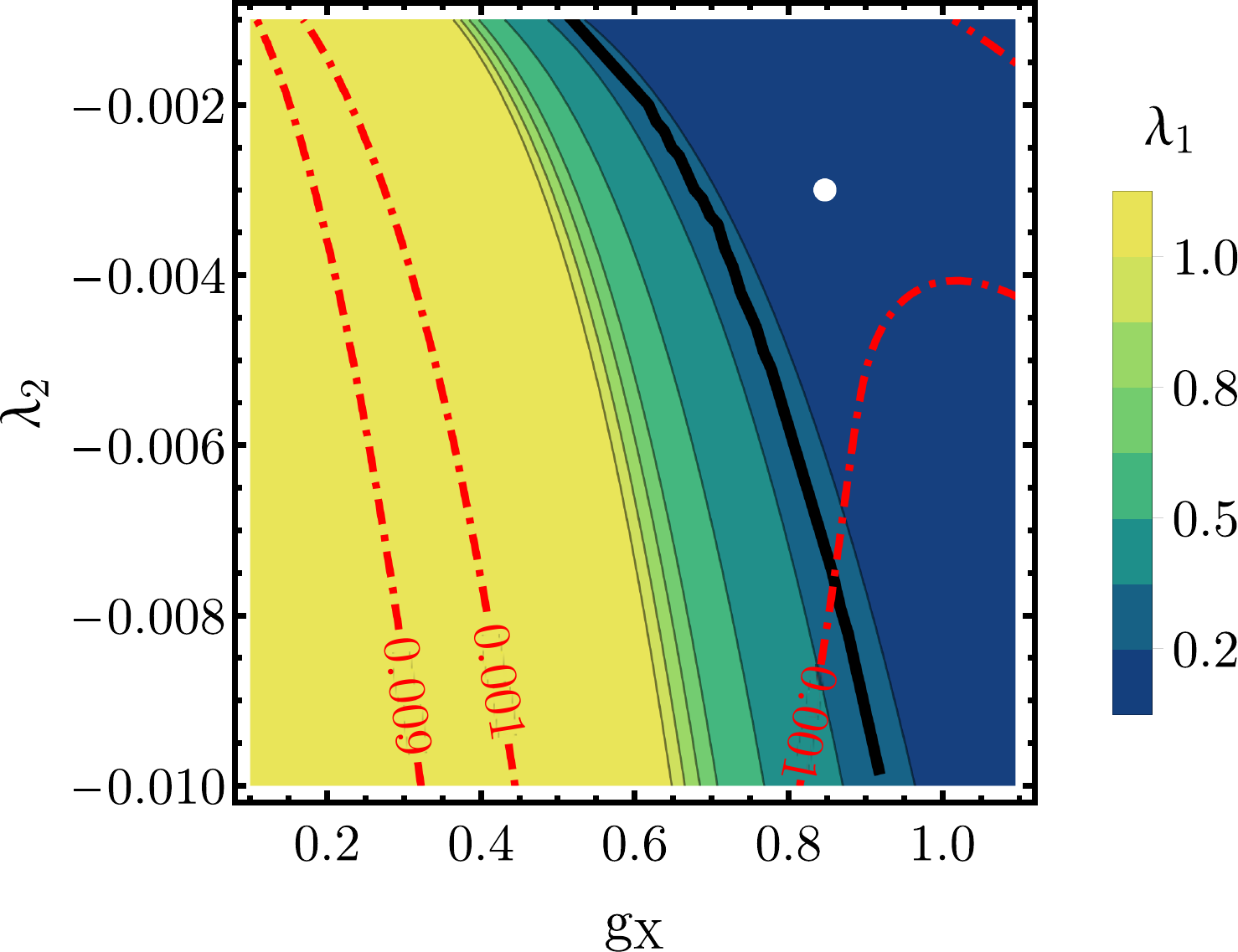}\hspace{.3cm}
\includegraphics[height=.35\textwidth]{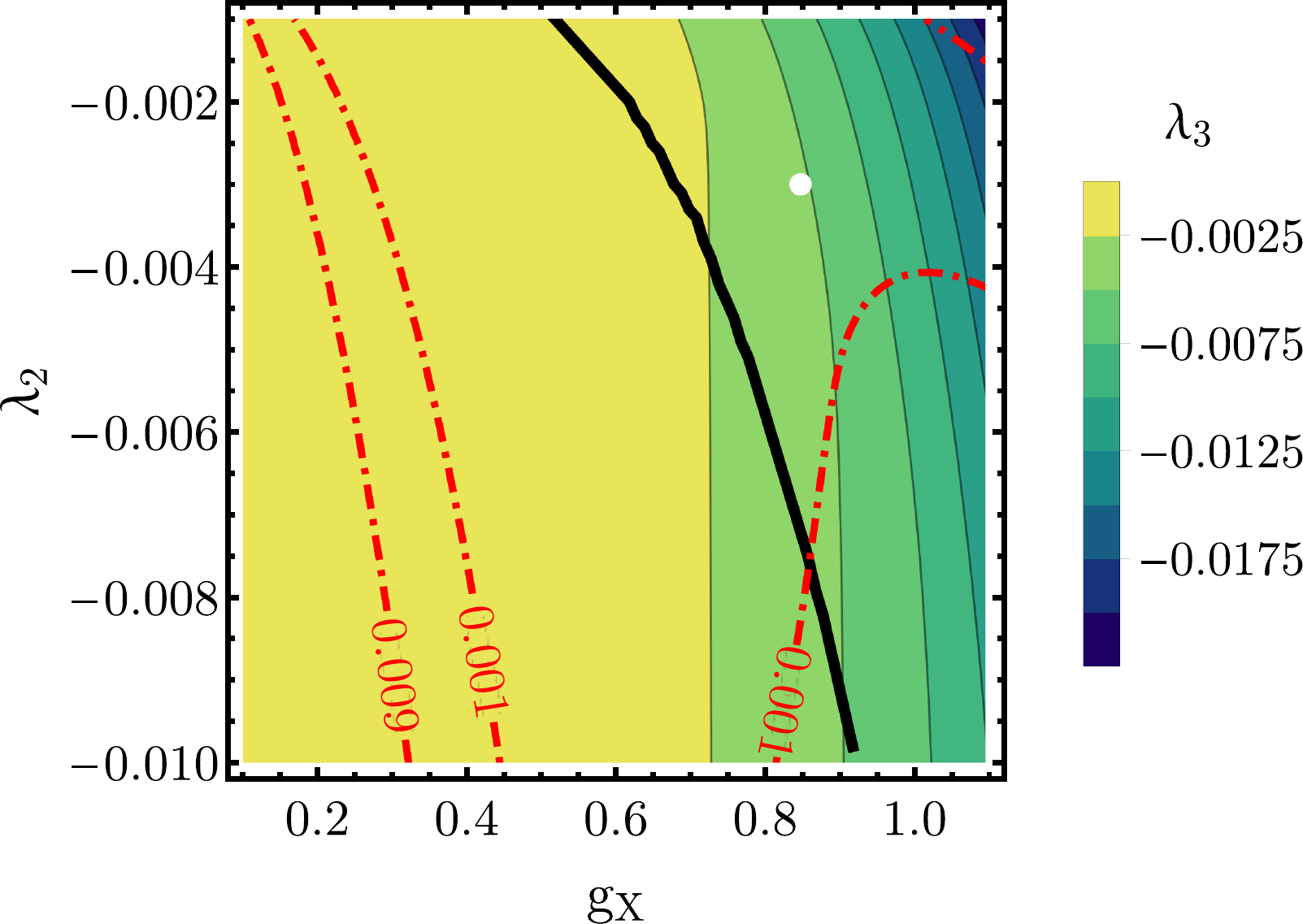}
\caption{Results of the scan of the $(\lambda_2, \gcw)$ parameter space for the mass hierarchy $M_H<M_S$. Upper panel: contour plots of the values of the mass of the extra scalar  $M_S$ (left) and of the new gauge bosons $M_X$ (right). The orange short-dashed lines show the decimal logarithm of the energy scale at which a Landau pole appears, the thick black line represents the boundary of the region where the potential is bounded from below  at the Planck scale (to the right of the curve). The long-dashed red lines represent the contours of constant values of $\cos\theta$, see eq.~\eqref{eq:mixing}. Middle panel: values of the VEV of the $\f$ field, $w$. Lower panel: values of $\la$ and $\lc$. The dot-dashed lines correspond to contours of constant values of the ratio given in eq.~\eqref{eq:scalar-approx}. The white dot represents the benchmark point from table~\ref{tab:BM}.\label{fig:g4-1}}
\end{figure}

Figure~\ref{fig:g4-1} shows the results of the scan for the $M_H<M_S$ scenario. In the upper panel possible values of the mass of the new scalar particle, $M_S$ (left) and of the new gauge bosons $M_X$ (right) are shown as functions of the couplings $\lb$ and $\gcw$.  On top of the mass contour plots we show the values of the decimal logarithm of the scale at  which a Landau pole appears as short-dashed orange contours. If there are no Landau poles below the Planck scale, we check whether positivity conditions are fulfilled at large energy scales (we take the scale at which the couplings are evaluated to be the Planck scale). If this is the case we consider the potential to be stable and mark the boundary of that region by the thick black line (the region of stability corresponds to the region above the curve). As explained before, this curve marks the boundary of applicability of the model and thus is also shown in the remaining plots. The shape of the boundary agrees qualitatively with the results presented in ref.~\cite{Carone:2013}. The red long-dashed lines show the contours of $\cos\theta$ (see eq.~\eqref{eq:mixing}). The stable region features the values of $\cos\theta$ in agreement with experimental bounds, the line with $\cos\theta=0.99$ lies also well within the stable region so even with improved precision of measurements this model will not be ruled out because of the mixing constraints. 

To sum up, from the results presented in the upper panel of figure~\ref{fig:g4-1} it is clear that the SU(2)cSM can account for RSB, giving the correct mass of the Higgs particle (and the other SM particles) and preserving stability and perturbativity up to the Planck scale. We can see that the extra scalar cannot be very heavy, with mass up to around $400\g$. The new gauge bosons, the $X$ particles, can have masses up to around $2\, \mathrm{TeV}$. The rescaling of the Higgs boson couplings due to mixing is within experimental bounds. 

In the middle panel of figure~\ref{fig:g4-1} we show the values of the VEV of the $\f$ field, $w$. In the stable region it ranges from around 1.5\,TeV to more than 3\,TeV. It is thus generically much, by a factor of $\mathcal{O}(10)$, higher than the VEV of the $h$ field. This is the complication of multi-scalar models discussed above --- there need not exist a common scale for both of the fields. Such discrepancy in the magnitudes of the two VEVs might (but does not necessarily) lead to complications in perturbative treatment of RSB. Since we have fixed $\mu$ to be equal to $v$, the contributions from the $X$ bosons to the effective potential may become sizeable. We will  discuss this issue later.

As we have underlined above, it is important to keep the hierarchy of couplings in mind. Therefore in the lower panel of figure~\ref{fig:g4-1} we show the values of $\la$ (left) and $\lc$ (right) obtained within the scan of the parameter space. One can see that $\la$ can acquire rather large values, above 1 (in fact the values in the lower part of the plot can be unreasonably large, even above 100). But comparing with the upper panel of figure~\ref{fig:g4-1} one realises that, not surprisingly, this is exactly the region where Landau pole(s) appear at low energies and these points are rejected.  In the stable region, i.e.\ in the upper right part of the plot the values of $\la$ are moderate and rather close to the SM value. The parameter $\lambda_3$ acquires very small values in the whole parameter space.

On top of the plots showing the values of the scalar couplings we present contours (red dot-dashed lines) of constant values of the ratio 
\be
\frac{V^{(1)}_{\textrm{scalar}}(v,w)}{V^{(1)}(v,w)},\label{eq:scalar-approx}
\ee
 where $V^{(1)}_{\textrm{scalar}}(v,w)$ is the value of the one-loop scalar contribution to the effective potential evaluated at the minimum while $V^{(1)}(v,w)$ represents the contributions from the remaining particles, i.e.\ the one-loop effective potential used in the numerical analysis. The value of this ratio tells us whether the approximation to the effective potential with no scalar contributions is justified. From the results shown in the plot it is clear, that with the values of $\frac{V^{(1)}_{\textrm{scalar}}(v,w)}{V^{(1)}(v,w)}$  smaller than 0.01 (surprisingly even in the region where $\lambda_1$ is rather large) this approximation is perfectly valid in the studied case.
 
For discussions presented in the next parts of this article it will be convenient to know what the behaviour of couplings and different contributions to the effective potential is. For this sake we analyse one benchmark point, marked in figure~\ref{fig:g4-1} with a white dot, the values of the couplings and relevant contributions are displayed in table~\ref{tab:BM}. The values of $\lb$ and $g_X$ presented in the first row of table~\ref{tab:BM} (for $\mu=246\g$) correspond to the values on the axes of the plot in figure~\ref{fig:g4-1}, the values in the remaining rows correspond to values at different scales, as for the other couplings.  The GW scale in the second row is the scale at which the tree-level potential develops a flat direction and at that scale computations are performed when GW method is applied (this is discussed in more detail in section~\ref{sec:SU2-GW}). The third row of table~\ref{tab:BM} corresponds to the RG scale, which is given by $\mu_0 e^{ t_*}|_{h=v,\f=w}=246\g e^{ t_*}|_{h=v,\f=w}$, where $t_*$ is defined in eq.~\eqref{eq:t*-def}. From the presented values it is clear that the $\lambda_1$ and $\lambda_2$ couplings run slowly around the discussed scales (the running of $\lb$ is only visible at further decimal positions), while $\lc$ can change by 1--2 orders of magnitude. It is striking to see how dramatically the ratio $\Vone/\Vtree$ can change with the scale. This however can be understood easily if one remembers that the scales considered are special: at the GW scale the tree-level potential vanishes at the minimum (therefore the ratio $\Vone/\Vtree$ is not infinity only because of numerical inaccuracies), while at the RG scale the one-loop correction vanishes (again numerical uncertainties lead to non-vanishing ratio $\Vone/\Vtree$).
\renewcommand{\arraystretch}{1.5}
\setlength{\tabcolsep}{5pt}
\begin{table}[h!]
  \begin{center}
\begin{tabular}{c|cccr@{.}lcccr@{$\;\cdot\;$}lc}
&$\mu\,[\textrm{GeV}]$&$\la$&$\lb$&\multicolumn{2}{c}{$\lc$}&$g_X$&$w\,[\textrm{GeV}]$&$\Vone_{\textrm{SM}}[\textrm{GeV}^4]$&\multicolumn{2}{c}{$\Vone_{\textrm{X}}[\textrm{GeV}^4]$}&$\Vone/\Vtree$\\\hline
CW&246&0.1236&-0.0030&-0&0047&0.8500&2411&$2.38\cdot10^7$&3.18&$10^{10}$&0.802\\
GW&940&0.1055&-0.0030&\multicolumn{2}{c}{$2\cdot10^{-5}$}&0.8141&2722&$6.28\cdot10^7$&-1.08&$ 10^{10}$&551\\
RG&734&0.1085&-0.0030&-0&0008&0.8204&2680&$5.82\cdot10^7$&-5.79&$10^7$&$-3\cdot 10^{-5}$
\end{tabular}
\end{center}
\caption{Values of the couplings, the $\phi$ VEV, the one-loop corrections from the SM sector and the hidden sector to the effective potential evaluated at the minimum and the ratio of the one-loop to tree-level contributions to the effective potential at the minimum at various renormalisation scales (corresponding to the $h$ VEV, the GW scale $\mgw$ and the RG scale) for the benchmark point shown in figure~\ref{fig:g4-1}.\label{tab:BM}}
\end{table}

Using the values presented in table~\ref{tab:BM} we can assess the magnitudes of different contributions to the stationary-point equations~\eqref{eq:stat1-g4}--\eqref{eq:stat2-g4}. In the first equation the term proportional to $\lambda_2$ is enhanced by the large ratio of the two scalar VEVs, resulting in a contribution with magnitude around -0.29, which is comparable to the $\lambda_1$ contribution. On the other hand, the loop contribution is of the order of 0.02, providing only a small correction to the tree-level result. In the second minimisation condition, eq.~\eqref{eq:stat2-g4}, the term proportional to $\lb$ is suppressed by the inverse ratio of the VEVs squared, giving a contribution of approximately $-3\cdot 10^{-5}$, which is negligible in comparison with both $\lc$ and the loop contribution, which is approximately equal to 0.005.  This suggests that the sequential approach, discussed in section~\ref{sec:seq+} is valid in the present case. In fact, normally the sequential analysis is performed at $\mu=w$ which allows to eliminate the dependence on $v$ from the second minimisation condition. The sizes of the contributions at that scale can be different from the ones at $\mu=v$, therefore for a final conclusion one should verify the magnitudes of the contributions at the scale corresponding to $w$. Moreover, one should be aware, that the applicability of this approximate method relies crucially on the ratio of the VEVs which can be determined only after the location of the minima is known. This means that one has to be careful if one wants to use the sequential approach to search for the minima of the potential and check the validity of the applied approximation a posteriori. Moreover, this example confirms that in the study of RSB in models with more scalar fields not only the hierarchy of couplings but also the hierarchy of VEVs matters.

\begin{figure}[h!]
\center
\includegraphics[height=.35\textwidth]{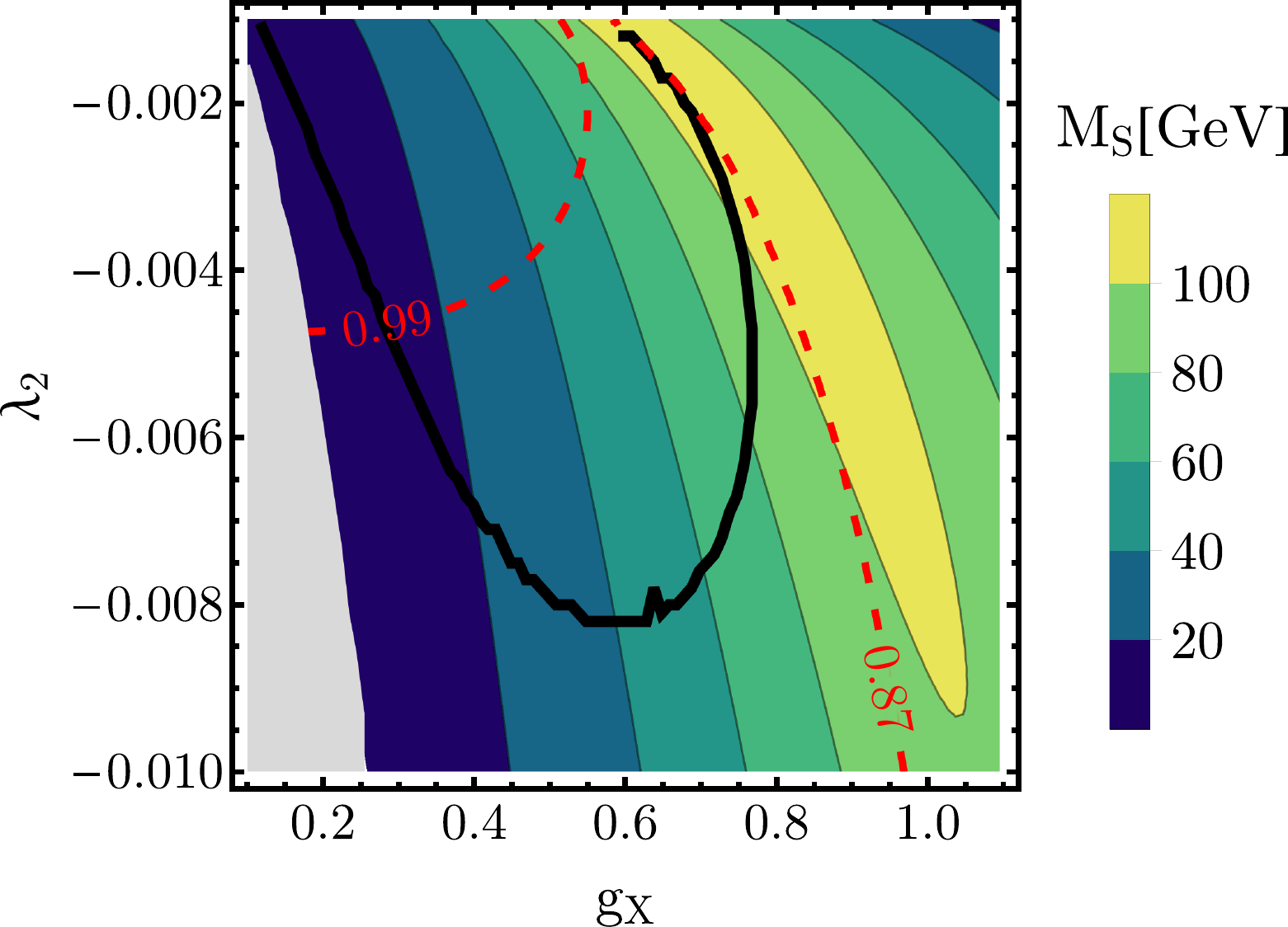}
\includegraphics[height=.35\textwidth]{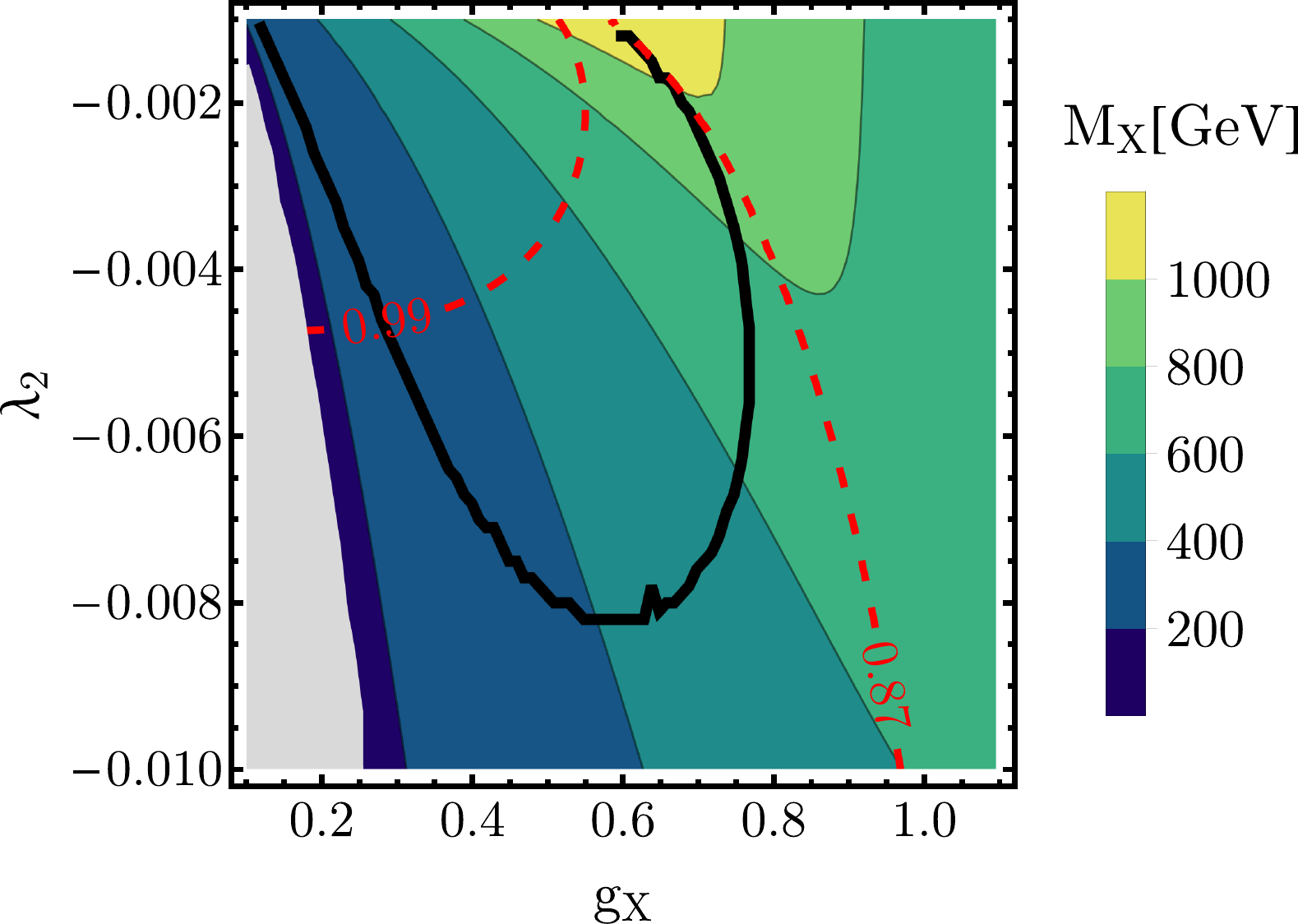}\\[5pt]
\includegraphics[height=.35\textwidth]{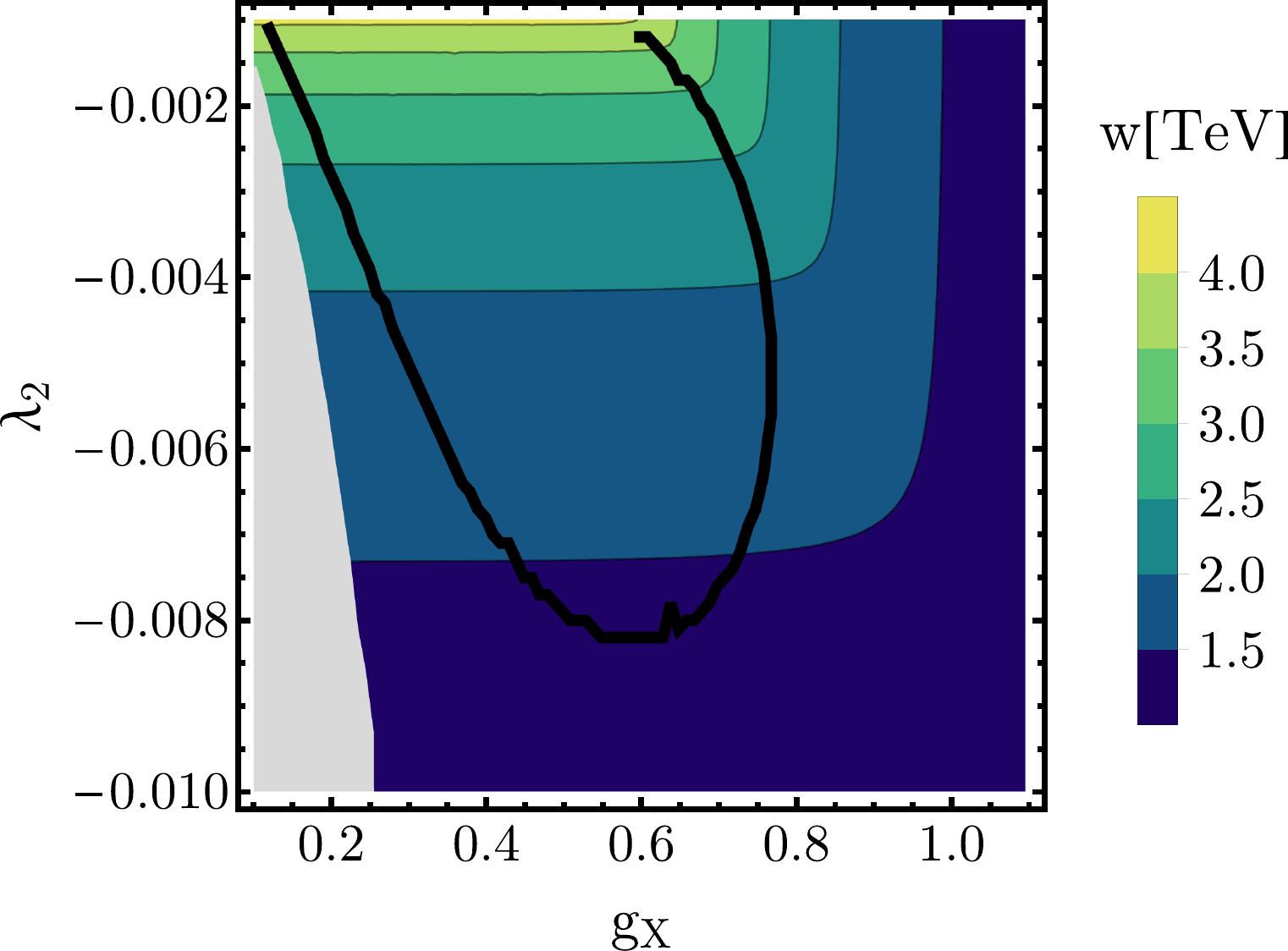}\\[5pt]
\includegraphics[height=.35\textwidth]{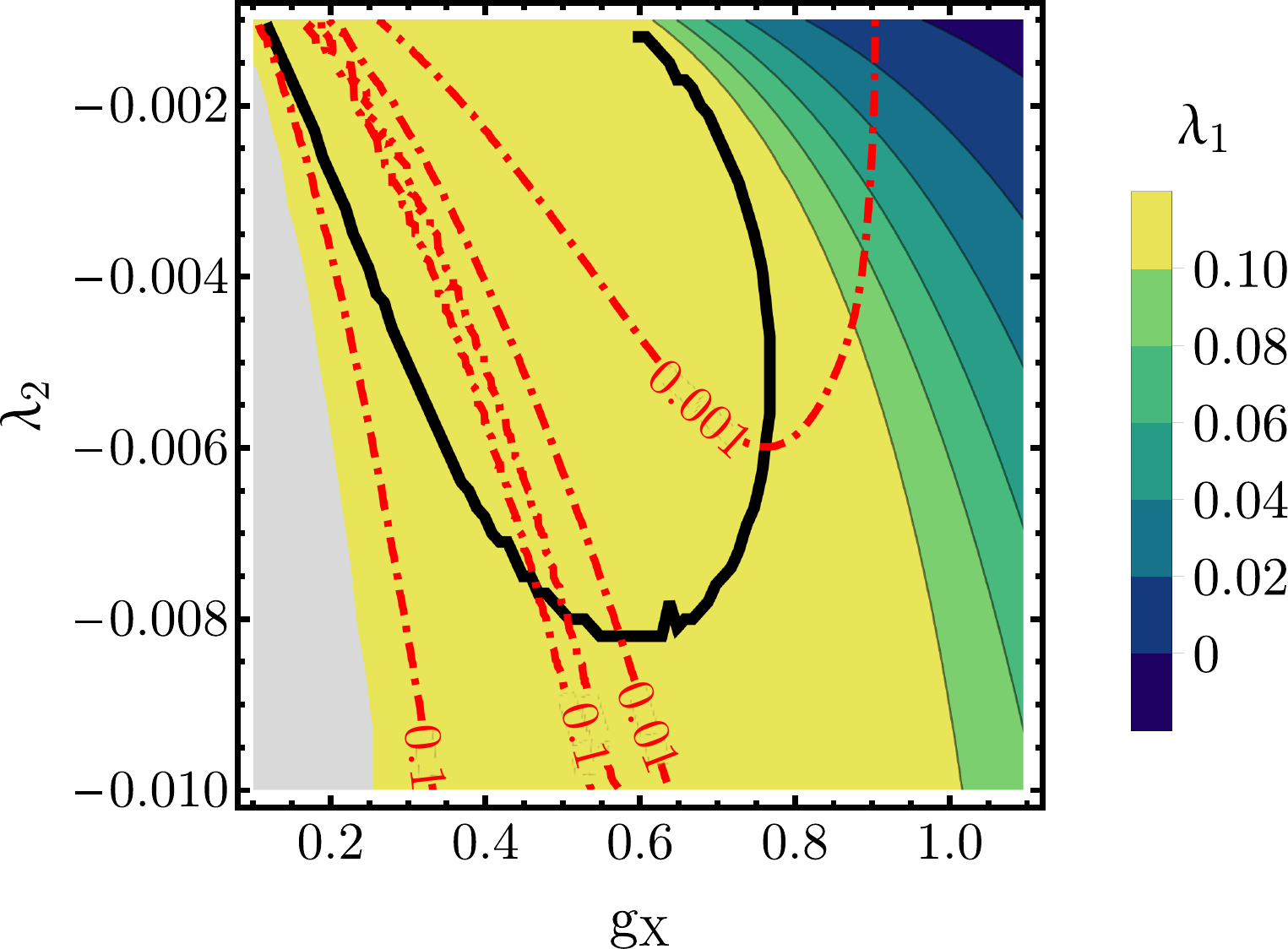}
\includegraphics[height=.35\textwidth]{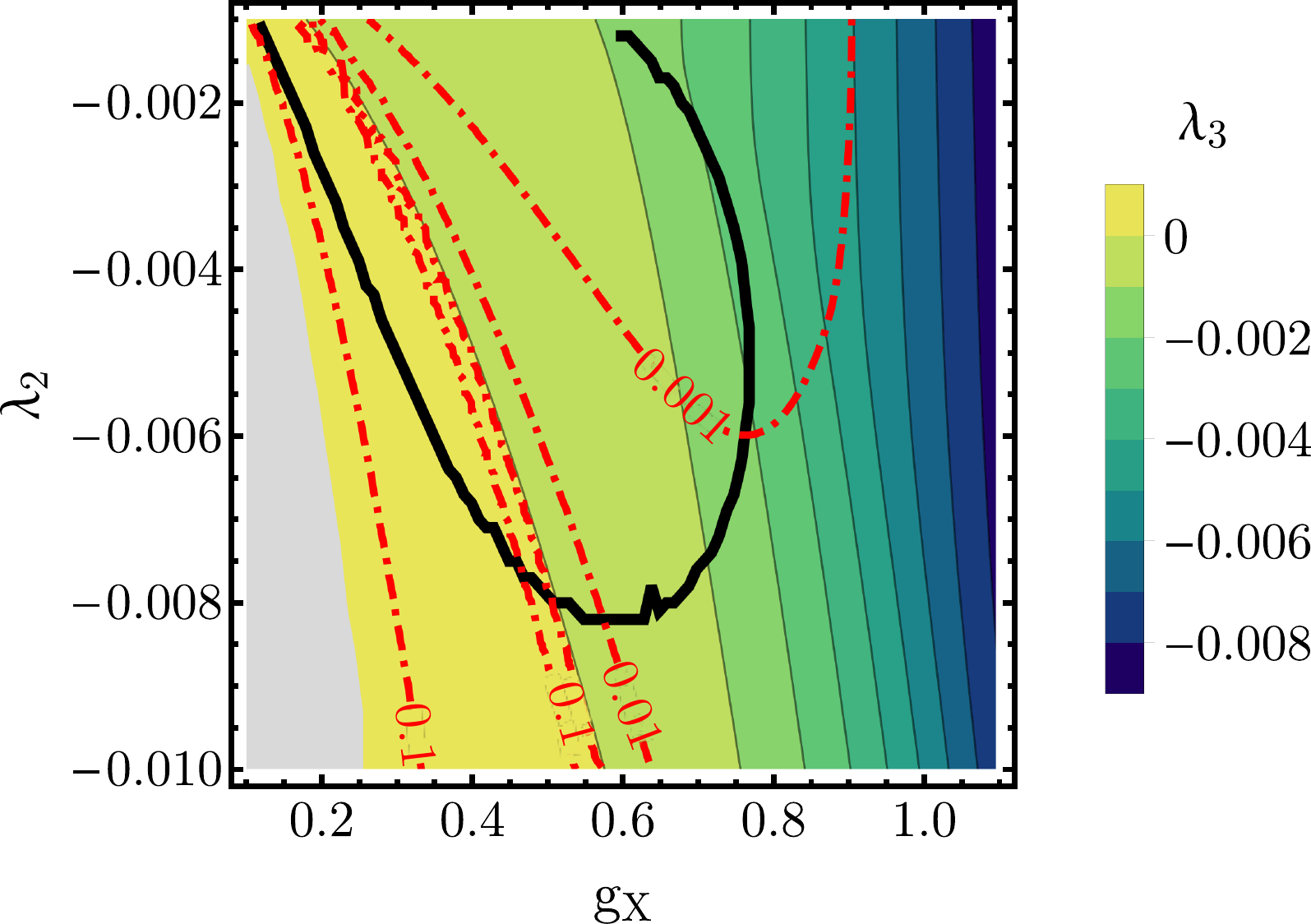}
\caption{Results of the scan of the $(\lambda_2, \gcw)$ parameter space for the mass hierarchy $M_H>M_S$. Upper panel: contour plots of the values of the mass of the extra scalar  $M_S$ (left) and of the new gauge bosons $M_X$ (right). The thick black line shows the boundary of the region where the potential is bounded from below (above the curve), the long-dashed red lines represent the contours of constant values of $|\sin\theta|$, see eq.~\eqref{eq:mixing}. Middle panel: the values of the VEV of the $\f$ field, $w$. Lower panel: the values of $\la$ and $\lc$.   The red dot-dashed lines in the lower panel correspond to contours of constant values of the ratio given in eq.~\eqref{eq:scalar-approx}. The grey region is excluded since there is no stable minimum in this region.\label{fig:g4-2}}
\end{figure}

The results of the numerical scan of the parameter space for the scenario where the extra scalar $S$ is lighter than the Higgs boson  are presented in figure~\ref{fig:g4-2}. We have found that the running coupling constants of the scalar sector do not develop Landau poles below the Planck scale. That is why there are no analogs of the orange dashed lines from figure~\ref{fig:g4-1} in this plot. The potential is stable up to the Planck scale above the thick black line (the shape of the line is in qualitative agreement with the results of ref.~\cite{Carone:2013}). In the grey region the stationary point found in the numerical procedure does not correspond to a stable minimum. Again, in the region that guarantees stability of the effective potential the mixing is small, the experimental bound $|\sin\theta|\geqslant0.87$ does not introduce new constraints on the parameter space. Even extremely small mixing corresponding to $|\sin\theta|=0.99$ could be accommodated within the stable region.

The possible range of mass values for the new mass-degenerate gauge bosons $X$ is narrower than it is in the $M_H<M_S$ case, ranging from above $200\g$ to around $1\,$TeV. The range of the $\f$ VEV is only slightly narrower than in the previous case, with $w$ between around 1.5\,TeV and 4\,TeV. The plots illustrating the values of the scalar couplings (lower panel of figure~\ref{fig:g4-2}) show that in the majority of the parameter space and in particular in the most interesting region where the potential is stable, $\la$ has SM-like values (the maximal value of $\la$ is 0.12). In the remaining parameter space it is very small. The values of $\lc$ are small in the whole parameter space analysed here. The approximation of neglecting scalar contributions to the effective potential is well valid in a big part of the stable region. In a narrow band it becomes greater that 0.1, which is most likely related to IR divergences coming from Goldstone contributions --- since the scalar contributions are suppressed by terms of order $\lambda_i^2$, while all scalar couplings are small, a big contribution from scalars can only come from the logarithmic terms.

Summing up, in the regime $\lb\in[-0.01,-0.1]$ SU(2)cSM can reproduce the correct mass spectrum of the observed particles and predicts the existence of an extra scalar particle which can be  lighter or heavier than the Higgs boson, with mass up to around $500\g$ and of three gauge bosons of equal masses  from around $200\g$ to above $1\,$TeV. 

Since the gauge bosons can in principle be good candidates for dark matter particles, this mass range could be constrained by requiring that the relic abundance of $X$ is in agreement with the measurements, however this topic is out of the scope of the present analysis. For related studies of freezing-out DM see refs.~\cite{Hambye:2013, Carone:2013, DiChiara:2015} and ref.~\cite{Hambye:2018} where a new mechanism of generation of the DM relic density was introduced. In ref.~\cite{Carone:2013} it has been shown that the scenario with light extra scalar, $M_S<M_H$ cannot reproduce the correct DM relic abundance via the freeze-out scenario therefore in the following where we compare different methods of studying RSB we focus on the case with the $M_H<M_S$ mass ordering.

\subsubsection{Intermediate and stronger portal coupling\label{sec:g2}}
For the sake of completeness we performed a scan of the parameter space also for $\lb$ in the following ranges $[-0.1,\;-0.01]$, $[-1,\;-0.1]$. We followed the same procedure as described in section~\ref{sec:method-numerical}. 

For $\lb\in[-0.1,\;-0.01]$ we found a small region of viable (i.e.\ perturbative and stable up to Planck scale) parameter space. In the case with $M_H<M_S$ mass ordering, the only acceptable parameter region is for small $|\lb|$ and large $g_X$ (approximately $|\lambda_2|\lesssim 0.02$ and $g_X\gtrsim 0.95$), which corresponds to extending the viable region presented in figure~\ref{fig:g4-1} in the lower right part. We have also checked that the approximation in which no scalar contributions to the one-loop potential are included is perfectly valid in the stable region. In the case with opposite mass ordering, $M_H>M_S$, no region where the potential is stable was found. This is in agreement with the intuition that if such region exists, it should be an extension of the valid region presented in figure~\ref{fig:g4-2}. We note, however, that in this case small regions appear where neglecting scalar contributions to $\Vone$ is not fully justified (scalar contributions are more than 0.1 of the included vector and fermion contributions).

For $\lb\in[-1,\;-0.1]$ we found no viable region of the parameter space --- either no solution for a minimum exists or the couplings develop Landau poles at low energies, or the potential is not stable. Neglecting scalar contributions is, however, not a reliable approximation in this regime, thus these conclusions are not fully trustable. 

\subsection{Comparison with the Gildener--Weinberg method\label{sec:SU2-GW}}
In this section we study how the results presented in section~\ref{sec:g4} would change if we applied the GW method~\cite{Gildener:1976} described briefly in section~\ref{sec:GW-SU2}. For concreteness, we focus on the scenario where $M_H<M_S$. 

The strategy of the analysis is as follows. We start from the sets of parameters $(\la, \lb, \lc, g_X)$ determined in section~\ref{sec:g4}, so for each pair $(g_X,\lb)$, the values of $\la$ and $\lc$ are fixed using the minimisation condition, requiring that the VEV of the $h$ field is $v=246\g$ and that the mass of the Higgs particle is $125\g$. All the couplings  and masses obtained in this way are defined at the electroweak scale, $\mu=246\g$. This completely defines the one-loop effective potential. Then we study this potential with the use of the GW method (see section \ref{sec:GW-SU2}). We solve the RG equations (see appendix~\ref{app:betas}) for the couplings and find the scale at which the GW determinant vanishes, see eq.~\eqref{eq:GW-det}, we refer to this scale as $\mu_{GW}$. Then the masses and VEVs are determined using the GW prescription, at the GW scale. We are interested in quantifying the differences between the GW and the full one-loop results. For the sake of conciseness we refer to the latter as the CW method, since the CW potential is used. Of course, we cannot expect that the GW procedure will yield exactly $M_H=125\g$ and $v=246\g$ because the parameters have been chosen such that these values are reproduced with the use of the full one-loop potential at the scale $\mu=246\g$. It is nonetheless instructive to quantify the differences between the two methods, since they are both used in the literature and the differences between them are not generally appreciated.

\begin{figure}[h!]
\center
\includegraphics[height=.38\textwidth]{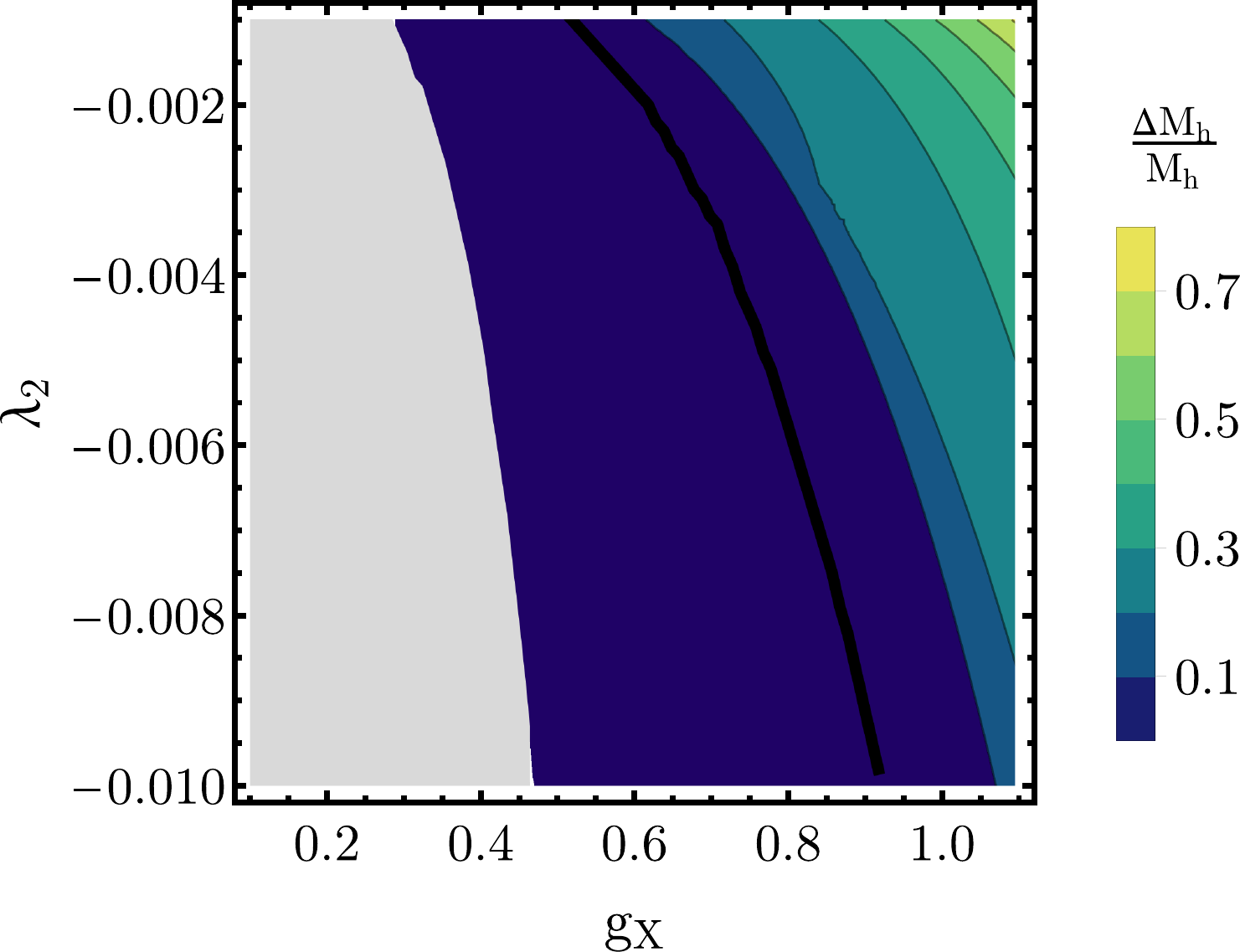}\hspace{.1cm}
\includegraphics[height=.38\textwidth]{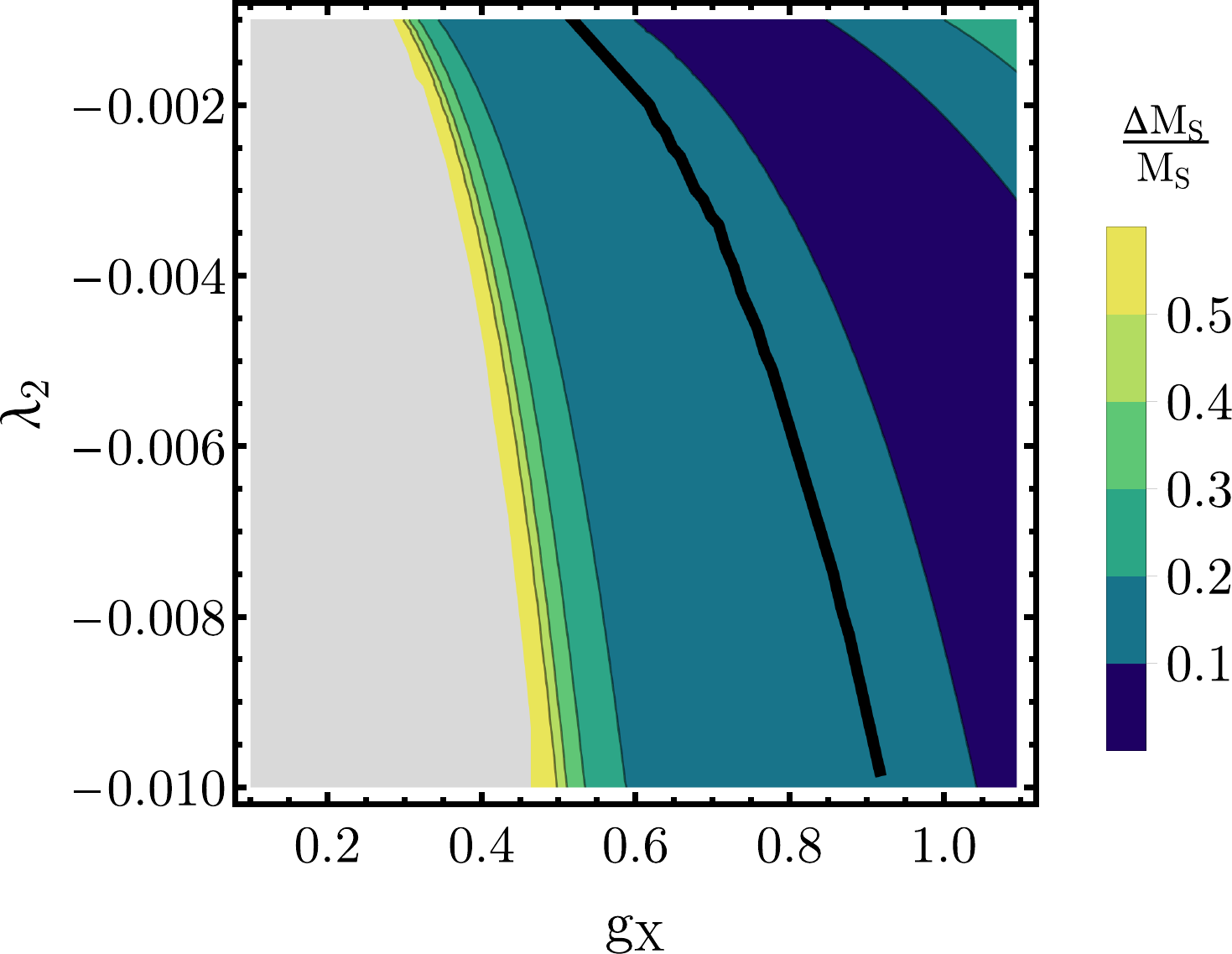}\\
\includegraphics[height=.38\textwidth]{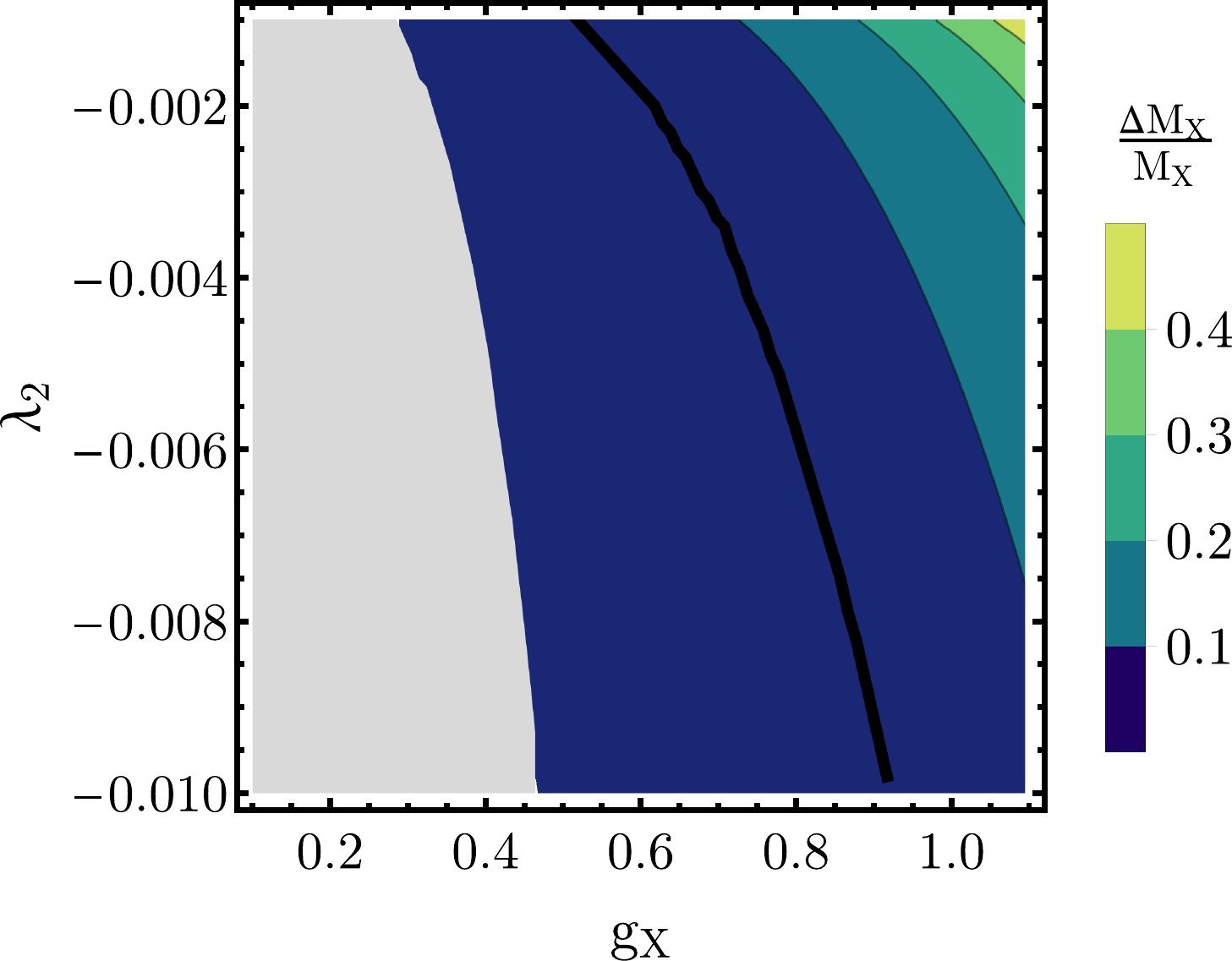}\\
\includegraphics[height=.38\textwidth]{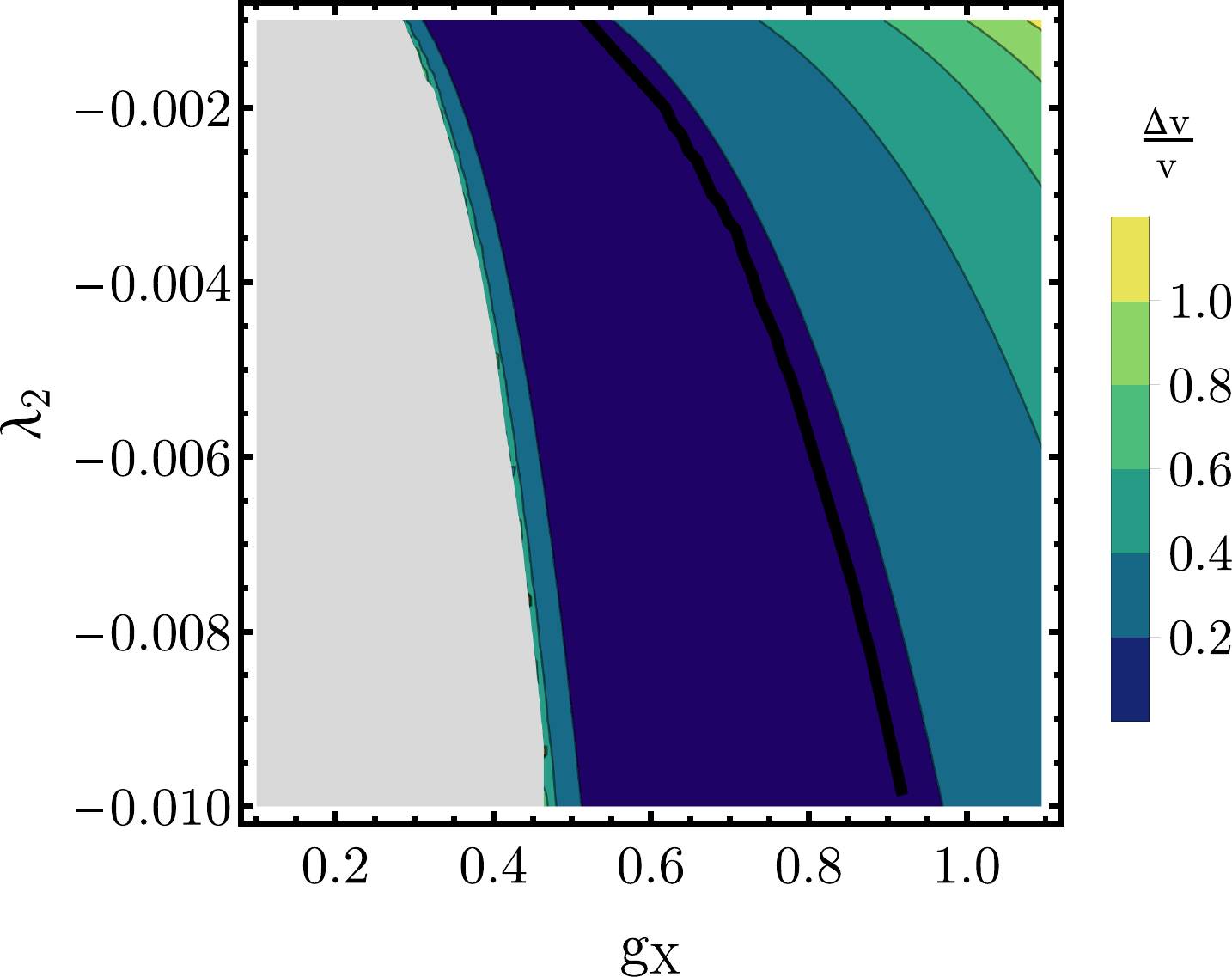}\hspace{.1cm}
\includegraphics[height=.38\textwidth]{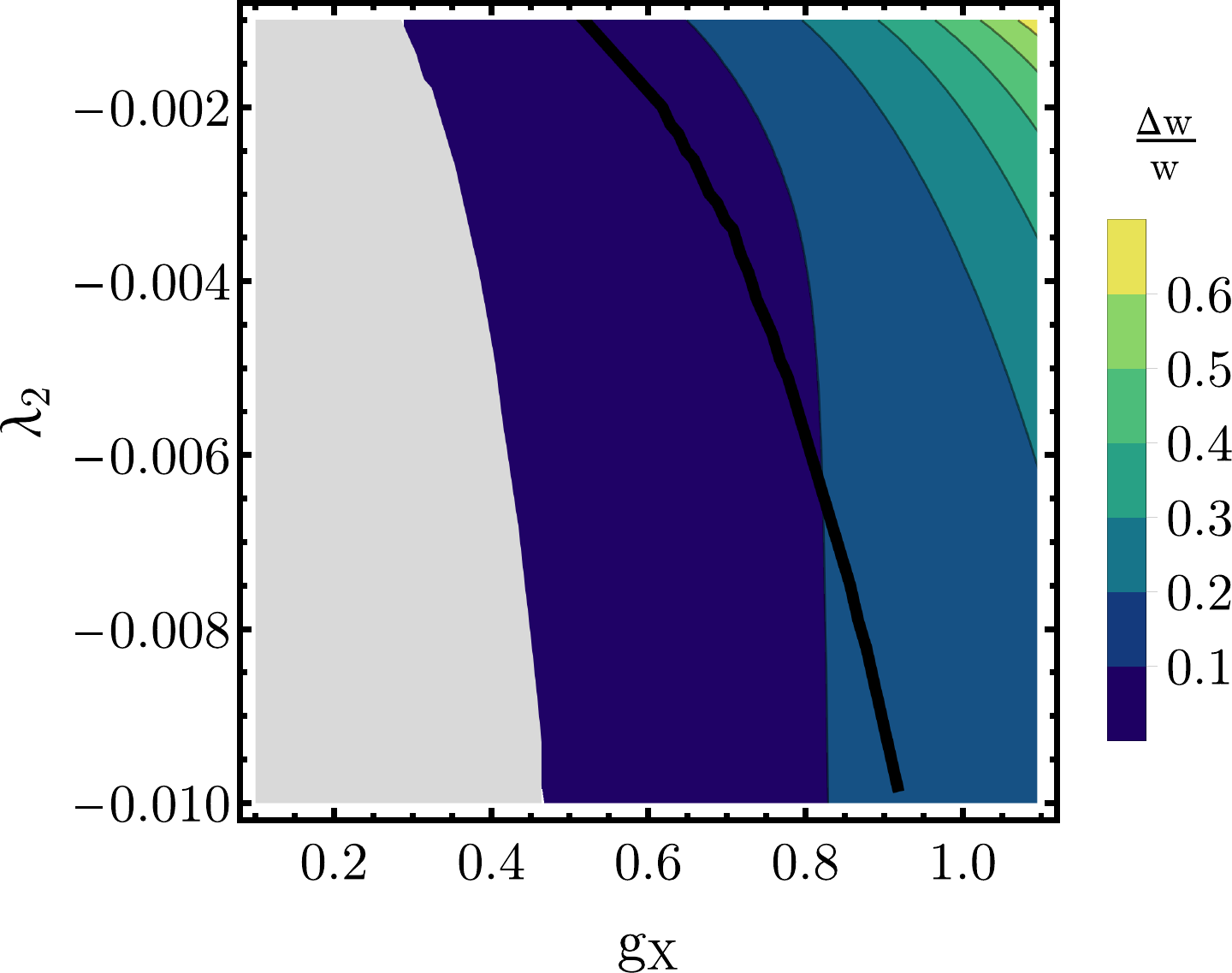}
\caption{Relative differences between masses and  scalar VEVs computed using the GW method and the one-loop potential as described in section~\ref{sec:method-numerical}. The relative differences are defined in eq.~\eqref{eq:rel-dif}. The thick black line shows the boundary of the stable region. The grey region is exluded. \label{fig:GW-comparison}}
\end{figure}

Figure~\ref{fig:GW-comparison} shows  contour plots representing relative differences between the results obtained with the one-loop effective potential and with the GW method. The relative quantities are defined as follows
\be
\frac{\Delta \alpha}{\alpha}=\frac{\left|\alpha_{\mathrm{GW}}-\alpha_{\mathrm{CW}}\right|}{\alpha_{\mathrm{CW}}},\label{eq:rel-dif}
\ee
where the subscript indicates the method used to compute the given quantity $\alpha$ --- $\alpha_{\mathrm{GW}}$ is computed using the GW method at $\mgw$, while $\alpha_{\mathrm{CW}}$ using the CW method at $\mu=246\g$ (as described in section~\ref{sec:method-numerical}). The grey region is excluded --- one of the scalar masses becomes complex signalling that the corresponding point is not a minimum of the effective potential.

It is striking to see that, while the differences in the mass of the new scalar are modest (within the stable region), at the level of 10--20\%, the predictions for the Higgs boson mass as well as the hidden gauge boson $X$ can be off by up to 50\%. At the same time the values of the VEVs computed with the use of the two different methods can disagree by up to 100\% in the case of $h$ and by up to 60\% when $\f$ is concerned. In all cases the upper right corner, corresponding to large $g_X$ and small $|\lb|$ is problematic.

Of course, the quantities discussed above are the running masses and VEVs which makes them RG scale dependent and thus not physical. To obtain the physical (pole) masses one would need to compute the self-energy corrections, which account for the non-zero external momentum corrections (recall that the effective potential which was used to compute these masses is the zero-momentum part of the effective action) and compensate for the $\mu$-dependence. It is reasonable to expect that after the inclusion of the self-energy corrections to masses the results obtained with the one-loop CW potential and the GW method would come closer, diminishing the dissonance between the two methods. Yet, we would like to emphasise, that this is not what is commonly discussed in the literature. Thus, we can conclude, that with the methods commonly used, one can obtain vastly differing results for a single point in the parameter space. Given that some observations can be already very constraining, e.g.\ the DM relic density measurements allow only a narrow band in the $(g_X,\lb)$ parameter space when the $X$ boson is interpreted as a DM candidate (see figure~3 from ref.~\cite{Carone:2013}), this can have a dramatic impact on the predictions of different analyses.

\begin{figure}[h!]
\center
\includegraphics[height=.38\textwidth]{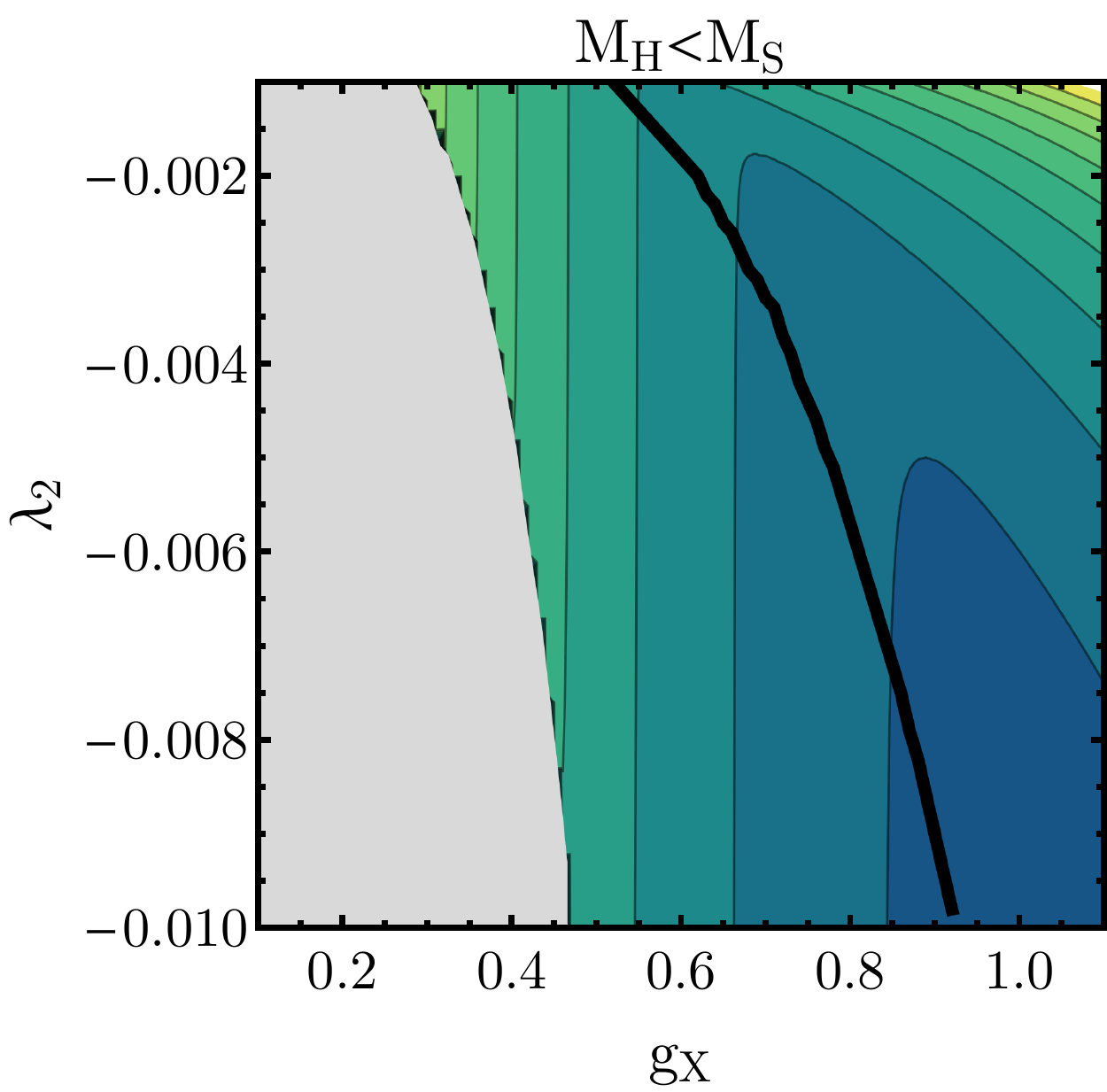}\hspace{.4cm}
\includegraphics[height=.38\textwidth]{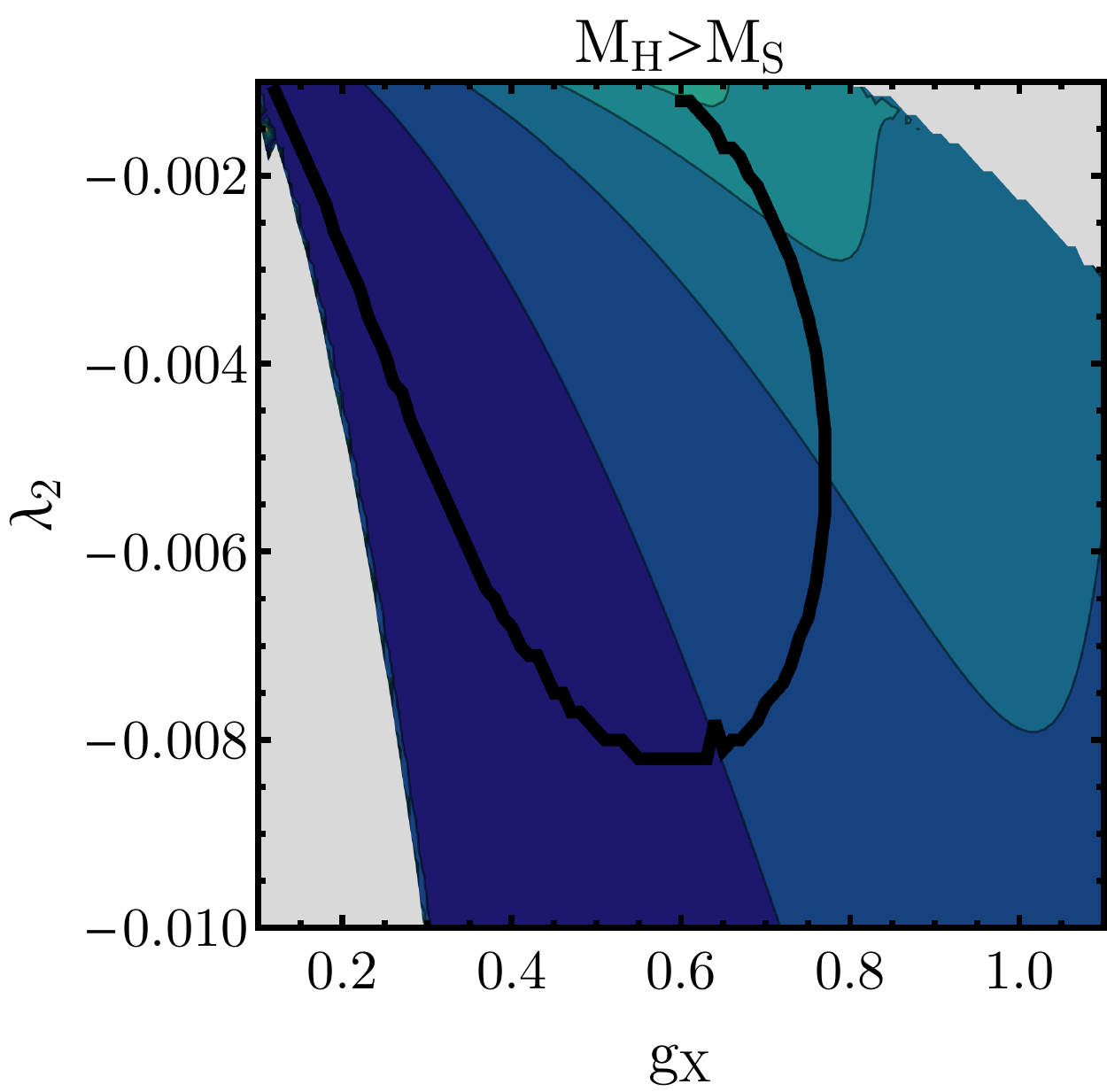}\\
\hspace{.8cm}\includegraphics[width=.8\textwidth]{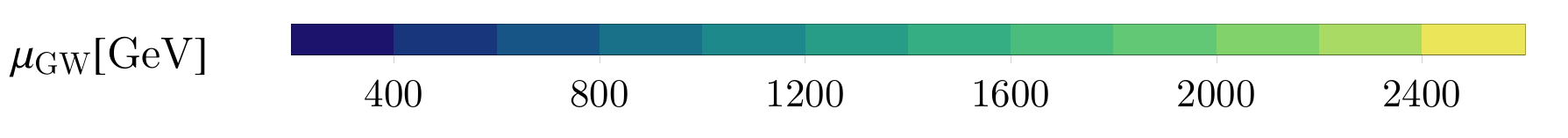}
\caption{The scale $\mgw$ at which the GW determinant vanishes. Left panel: $M_H<M_S$, right panel: $M_H>M_S$. The light grey shaded regions are excluded since there at least one of the scalar masses becomes complex (the GW method does not find a minimum). The black lines denote the boundary of the region where potential is stable up to the Planck scale (in the left panel to the right of the curve, in the right panel above the curve). The grey region is excluded, the points there do not correspond to stable minima. \label{fig:mu-GW}}
\end{figure}

The difference between the two discussed methods originates mostly from the difference in scales at which the computations are performed. Figure~\ref{fig:mu-GW} shows the values of the scale $\mgw$ at which the GW determinant vanishes. The $\mgw$ scale  in the case discussed above ($M_H<M_S$, left panel) acquires values up to around $2500\g$, an order of magnitude above the scale of the SM scalar VEV, $\mu=246\g$. This has an effect on the magnitudes of couplings, which can change their values significantly running between these two scales. Moreover, this also influences the sizes of the logarithms present in the one-loop correction to the effective potential. We note that in the case of inverted mass hierarchy in the scalar sector, when $M_S<M_H$, the GW scale is typically lower and also the predictions obtained with the CW and GW methods agree better (the deviations typically do not exceed 10\% with the exception of $v$ for which the predictions can differ by up to 30\% within the stable region). This also indicates the relevance of the choice of scale for the obtained results.

An important question arises, if the two methods presented above provide so vastly different results, which of them is more reliable? One can argue, that the EW scale is a more appropriate choice for the computation of masses of particles which are comparable in size with this scale (the relevant SM particles). The one-loop corrections to masses originating from the effective potential contain logarithms of ratios of the tree-level masses computed at the minimum to the renormalisation scale. If the scale is close to these masses, the loop corrections are small and the perturbative computations are reliable. Therefore, the scale of the VEV is a first natural choice for the renormalisation scale. Of course, in the case of models with more than one scalar field, the VEVs of the different scalars can be vastly different, which is also the case in the discussed model, and thus the renormalisation scale cannot be (generically) close to both of the VEVs. Thus some intermediate scale might be more appropriate. The issue of scale dependence of the obtained results can be alleviated by the use of the RG improved potential discussed in ref.~\cite{Chataignier:2018}. This is the subject of the following section.

\begin{figure}[ht]
\center
\includegraphics[height=.38\textwidth]{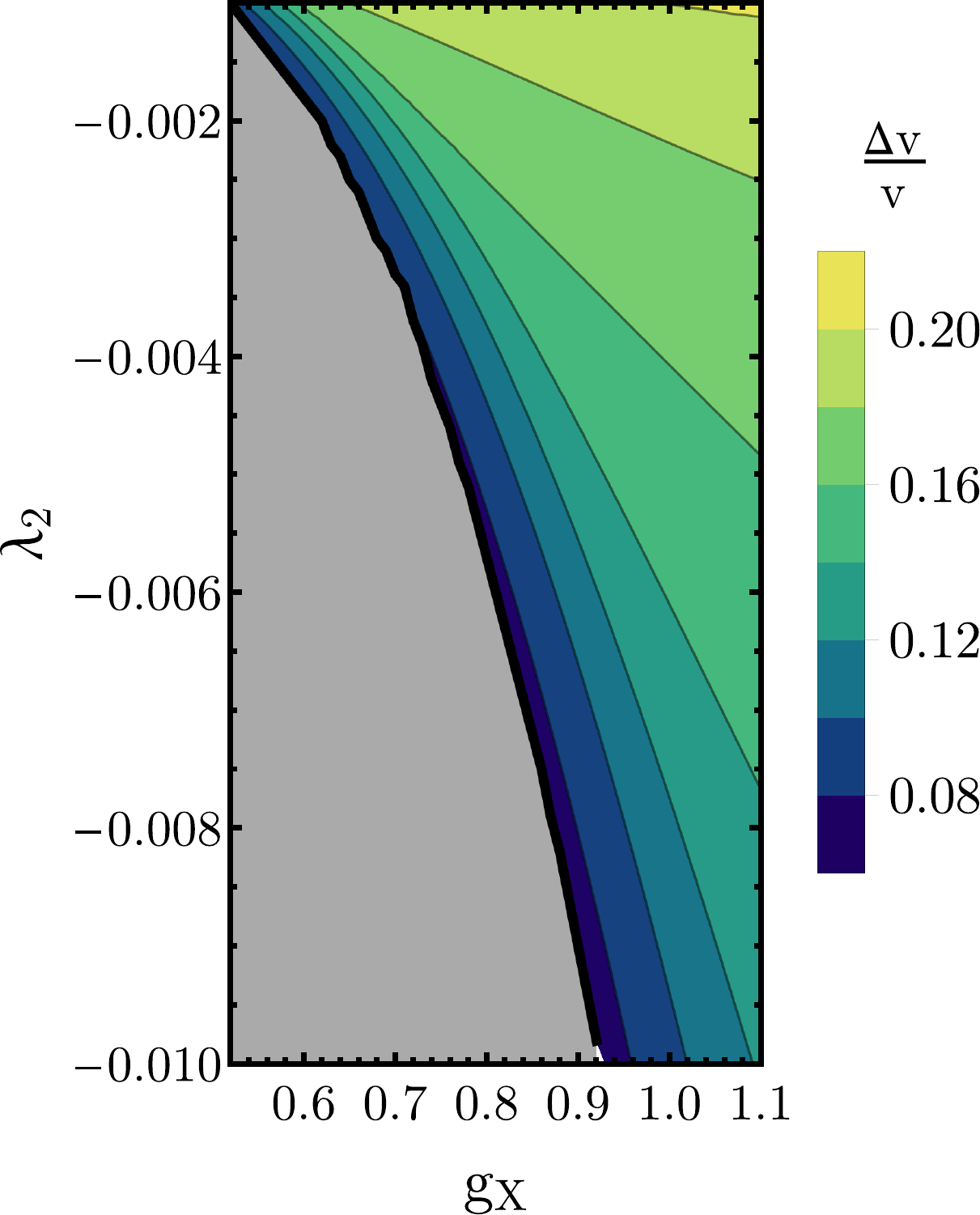}\hspace{.3cm}
\includegraphics[height=.38\textwidth]{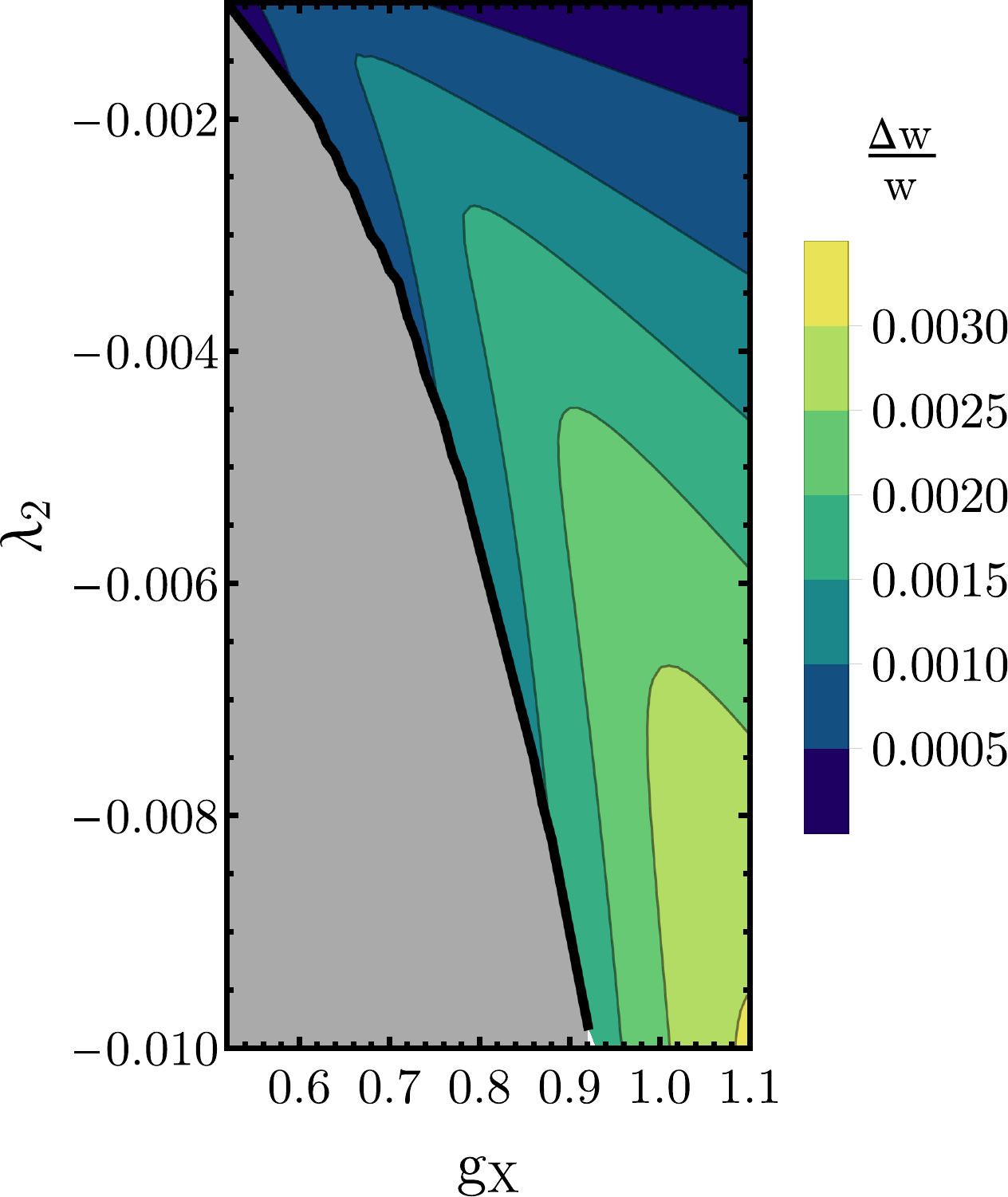}\hspace{.3cm}
\includegraphics[height=.38\textwidth]{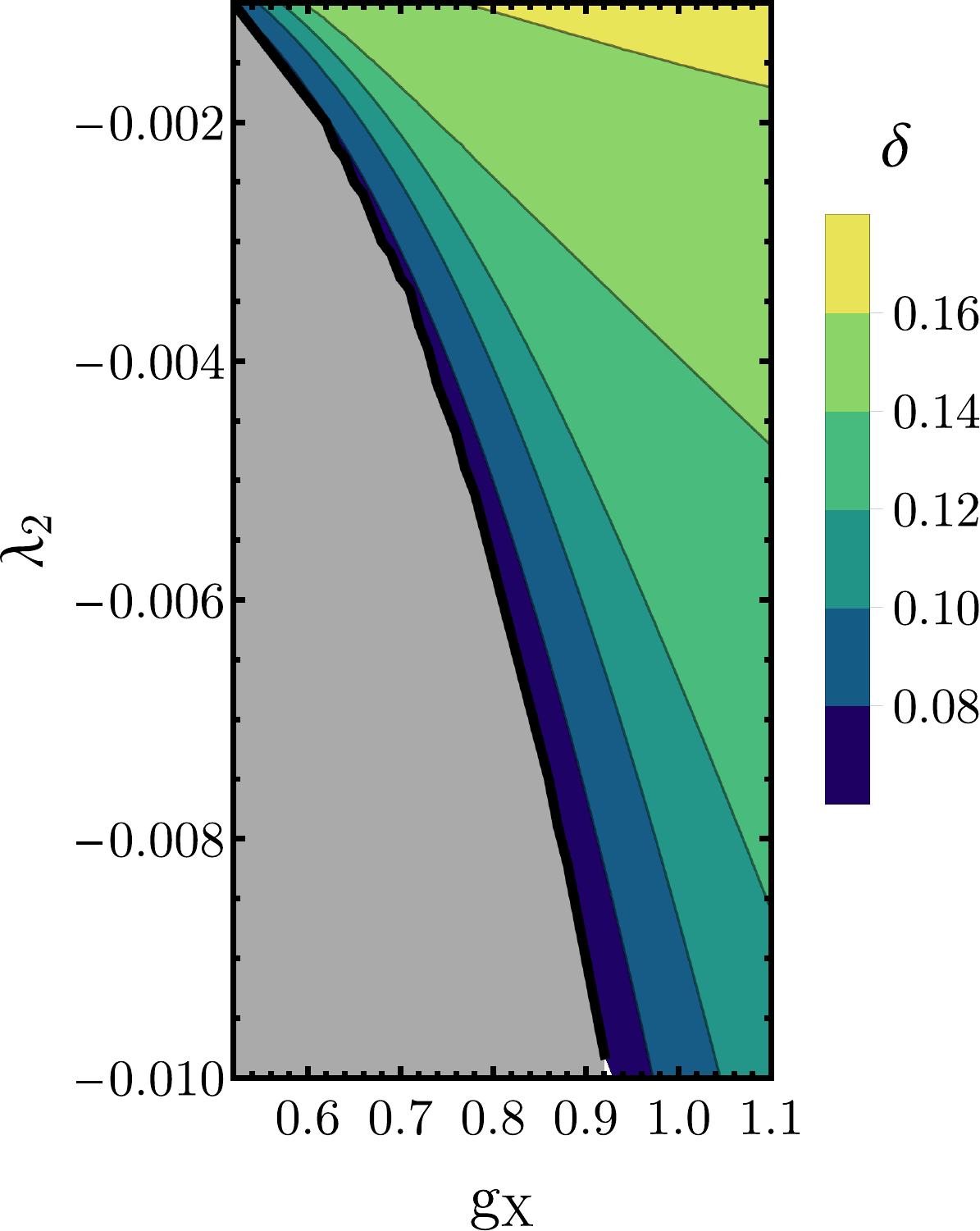}
\caption{The  relative differences between the VEVs of the scalar fields, computed using the one-loop CW potential at $\mgw$ and the GW method (defined as in eq.~\eqref{eq:rel-dif} with the change that $v_{\textrm{CW}}$ and $w_{\textrm{CW}}$ are defined at $\mgw$).  The black curve is the boundary of the region where the potential is stable up to the Planck scale, the shaded grey region is not viable.\label{fig:GW-at-muGW}}
\end{figure}

Before entering the detailed discussion of scale dependence and RG improvement, let us perform one more check of the GW method --- we will compare the predictions of the GW method with the predictions obtained from the one-loop CW potential evaluated at the GW scale (in contrast to using the one-loop CW potential at $\mu=246\g$ before). The main point of the GW method is that at the GW scale the tree-level potential has a flat direction and a radiatively generated minimum forms approximately along this direction given by the ratio of the scalar couplings, see eq.~\eqref{eq:GW-min}. In fact, GW pointed out that the minimum is not located exactly at the flat direction, there is a small correction due to quantum effects, which is typically neglected. By comparing the location of the minimum of the full one-loop potential computed at the scale $\mgw$ to the location of the minimum given by the GW formulas, we can check to which extend the common approximation of the GW method is valid.  This is done by comparing the locations of the minima in the $h$ and $\f$ directions, see left and middle panels of figure~\ref{fig:GW-at-muGW}. The relative differences between the VEVs are defined as in eq.~\eqref{eq:rel-dif}, with the important change that now $v_{\textrm{CW}}$ and $w_{\textrm{CW}}$ are defined at the GW scale. The plot in the right panel shows the relative difference in the direction in the field space along which the minimum is located,  defined as
\be
\delta= 1-\frac{v_{\textrm{CW}}/w_{\textrm{CW}}}{v_{\textrm{GW}}/w_{\textrm{GW}}},\label{eq:delta-GW}
\ee
where again $v_{\textrm{CW}}$ and $w_{\textrm{CW}}$ are defined at $\mu=\mgw$.
The values of $\lambda_2$ and $\gcw$ are defined at $\mu=246\g$ as before and evolved to the $\mgw$ scale using their RG equations (the same applies to the input values for the other scalar couplings). We limit ourselves to the region of the parameter space which is stable up to the Planck scale (the black thick line is the boundary of this region, the shaded grey region is not viable since Landau poles appear below the Planck scale there).  

From figure~\ref{fig:GW-at-muGW} it is clear that the GW method reproduces the location of the minimum along the $\f$ direction very well. The accuracy for $v$ is less impressive but still far better than what has been shown in figure~\ref{fig:GW-comparison}. The direction in the field space along which the minimum occurs coincides with the tree-level flat direction at the level of $\mathcal{O}(0.1)$. This demonstrates that the GW method provides a fairly accurate approximation to the location of the radiatively generated minimum at $\mgw$. We also computed the scalar masses from the one-loop CW potential at the GW scale and performed a comparison analogous to that of figure~\ref{fig:GW-comparison}. It turns out, that the relative differences obtained this way do not exceed 10\% in the case of the mass of the Higgs boson, while being below 7\% for the new scalar's mass. This supports the hypothesis that the difference between the CW analysis at $\mu=246\g$ and the GW results stems mostly from the scale dependance of the considered quantities. 

\begin{figure}[ht]
\center
\includegraphics[height=.38\textwidth]{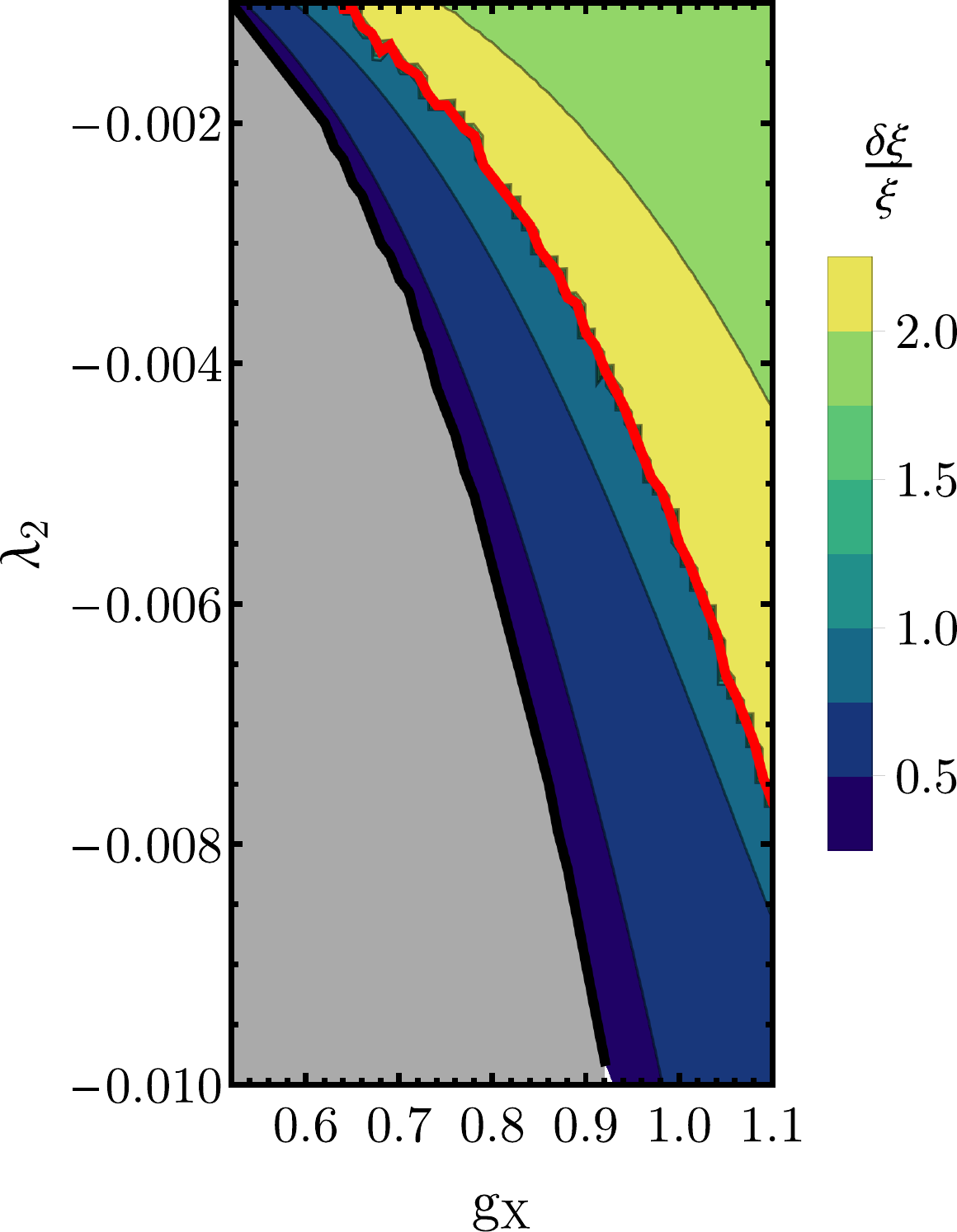}\hspace{1.5cm}
\includegraphics[height=.38\textwidth]{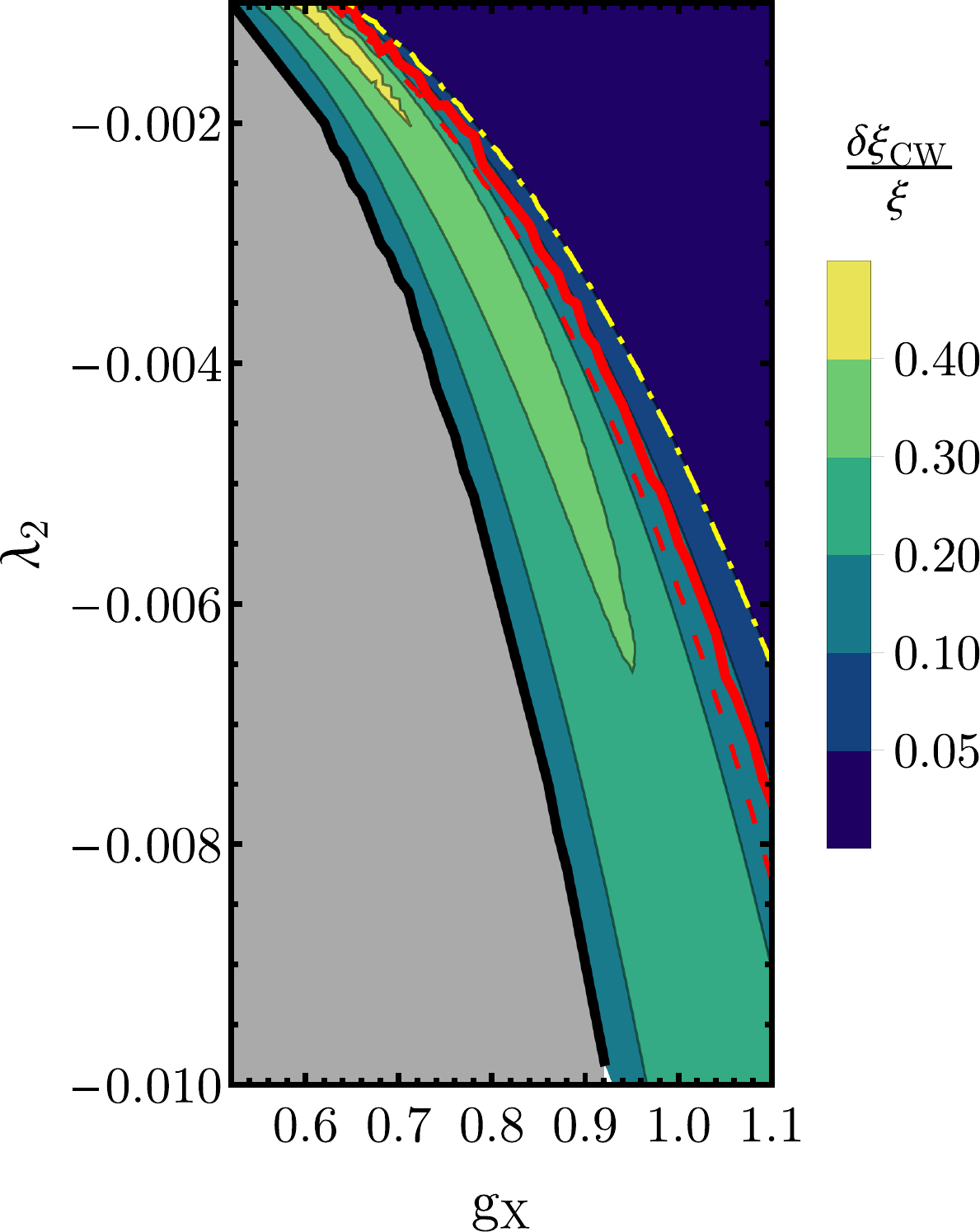}
\caption{The  relative differences between the rescalings of the Higgs couplings. Left panel: relative difference defined in eq.~\eqref{eq:delta-xi-GW} expressing the difference in the predictions of the GW method and the CW method at the GW scale, $\mgw$. Thick red line: boundary between the region where the Higgs boson has loop-generated mass (to the left of the line) and where it acquires the tree-level mass (to the right of the line). Right panel: the relative difference defined in eq.~\eqref{eq:delta-xi-CW} quantifying the running of the mixing between mass eigenstates between the electroweak and the GW scale. Red dashed line: experimental exclusion line of $\cos\theta\geqslant 0.93$ (to the right of the line) computed at $\mu=246\g$. Yellow dot-dashed line: experimental exclusion line of $\cos\theta\geqslant 0.93$ (to the right of the line) computed at $\mu=\mgw$.\label{fig:mixing-GW}}
\end{figure}

The GW method gives not only approximate values of the scalar masses, but also of the mixing between the mass eigenstates, which modifies the couplings of the Higgs boson with respect to the SM (for a recent discussion of the mixing with the use of the GW method see ref.~\cite{Loebbert:2018}). Since one of the mass eigenvalues at tree-level is zero, the mixing angle coincides with the angle at which the tree-level flat direction is formed, see eqs.~\eqref{eq:mixing-GW}--\eqref{eq:tan-alpha}. When loop corrections are added to the effective potential, the mixing is modified since a new non-zero loop-generated mass arises.  The Higgs boson can correspond to the state with mass generated at tree-level (denoted by $\phi_1$), see eq.~\eqref{eq:GW-tree-mass}, or at one-loop-level, denoted by $\phi_2$ with mass given by eq.~\eqref{eq:GW-loop-mass}. Due to the mixing the Higgs couplings are rescaled by $\xi_{\textrm{GW}}$, which is equal to $-\sin\alpha$ or $\cos\alpha$, depending on whether the Higgs boson corresponds to $\phi_1$ or $\phi_2$, respectively. Left panel of figure~\ref{fig:mixing-GW} shows the relative difference between the mixing predicted by the GW method, $\xi_{\textrm{GW}}$, and computed from diagonalising the full one-loop mass matrix at $\mu=\mgw$ ($\xi_{\textrm{CW}}=\cos\theta$), normalised to the CW result,
\be
\frac{\delta \xi}{\xi}=\frac{\xi_{\textrm{CW}}(\mgw)-\xi_{\textrm{GW}}}{\xi_{\textrm{CW}}(\mgw)}.\label{eq:delta-xi-GW}
\ee
The parameter space is divided into two regions --- to the left from the red thick line the Higgs corresponds to $\phi_2$ (has loop-generated mass) and to the right to $\phi_1$ (tree-level-generated mass). As can be seen, the GW approximation seems to fail dramatically in the $\phi_1$ region. This should be expected since we assumed that the Higgs is the lighter scalar particle and we do not expect the GW method to work correctly if the loop-generated mass is greater than the tree-level one. However, interestingly such a big difference comes mostly from the sign difference between the $\xi_{\textrm{CW}}$ and $\xi_{\textrm{GW}}$. If one is interested in absolute values of these quantities they come to good agreement in the $\phi_1$ region, with relative differences not exceeding 10\% and even being below 1\% in a big part of the parameter space. In the $\phi_2$ region the signs of $\xi_{\textrm{CW}}$ and $\xi_{\textrm{GW}}$ agree, nonetheless the differences between them stay significant proving that the mixing angle between the mass eigenstates is significantly modified by inclusion of loop corrections.

In the light of the earlier discussion of the scale dependence of the results it is interesting to see how the rescaling of the Higgs coupling computed from the full one-loop CW potential is affected by going from the scale $\mu=246\g$ to the GW scale. This is illustrated in the right panel of figure~\ref{fig:mixing-GW} with the use of the relative difference defined as 
\be
\frac{\delta \xi_{\textrm{CW}}}{\xi}=\frac{\xi_{\textrm{CW}}(246\g)-\xi_{\textrm{CW}}(\mgw)}{\xi_{\textrm{CW}}(246\g)}.\label{eq:delta-xi-CW}
\ee
The relative difference can exceed 40\%. In figure~\ref{fig:mixing-GW} one can also see the experimental exclusion lines of $\cos\theta\geqslant 0.93$ for $\mu=246\g$ (dashed red line) and $\mu=\mgw$ (dot-dashed yellow line) --- it is clear that the running of the mixing angle can influence the size of the excluded region. One should also note, that the region where GW gives inaccurate predictions for the absolute value of the rescaling of the Higgs couplings is anyway excluded by the experiment.

The conclusions that can be drawn from the considerations above are that the hierarchy of different contributions that was studied in section~\ref{sec:systematic-SU2} is not of crucial importance for the SU(2)cSM model --- even though the loop contributions to stationary point equation from the dark sector are of the same magnitude as the tree-level contributions (at $\mu=246\g$), the GW method gives the location of the global minimum and the values of the scalar masses with reasonable accuracy when compared to the results obtained from the one-loop potential evaluated at $\mgw$ (this is not true, however, for the mixing between the mass eigenstates). Nonetheless, the dependence on the scale that was discussed in section~\ref{sec:scale-dep} is very important and introduces significant uncertainties to the parameters such as the scalar masses (see figure~\ref{fig:GW-comparison}) and the mixing angle (see figure~\ref{fig:mixing-GW}).

\subsection{Analysis of the one-loop RG improved potential}

The perturbative loop expansion of the effective potential might not be a good framework for studying quantum corrections in some cases. As we discussed in section~\ref{sec:RG-improvement}, in models with more scalar fields the minimum of the effective potential may reside at a point in the parameter space where the VEVs of various scalar fields significantly differ in magnitude. This can lead to large logarithms, since there is just one renormalisation scale. In such a case RG improvement of the effective potential is needed, which reorganises the perturbative series in such a way that the dominant terms are resummed and the remaining series is perturbative.

As we have shown analysing the SU(2)cSM model with weak $\lb$ coupling and the $M_S>M_H$ mass ordering  in section~\ref{sec:g4}, the VEV of the new scalar field $\f$ typically is $\mathcal{O}(10)$ times greater than the one of the SM Higgs, $v=246\g$. This does not necessarily lead to the breakdown of perturbative expansion in terms of loops,  however it weakens the accuracy of a given truncation of the perturbative series. In~particular, the VEVs of the scalar fields computed with the use of the one-loop effective potential are strongly scale-dependent, which is not a desirable feature. The procedure of RG improvement renders the potential scale independent, up to higher loop effects (see ref.~\cite{Chataignier:2018} for a detailed discussion and examples) thus also attenuating the running of the VEVs. In this section we analyse the effect of RG improvement on the one-loop predictions for the SU(2)cSM analysed before. We use the method of RG improvement described in~\cite{Chataignier:2018} and briefly outlined in section~\ref{sec:RG-improvement}. In our analysis we find numerically the value of full $t_*$ given by eq.~\eqref{eq:t*-def}  to evaluate the RG improved potential. Studying benchmark points, we found the ratio $\Vone(\bar{\mu}(t_*^{(0)}),\bar{\lambda}_j(t_*^{(0)}),\bar{\f}_i(t_*^{(0)}))/\Vtree(\bar{\lambda}_j(t_*^{(0)}),\bar{\f}_i(t_*^{(0)}))$ evaluated  at the minimum to be typically $\mathcal{O}(0.1)$ but possibly reaching 30\%.  For example, for the benchmark point discussed in table~\ref{tab:BM} this ratio is equal to 0.12. Applying the numerically evaluated $t_*$ we are able to push this ratio down to $\mathcal{O}(10^{-5})$--$\mathcal{O}(10^{-6})$. The cost of the improvement of accuracy is the computing time, which increases significantly with the use of $t_*$. Therefore $t_*^{(0)}$ might be still more convenient for some applications when the speed of calculation is more needed than increased accuracy.

\begin{figure}[ht]
\center
\includegraphics[width=.48\textwidth]{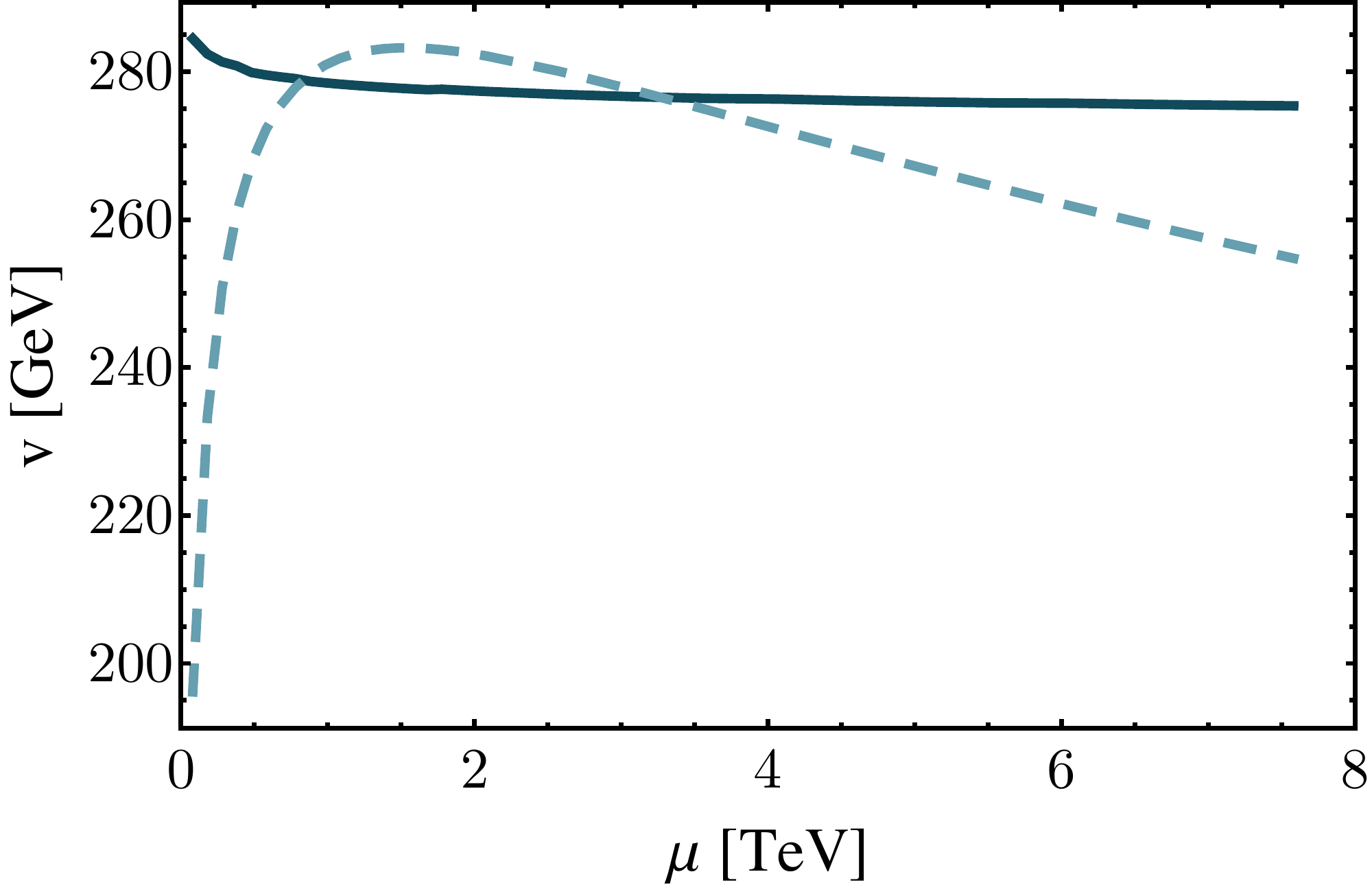}\hspace{.4cm}
\includegraphics[width=.48\textwidth]{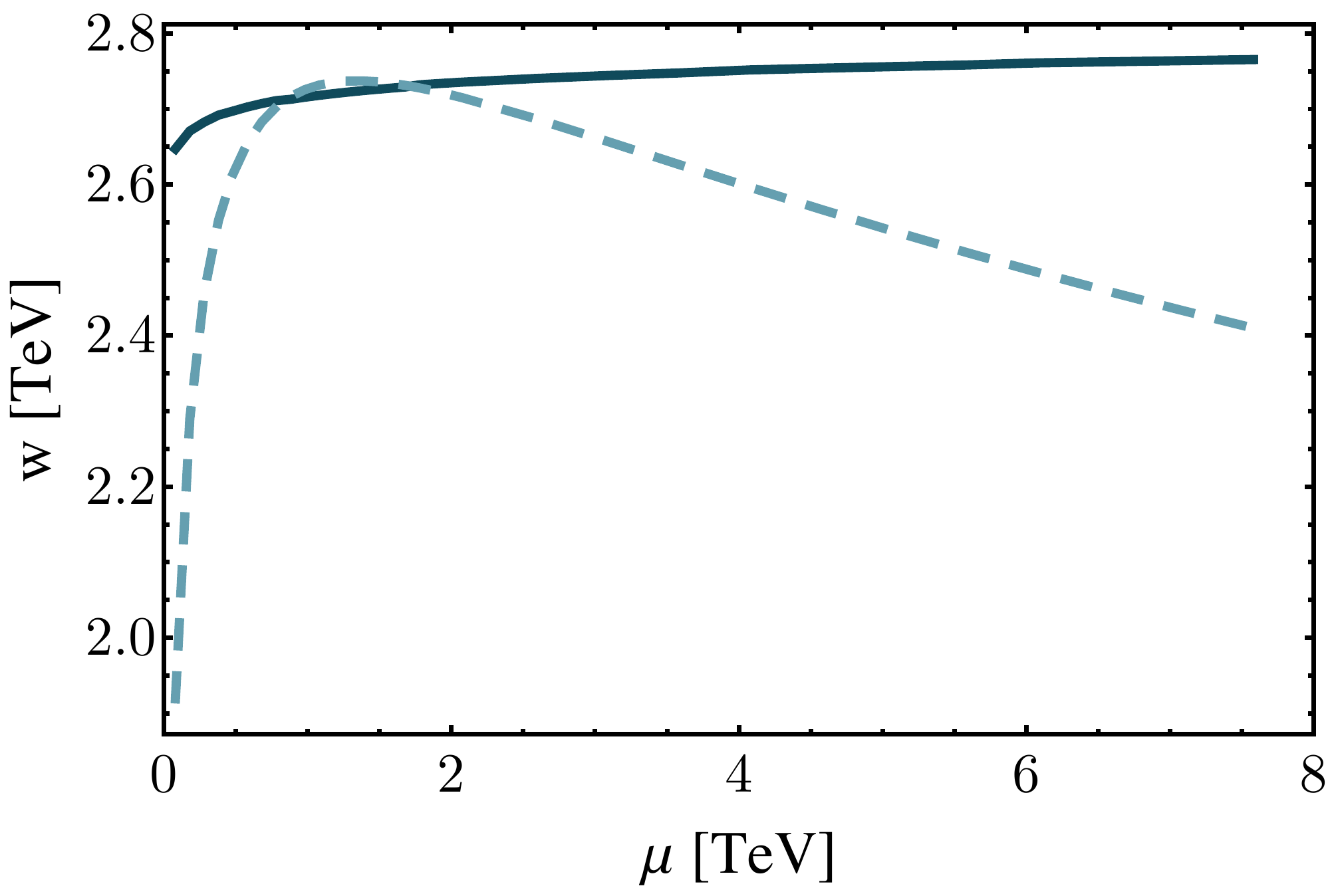}
\caption{The VEVs of the scalar fields $h$ (left panel) and $\f$ (right panel) as functions of the RG scale $\mu$ computed with the use of the one-loop effective potential (dashed lines) and RG-improved potential (solid lines). The plots are obtained for the benchmark point from table~\ref{tab:BM}.\label{fig:running-vev}}
\end{figure}

Let us start by showing that implementing the RG improved potential can indeed improve the obtained results. To this end we study the dependence of the VEVs of the $h$ and $\f$ scalar fields on the renormalisation scale using both RG improved and unimproved effective potential. Figure~\ref{fig:running-vev} represents the VEV of the $h$ field (left panel) and the $\f$ field (right panel) as functions of the RG scale $\mu$.  The results are obtained for the benchmark point defined in table~\ref{tab:BM}. We have checked that for several other points  within the stable region of the parameter space  the dependence of the VEVs on $\mu$ looks qualitatively the same, however the dependence on $\mu$ can be more or less significant. It is clear that the implementation of RG-improvement changes the behaviour of the running VEVs drastically. The VEVs computed with the use of the RG-improved potential exhibit only mild $\mu$-dependence at low scales, while at high scales they are almost constant. On the other hand, the VEVs following from the one-loop potential change significantly with the scale. Since physical observables do not depend on the arbitrary RG scale, it is clear that the RG improved potential provides a better approximation to physical quantities. Plots of figure~\ref{fig:running-vev} also help to understand the discrepancies between the CW (at 246$\g$) and GW approaches, discussed in the previous section --- it is clear that the VEVs computed from the unimproved potential run significantly between the two considered scales.

\begin{figure}[ht]
\center
\includegraphics[width=.24\textwidth]{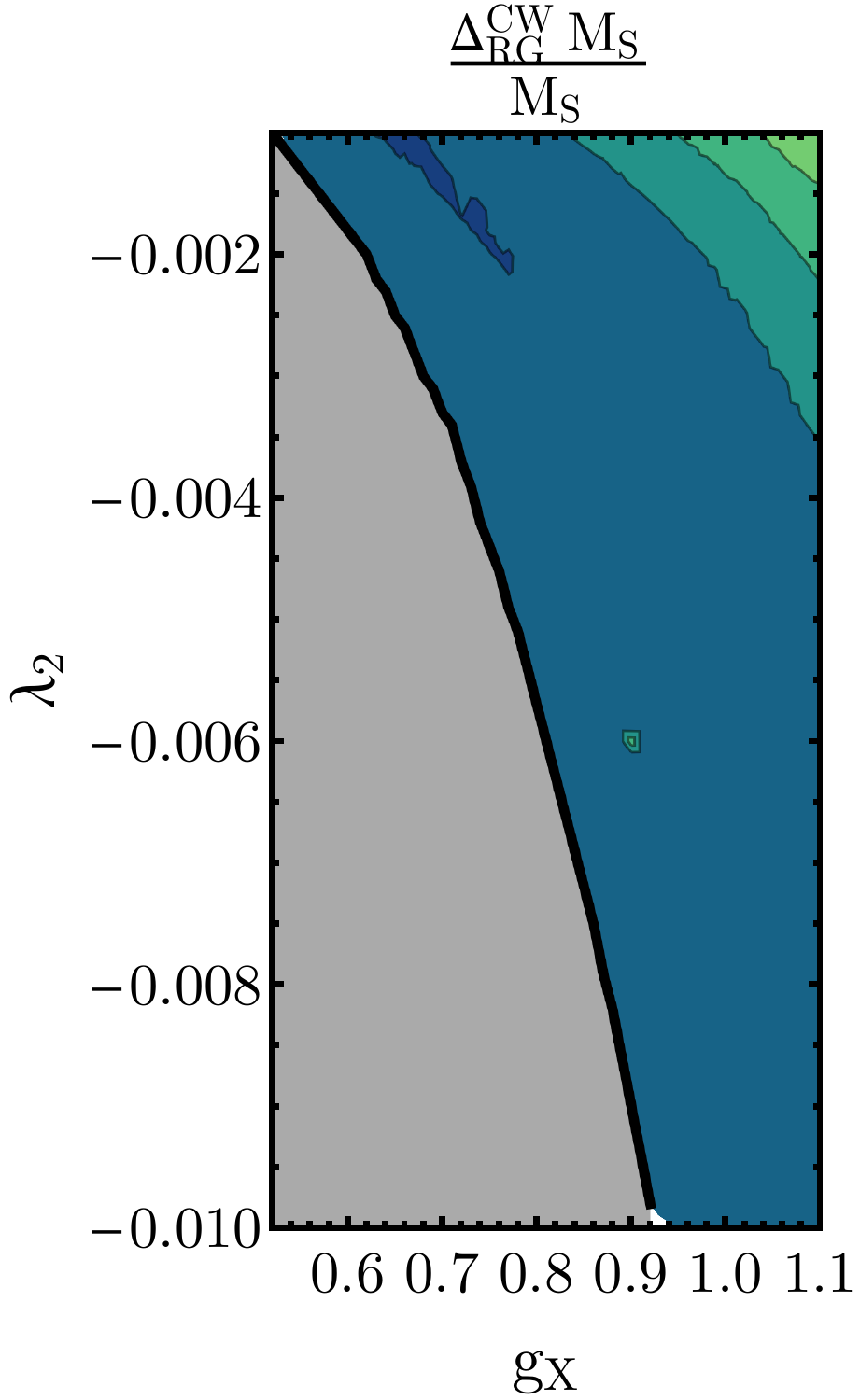}
\includegraphics[width=.24\textwidth]{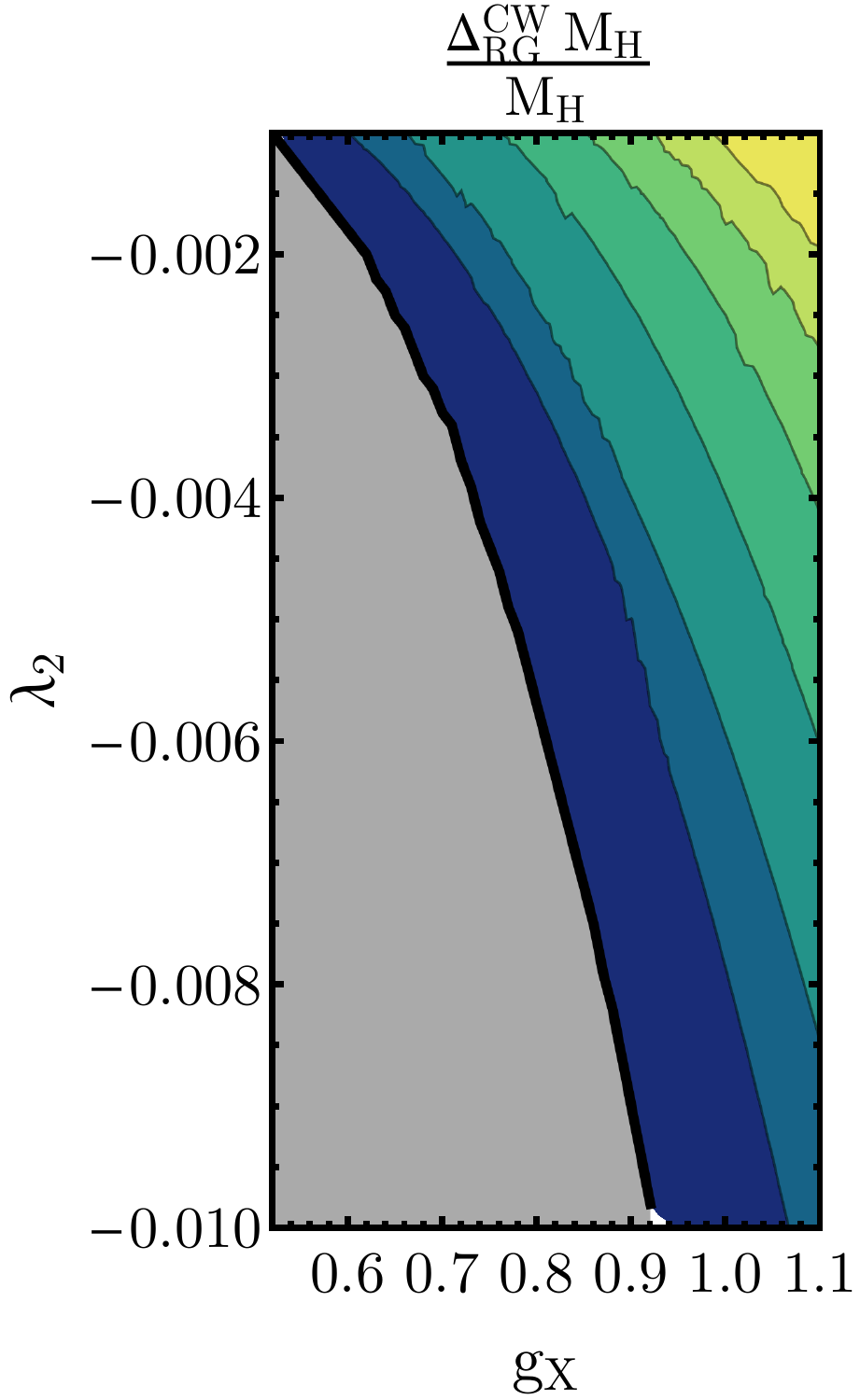}
\includegraphics[width=.24\textwidth]{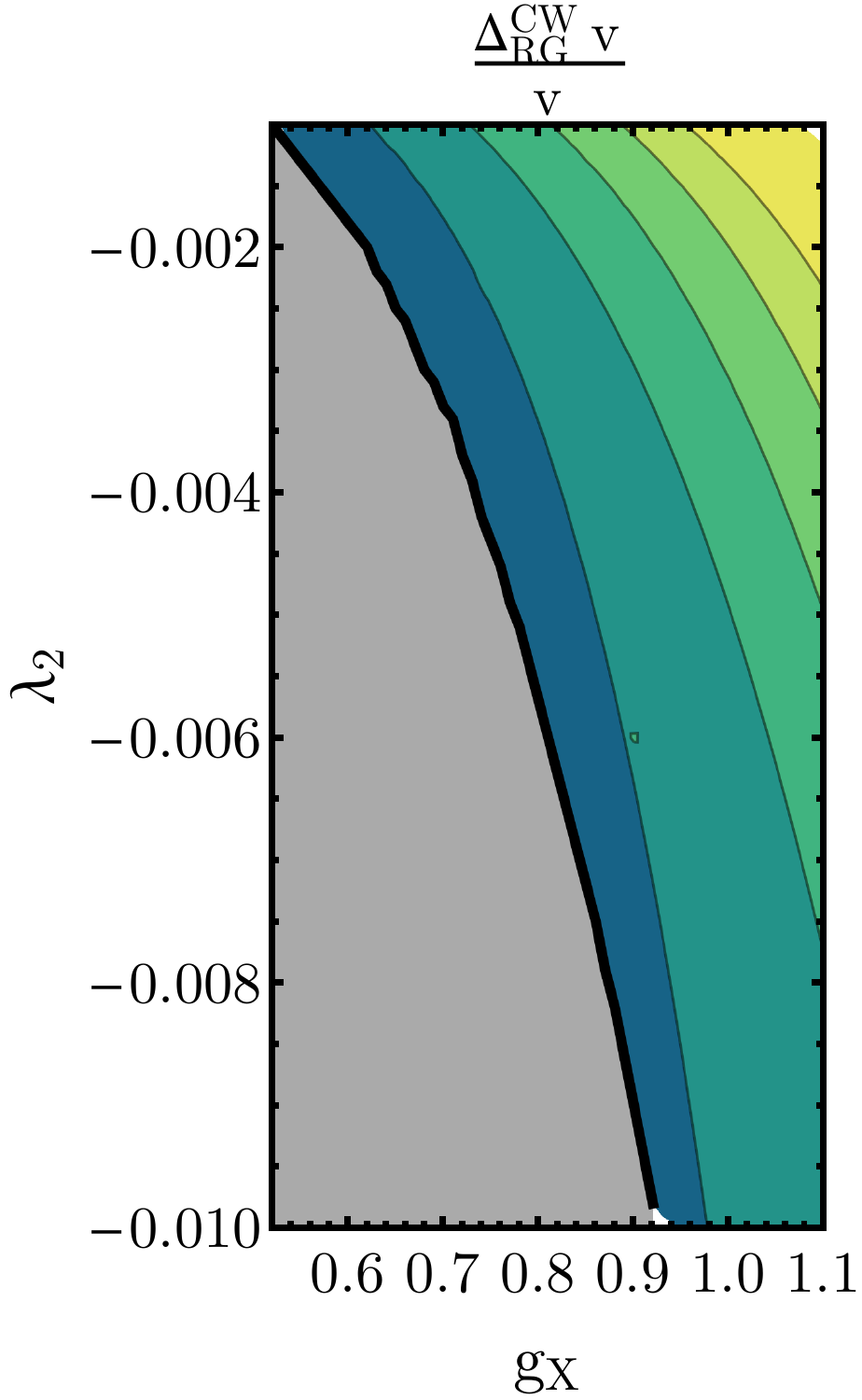}
\includegraphics[width=.24\textwidth]{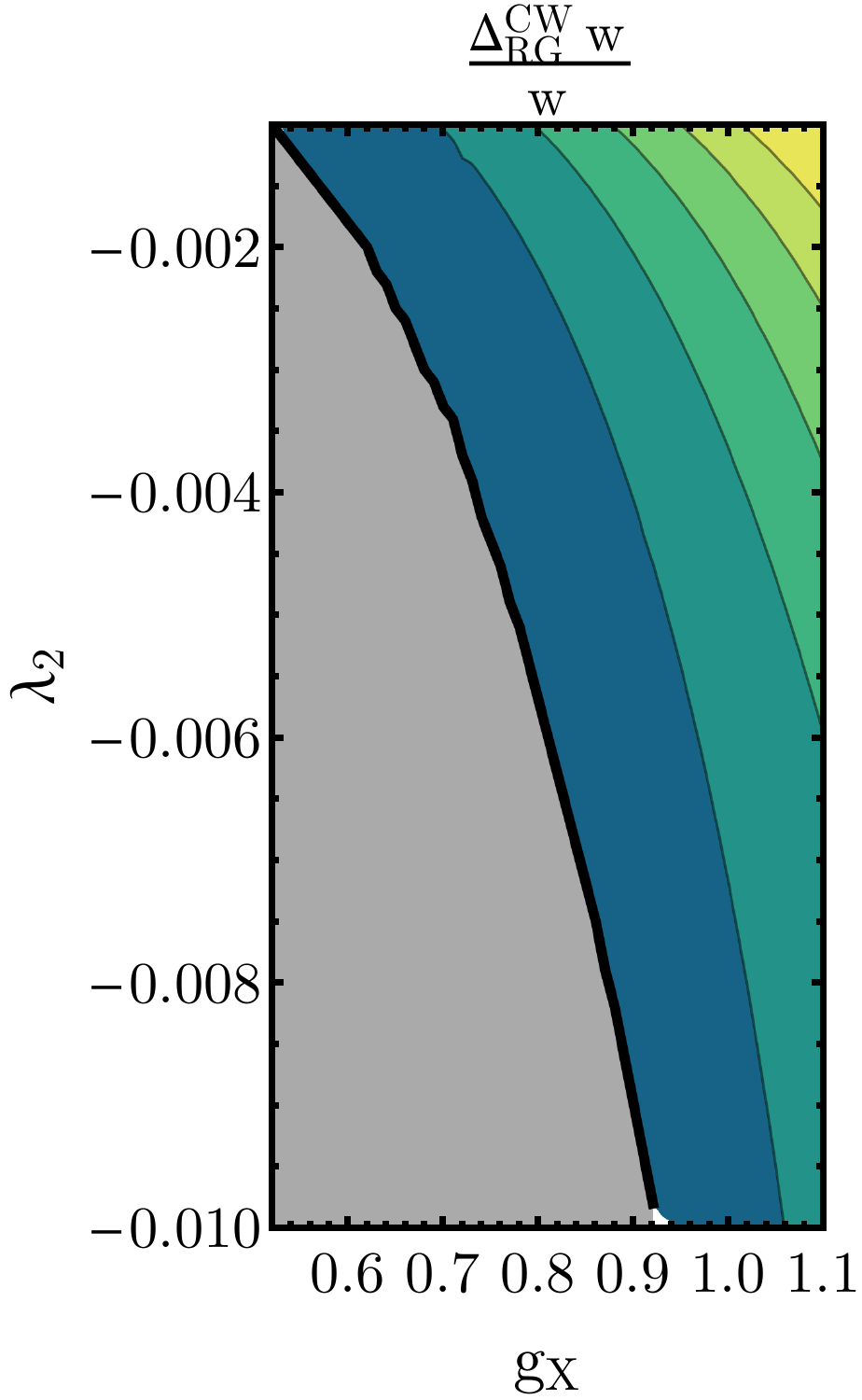}\\
\includegraphics[width=.7\textwidth]{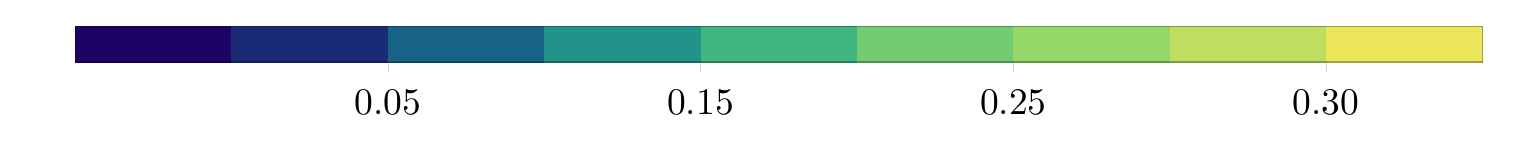}\\
\includegraphics[width=.24\textwidth]{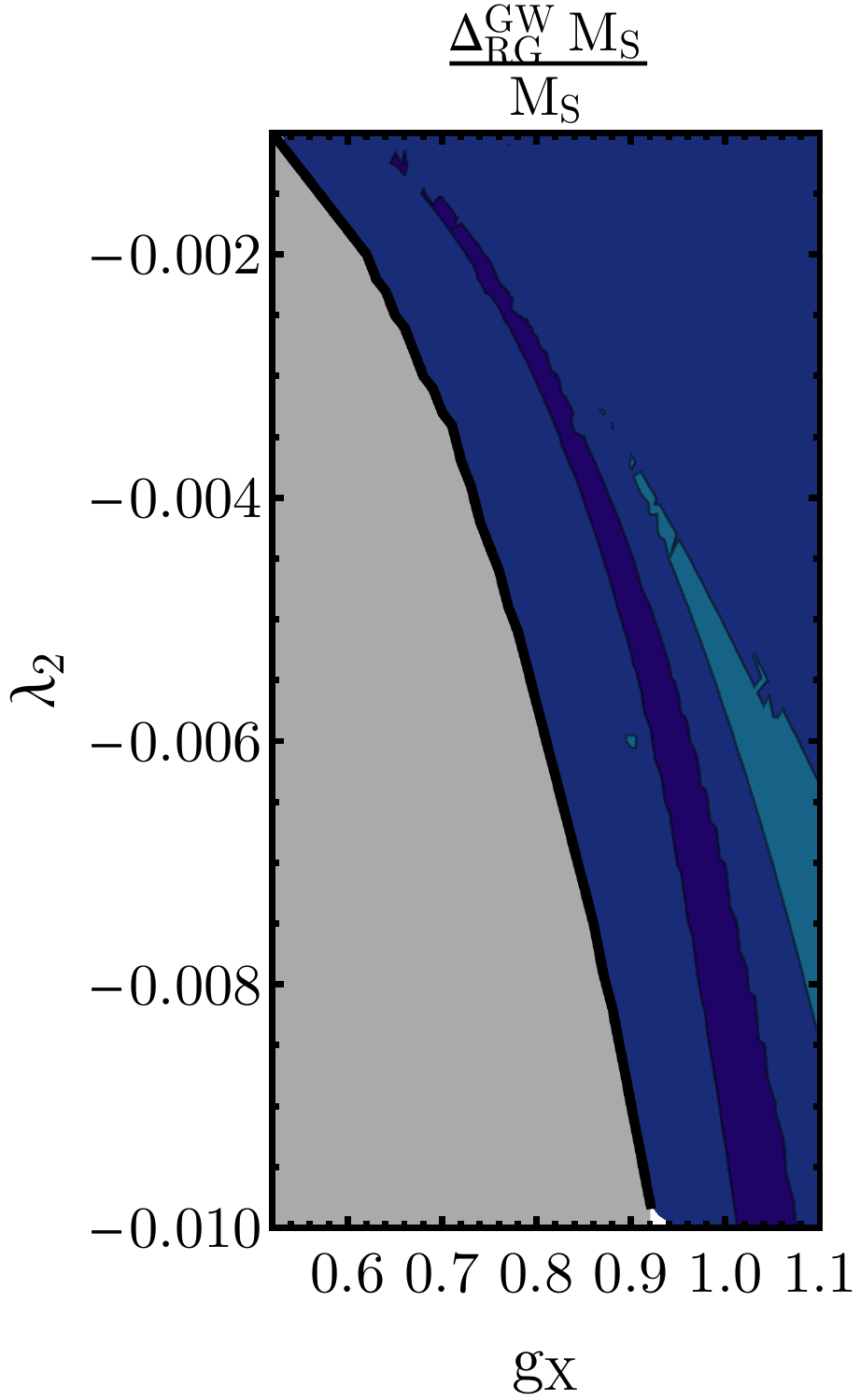}
\includegraphics[width=.24\textwidth]{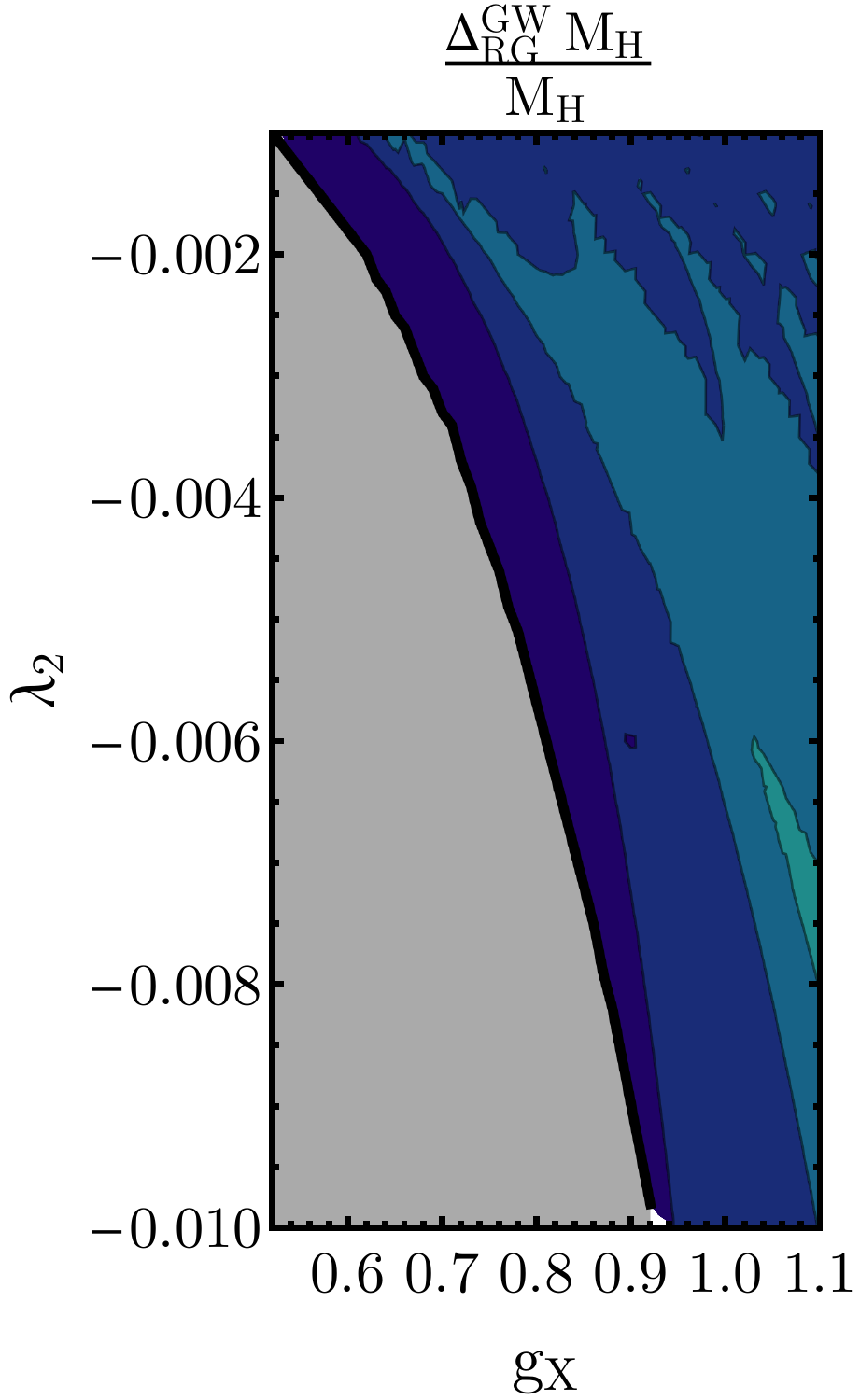}
\includegraphics[width=.24\textwidth]{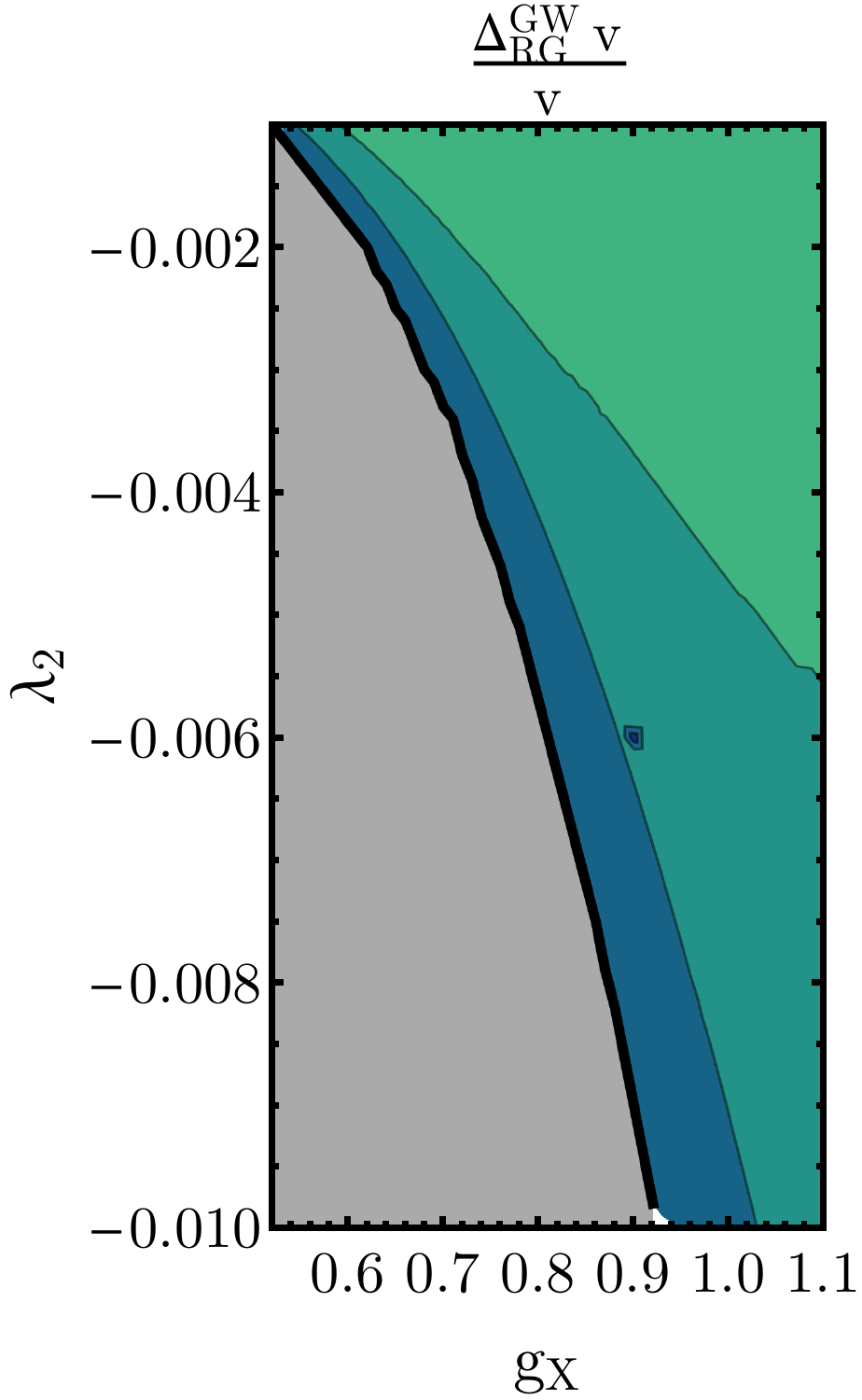}
\includegraphics[width=.24\textwidth]{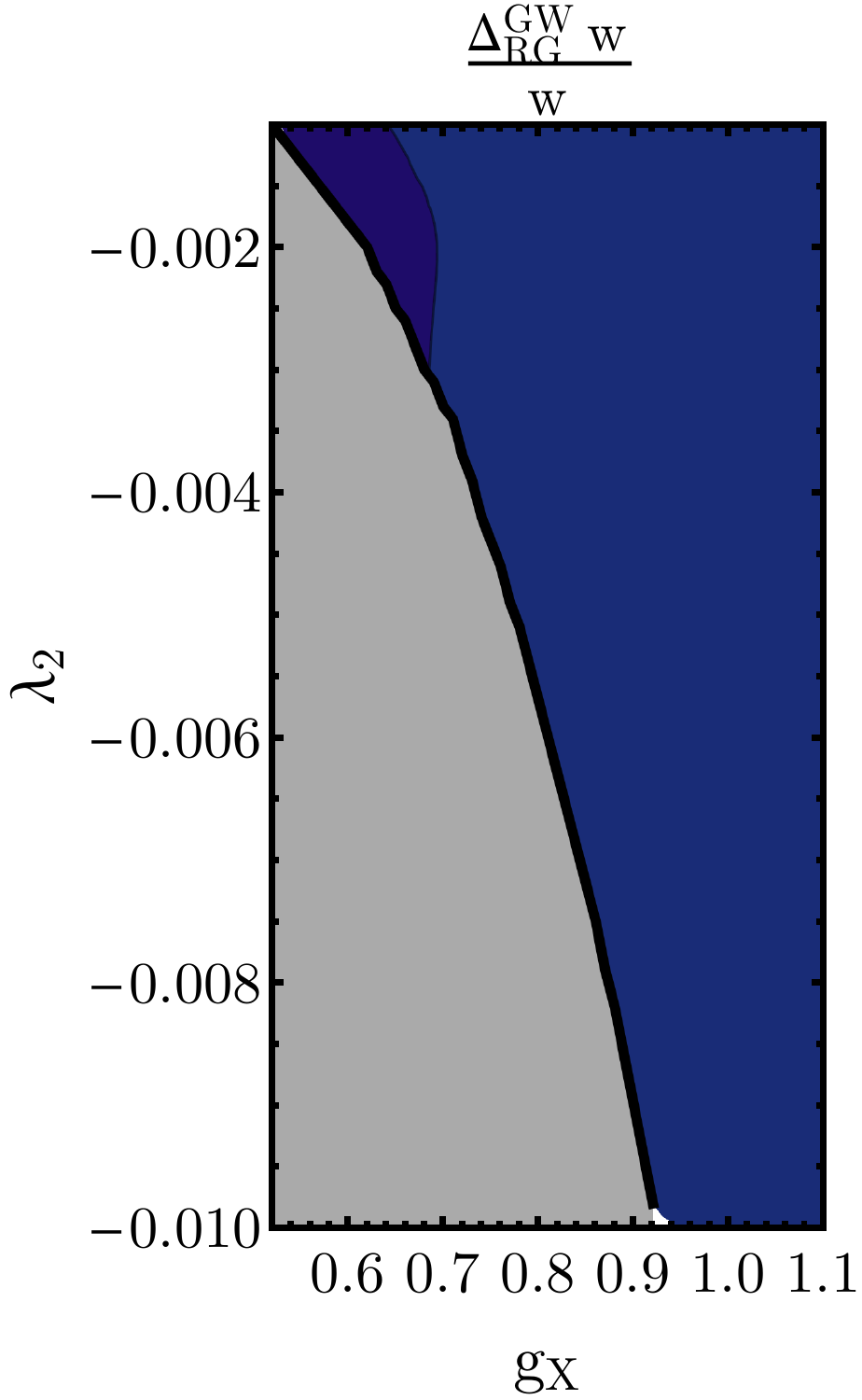}
\caption{The relative differences between the results obtained from one-loop perturbative approximations to the effective potential with respect to the outcome of the analysis using RG-improved potential, defined in eq.~\eqref{eq:delta-RG}. Upper panel: for one-loop effective potential at $\mu=246\g$. Lower panel: for GW method. The colour coding is common for all the plots. The black curve is the boundary of the region where the potential is stable up to the Planck scale.\label{fig:RG-scan}}
\end{figure}

With the results above it is clear that the RG improvement procedure can help to overcome some of the weaknesses of the effective potential. Let us then study the parameter space of the SU(2)cSM model with the use of the RG improved potential, as we did before with the use of the general one-loop potential (section~\ref{sec:g4}) and the GW method (section~\ref{sec:SU2-GW}).  To avoid problems with running coupling constants we limit the analysis to the region of the parameter space where there are no Landau poles up to the Planck scale and the potential is stable. For the sake of completeness we compare both, the results obtained from the one-loop potential at $\mu=246\g$ and the GW results with the predictions from the RG-improved potential. In analogy to the previous considerations, we define relative differences as
\be
\frac{\Delta_{\textrm{RG}}^{\textrm{CW/GW}}\alpha}{\alpha}=\frac{|\alpha_{\textrm{CW/GW}}-\alpha_{\textrm{RG}}|}{\alpha_{\textrm{RG}}},\label{eq:delta-RG}
\ee
where the subscript on $\alpha$ on the right-hand side of the equation indicates the method of computing $\alpha$.
The results are shown in figure~\ref{fig:RG-scan}, in the upper panel for the one-loop potential at $\mu=246\g$ (CW) and in the lower panel for the GW method.\footnote{For figure~\ref{fig:RG-scan} we used fewer data points than for the previous plots due to numerical complexity.} It is clear that in a wide range of the parameter space the perturbative one-loop approximations give reasonably accurate answers (assuming that the RG improved results are accurate), with differences with respect to the RG improved case at the level of up to 20\%. However, in the region of large $g_X$ and small $|\lb|$, the inaccuracies can exceed 30\%. This manifestly shows that the perturbative one-loop effective potential at fixed scale $\mu=246\g$ is not a perfect tool to study RSB. 

One should note, that the one-loop potential at $\mu=246\g$ typically provides a poorer approximation of the RG-improved results than the GW method. This can be understood by analysing table~\ref{tab:BM}. At~$\mu=246\g$, the contribution from the SM sector is small, since the scale is close to the SM masses, while the contribution from the hidden sector is large, since the VEV of the $\f$ field is generically large, giving rise to a rather large logarithm. At the GW scale, which is typically closer to the scale of $w$, the proportion between the SM and the X contributions is approximately reversed. On the other hand, at the RG scale\footnote{Note that the RG scale is field dependent therefore the value listed in table~\ref{tab:BM} is only valid for the point corresponding to the minimum.} the SM and the X contributions are more balanced since in total the one-loop correction at the RG scale is required to vanish. It is worth noting, that at the RG scale not only the one-loop correction to the effective potential is cancelled, but also the individual contributions from different sectors are smaller than at the other scales considered. Intuitively, since $w$ is significantly greater than $v$, the ``optimal'' scale is closer to $w$ than to $v$ in order to balance the large contribution of the hidden vectors. This is the reason why the GW method can give results that are closer to the RG improved ones. 

The considerations presented above indicate than when studying RSB in models with extended scalar sector it is not the best strategy to fix the RG scale to one of the VEVs or the GW scale. Rather, one should use the RG-improved potential to assure that the scale is chosen in such a way, that the perturbative treatment remains valid. It is worth underlining that an advantage of the RG improved effective potential is that there is no need to fix the RG scale arbitrarily before the hierarchy of the VEVs is known, because the optimal scale is selected automatically through the requirement that the one-loop correction to the effective potential vanishes.\footnote{There is a residual scale dependence, since one needs to fix the initial scale denoted by $\mu$ in eq.~\eqref{eq:t0}, however the dependence on this scale is only of higher order.} Fixing the scale to the value of one of the VEVs or the GW scale, as is commonly done in the literature, can on the one hand lead to uncertainties coming from large logarithmic contributions and on the other introduces significant differences between different analyses. The use of the RG improved potential alleviates these difficulties.

\section{Summary and conclusions\label{sec:summary}}

The mechanism of spontaneous symmetry breaking and mass generation can be well implemented within the framework of RSB. However, to this end typically new scalar fields are needed. This introduces several complications to the analysis of symmetry breaking pattern within perturbation theory. In the present paper we provide an overview of these issues discussing RSB within the context of arbitrary extensions of the conformal SM with scalar fields. Moreover, we thoroughly analyse RSB in the SU(2)cSM model, discussing in detail the main issues of the general analysis. 

First issue is the hierarchy between different contributions to the effective potential. S.~Coleman and E.~Weinberg showed in their seminal paper~\cite{Coleman:1973} that in the minimal model of RSB, which is scalar QED, the relation between the scalar coupling $\lambda$ and the gauge coupling $g$ is crucial for the occurrence of RSB within the region of validity of perturbation theory, namely the couplings should scale as $\lambda \sim \mathcal{O}(g^4)$, at the scale of the radiatively generated minimum. It is not straightforward to generalise this relation to models with multiple scalar fields, as in those models many scalar couplings are present, as well as many directions in the field space along which a minimum could form.  
In the present paper we discuss various hierarchies between tree-level and one-loop contributions and describe different scenarios in which different methods of studying RSB are valid. We have shown that not only the hierarchy of couplings, but also the hierarchy of the VEVs is relevant for the study of RSB.

Another issue important for the study of RSB is the dependence on the renormalisation scale $\mu$. In classically conformal models with a single scalar field through RSB there arises a single distinguished energy scale --- the scale of the VEV of the scalar field. It is thus natural to perform the computations at this scale, since it also enters the masses of all the particles. In the case of multi-scalar models there is no such a scale, since different scalar fields may acquire VEVs that differ by orders of magnitude. This poses a question at which scale the computations should be performed, since quantities computed from a fixed-loop-order approximation to the effective potential do depend on the renormalisation scale. Moreover, for different methods of studying RSB different scales are most convenient and because of that results obtained with different methods can diverge. Comparing such results with one another as well as with the physical quantities introduces significant inaccuracies. We have demonstrated that this is indeed the case within SU(2)cSM --- performing computations at different scales can lead to significantly different results, demonstrating the weakness of perturbative methods.

An alleviation of the issue of scale dependence of the results, as well as the problem of perturbativity of the loop expansion in the presence of vastly different VEVs of scalar fields can come from RG improvement of the effective potential. In the present work we have implemented within the SU(2)cSM model the method of RG improvement recently developed in our paper~\cite{Chataignier:2018}. We have shown that the use of the RG improved potential indeed substantially reduces the dependence of the results on the scale, in particular it significantly mitigates the running of the scalar fields VEVs. The field-dependent scale used in the procedure of RG improvement lies in between the two scalar VEVs, thus mitigating the SM and the hidden sector contributions to the effective potential. An additional advantage of working with the RG improved potential is that we do not need to fix the scale before we know the numerical values of the VEVs and the hierarchy between them, as is conventionally done by fixing $\mu$ to one of the VEVs --- the scale is selected automatically in an optimal way, which allows to cancel the one-loop correction to the effective potential.

The multi-scalar classically conformal models constitute viable candidates to describe various physical phenomena. They can provide good dark matter candidates from the scalar sector or from new hidden gauge groups. Furthermore, the issue of thermal effects in such models is worth studying. Classically conformal models provide a suitable framework to model electroweak baryogenesis --- since the potential is flat around the origin (there is no negative $\phi^2$ term in the tree-level potential) thermal fluctuations can easily lead to a first order phase transition. Moreover, one can envisage multi-step phase transitions due to non-trivial vacuum structure in the multi-field space. On top of that, the gravitational wave signal from such phase transitions would be interesting to study. These important applications call for a framework of studying RSB that could provide reliable and precise results so we believe that the issues studied in the present work are of considerable significance for future research in this direction.

\section*{Acknowledgments}
B.{\'S}. is grateful to T.~Robens for her explanations of the analysis of the experimental constraints on the mixing angle and to D.~Soko{\l}owska for correspondence on this topic. T.P. and B.{\'S}. acknowledge funding from the D-ITP consortium, a program of the NWO that is funded by the Dutch Ministry of Education, Culture and Science (OCW). This work is part of the research programme of the Foundation for Fundamental Research on Matter (FOM), which is part of the Netherlands Organisation for Scientific Research (NWO). B.{\'S}. acknowledges the support from the National Science Centre, Poland, through the HARMONIA project under contract UMO-2015/18/M/ST2/00518 (2016-2019) and from the Foundation for Polish Science (FNP). L.C. acknowledges the support of Utrecht University (UU), the Netherlands, in the form of the Utrecht Excellence Scholarship (UES) received from September 2015 to June 2017.

\appendix
\section{RG functions\label{app:betas}}
The $\beta$ functions for the scalar couplings $\la$--$\lc$ and the new gauge coupling $g_X$ read~\cite{Khoze:2014, Carone:2013, Hambye:2013}
\begin{displaymath}
\begin{aligned}
\beta_1 &= \frac{1}{8\pi^2}\left[12\lambda_1^2+\lambda_2^2+\frac{\lambda_1}{2}\left(-9g^2-3g^{\prime 2}+12Y_t^2\right)+\frac{3g^4}{8}+\frac{3(g^2+g^{\prime 2})^2}{16}-3Y_t^4\right],\\
\beta_2 &= \frac{1}{8\pi^2}\left[6\lambda_1\lambda_2+2\lambda_2^2+6\lambda_2\lambda_3+\frac{\lambda_2}{4}\left(-9g^2-3g^{\prime 2}+12Y_t^2-9g^2_{X}\right)\right], \\
\beta_3 &= \frac{1}{8\pi^2}\left[\lambda_2^2+12\lambda_3^2-\frac{9}{2}\lambda_3g^2_{X}+\frac{9g^4_{X}}{16}\right],\\
\beta_{X} &= \frac{1}{16\pi^2}\left[-\frac{43}{6}g^3_{X} - \frac{1}{(4\pi)^2}\frac{259}{6}g^5_{X}\right].
\end{aligned}
\end{displaymath}

\bibliography{conformal-bib}

\end{document}